\documentclass[aps,prb,nolongbibliography,notitlepage,twocolumn,superscriptaddress]{revtex4-2}

\usepackage{amsmath}
\usepackage{amssymb}
\usepackage{bm}
\usepackage{bbm}
\usepackage{booktabs}
\usepackage{hyperref}

\usepackage[dvipsnames]{xcolor}
\usepackage{braket}
\usepackage{graphicx}
\usepackage{enumerate}  
\usepackage{CJKutf8}
\def\H{{\mathcal H}}
\def\K{{\mathcal K}}
\def\B{{\mathcal B}}
\def\Q{{\mathcal Q}}
\def\R{{\mathcal R}}

\def\D{{\mathcal D}}
\def\A{{\mathcal A}}
\def\O{{\mathcal O}}

\def\F{{\mathcal F}}

\def\balpha{{\bm{\alpha}}}
\def\bxi{{\bm \xi}}

\def\bk{{\bm k}}
\def\bu{{\bm u}}
\def\bv{{\bm v}}
\def\bq{{\bm q}}
\def\br{{\bm r}}
\def\bz{{\bm z}}
\def\bn{{\bm n}}

\def\bG{{\bm G}}
\def\bp{{\bm p}}
\def\ba{{\bm a}}

\def\bb{{\bm b}}
\def\bsigma{{\bm \sigma}}
\def\bDelta{{\bm \Delta}}
\def\bL{{\bm L}}

\let\oldv\v
\renewcommand{\v}[1]{{\boldsymbol{#1}}}

\begin{document}

\title{Anyon Dispersion in Aharonov-Casher Bands and Implications for Twisted MoTe${}_2$}
\author{Zihan Yan}
\affiliation{Department of Physics, Harvard University, Cambridge, MA 02138, USA}
\author{Qingchen Li}
\affiliation{Department of Physics, Harvard University, Cambridge, MA 02138, USA}
\author{Tomohiro Soejima (\begin{CJK*}{UTF8}{bsmi}副島智大\end{CJK*})}
\affiliation{Department of Physics, Harvard University, Cambridge, MA 02138, USA}
\affiliation{Center for Computational Quantum Physics, Flatiron Institute, New York, New York 10010, USA}
\affiliation{Center for Quantum Phenomena, Department of Physics, New York University, 726 Broadway, New York, NY, 10003, USA}
\author{Eslam Khalaf}
\affiliation{Department of Physics, Harvard University, Cambridge, MA 02138, USA}

\date{\today}
\begin{abstract}
The discovery of fractional quantum anomalous Hall (FQAH) states in two-dimensional heterostructures has opened the door to realizing phases of dispersing anyons. Here, we develop an analytically controlled theory of anyon dispersion in FQAH states realized in ideal or Aharonov-Casher (AC) bands by projecting interactions onto the space of Laughlin quasiholes. Constructing quasihole momentum eigenstates 
allows efficient evaluation of the single quasihole dispersion using Monte Carlo. We find that the quasihole bandwidth grows with increasing quantum-geometry inhomogeneity of the AC band and with increasing interaction screening length. For realistic parameters relevant to the bands of twisted MoTe${}_2$, the quasihole bandwidth is of order 1 meV and increases with increasing displacement field, suggesting that itinerant-anyon physics may play an important role in sufficiently clean samples.
Furthermore, we develop a microscopic Lagrangian framework in terms of a quasihole guiding-center coordinate, which reproduces the momentum-space formula for the dispersion. 
This approach reveals that quasihole dispersion originates from the combined effects of an interaction-generated periodic potential, arising from non-uniform quantum geometry of the single particle bands, and the quasihole \emph{many-body} Berry phase arising from the background magnetic field. The latter endows the guiding-center coordinate with a noncommutative structure, converting the periodic potential into a finite dispersion. Finally, we outline how this framework generalizes to multiple quasiholes, enabling a microscopic theory of charged excitations in FQAH systems that retains only the anyon degrees of freedom.
\end{abstract}
\maketitle

\section{Introduction}
The recent discovery of fractional quantum anomalous Hall (FQAH) states (zero-field fractional Chern insulators) marks a major advance in the study of strongly correlated topological phases. Experimental realizations in twisted MoTe${}_2$ \cite{xu2023observation, zeng2023thermodynamic, cai2023signatures, park2023observation} and rhombohedral pentalayer graphene on hexagonal
boron nitride \cite{Lu2024} have demonstrated that correlated flat-band systems can host fractionalized states analogous to those of the fractional quantum Hall (FQH) effect, despite the absence of an external magnetic field \cite{liuRecentDevelopmentsFractional2022,parameswaranFractionalQuantumHall2013,BergholtzReview2013,neupertFractionalQuantumHall2011,shengFractionalQuantumHall2011,regnaultFractionalChernInsulator2011,qi_generic_2011,parameswaranFractionalChernInsulators2012,wuBlochModelWave2013,kourtisFractionalChernInsulators2014, zhangNearlyFlatChern2019, tarnopolskyOriginMagicAngles2019,wu2019topological,morales-duran2024magic,ledwithFractionalChernInsulator2020a,wang2021exact,ledwith_vortexability_2023,TMDAharonocCasher,repellinFerromagnetismNarrowBands2020, meraEngineeringGeometricallyFlat2021,meraKahlerGeometryChern2021,ledwithFamilyIdealChern2022,lu2025electromagnetic,Tianhong2025,abouelkomsan2023quantum, abouelkomsanParticleHoleDualityEmergent2020a}. A central conceptual distinction between these two settings lies in the nature of translation symmetry. Conventional Landau levels possess Galilean invariance, or continuous magnetic translation symmetry (CMTS), which forbids dispersion of charged excitations. In lattice systems, CMTS is reduced to a discrete subgroup, permitting dispersion of fractional charged quasiparticles. This symmetry reduction opens a qualitatively new regime: anyons may acquire a finite effective mass and become itinerant. This is in sharp contrast to conventional FQH systems where anyons have infinite effective mass and get localized by arbitrarily weak disorder, leading to the famous FQH plateaus physics. Recent theoretical works \cite{shi2025doping, divic2024anyon, nosov2025anyonsuperconductivityplateautransitions, shi2025anyondelocalizationtransitionsdisordered,PICHLERanyonSC,han2025anyon,Pichlerspectral} suggest that itinerant anyons can collectively form a variety of phases, including superconductors, thereby reviving the classic idea of anyon superconductivity \cite{LaughlinAnyonSC, FetterHannaLaughlin, 
laughlin1988, WilczekWittenHalperinAnyonSC,PhysRevLett.63.903,PhysRevB.39.11413,PhysRevB.42.342,PhysRevB.41.240,doi:10.1142/S0217979291001607,PhysRevB.41.11101}. Intriguingly, superconductivity has indeed been observed in the vicinity of FQAH states in twisted MoTe${}_2$ \cite{MoTe2SC}, and several phenomenological features of these observations appear consistent with the framework of anyon superconductivity \cite{shi2025doping, nosov2025anyonsuperconductivityplateautransitions, shi2025anyondelocalizationtransitionsdisordered,PICHLERanyonSC}.

While anyon dispersion is allowed by symmetry, there is currently no scalable and analytically controlled method to compute it and clarify its microscopic origin. Some recent works \cite{bernevig2025ED, WuAnyon} used exact diagonalization (ED) to compute anyon dispersion, but these approaches are limited to small system sizes, which can limit their accuracy in the presence of sizable variations in the underlying quantum geometry\cite{shi2025effects}. Furthermore, the exponential scaling of ED makes it inadequate to address multi-anyon physics, including questions concerning multi-anyon bound states \cite{JainMolecularAnyons, BoYangAnyonClusters}, collective dispersions, and emergent phases of itinerant anyons. A controlled theoretical framework capable of accessing these properties in an efficient way is therefore needed.

Here, we develop an analytically-controlled scalable approach to compute anyon dispersion. Our approach exploits the ideal quantum geometry identified in the flat bands of twisted MoTe$_2$ and related systems. Under reasonable approximations, these bands can be mapped to an Aharonov–Casher (AC) Hamiltonian \cite{aharonov1979ground, morales-duran2024magic, TMDAharonocCasher}, whose flat bands are equivalent to the lowest Landau level (LLL) in a nonuniform magnetic field \cite{ledwithFractionalChernInsulator2020a, wang2021exact, ledwith_vortexability_2023, meraEngineeringGeometricallyFlat2021, meraKahlerGeometryChern2021, meraKahlerGeometryChern2021, meraRelatingTopologyDirac2021}. Hence, these bands retain much of the analytic structure of conventional LLL physics, while incorporating CMTS-breaking through non-uniform quantum geometry. This mapping enables the construction of exact Laughlin-like ground states for short-range interactions and their associated quasihole excitations \cite{ledwithFractionalChernInsulator2020a, wang2021exact, ledwith_vortexability_2023}. For interactions of the form $\alpha \hat V_{\rm TK} + \hat V$, where $\alpha$ is large and $\hat V_{\rm TK}$ is the Trugman-Kivelson pseudopotential \cite{trugmanExactResultsFractional1985a} at the filling fraction of interest \footnote{This form arises naturally when expanding any short range interaction in terms of pseudopotentials}, the space of quasiholes is defined by the zero modes of $\hat V_{\rm TK}$. The Hamiltonian in the space of quasiholes is obtained by projecting $\hat V$ onto that space. For a single quasihole, the dispersion is then simply obtained as $\epsilon_{\v{k}} = \langle \psi_{\v{k}}|\hat V|\psi_{\v{k}} \rangle$, where $|\psi_{\v{k}} \rangle$ are momentum quasihole wavefunctions, i.e. they are eigenstates of the discrete magnetic translation operator which are exact zero modes of $\hat V_{\rm TK}$. We construct the wavefunctions  $|\psi_{\v{k}} \rangle$ explicitly, enabling us to compute the dispersion efficiently using Monte Carlo sampling of the corresponding high-dimensional integrals. 

This approach parallels a similar approach introduced recently by some of us which enabled exact calculation of the quasihole dispersion in a toy model with non-uniform magnetic field on the sphere \cite{schleith2025anyon}. The current setup differs in certain subtle aspects associated with the properties of magnetic translation symmetry on the torus geometry and in the fact that we have to use Monte Carlo to numerically evaluate the dispersion.

\begin{figure*}
    \centering
    \includegraphics[width=\linewidth]{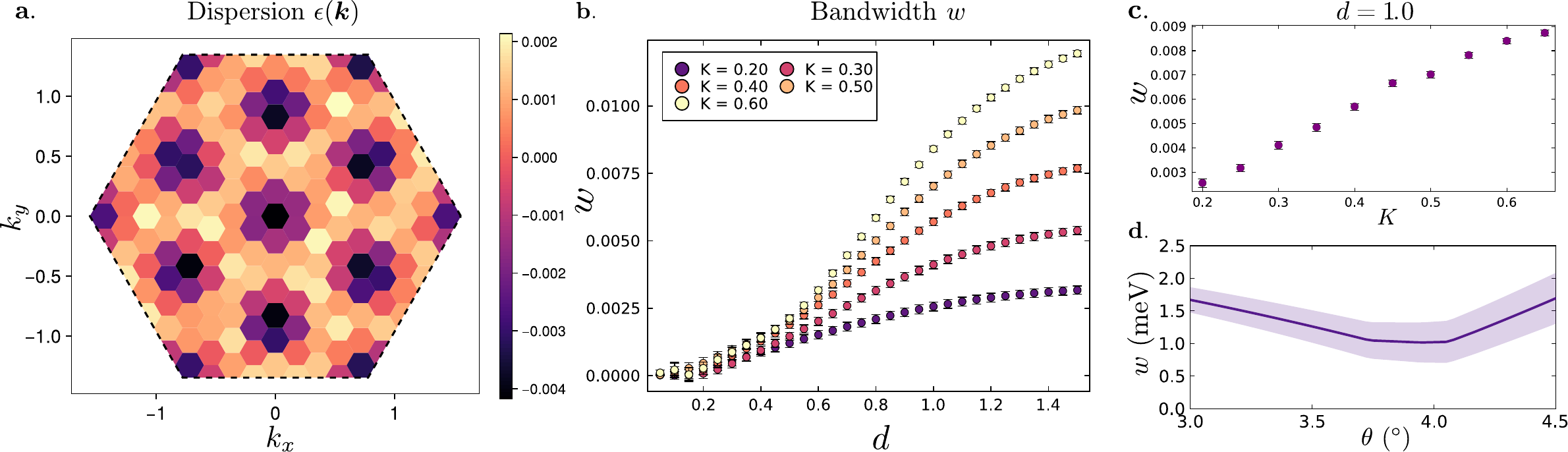}
    \caption{Summary of the numerical results for Laughlin quasiholes in first harmonic approximated AC bands and twisted MoTe${}_2$ (tMoTe${}_2$) AC bands with realistic parameters. \textbf{a}-\textbf{c}: The dispersion of quasihole wavefunction $\psi^{AC}_{\bk}(\{\bz_i\})$ defined using Eq.~(\ref{PsiAC}) and Eq.~(\ref{eq:momentum-state}), in an AC band with first harmonic approximation, $Q(z)=-2K\sum_{\bb_i}\cos(\bb_i\cdot\bz)$ where $\bb_i,i=1,2,3$ are the reciprocal lattice vectors of $\Lambda^{*}$ with $\bb_3=-\bb_1-\bb_2$, under screened Coulomb interaction Eq.~(\ref{V:sCoul-real}), measured in natural unit pseudopotential $V_1 = \int \frac{d^2 \bq}{(2\pi)^2} V(\bq) e^{-\bq^2}L_1(\bq^2)$. $K$ and $d$ control the non-uniformity and gate distance respectively. $\ell_B$ is set to 1. \textbf{a}. Dispersion plotted in the original unfolded Brillouin Zone for $N_e=56,\ K=0.4,\ d=1.0$, the $q^2=9$-fold degeneracy can be observed clearly, with minor deviations due to finite-size effects and Monte Carlo (MC) noise. Average energy has been subtracted. \textbf{b}. Dependence of the quasihole bandwidth $w$ for $N_e=16$ system on different $K$ and $d$. In the small-$d$ regime, the extracted bandwidth is numerically more delicate and should not be overinterpreted. \textbf{c}. Bandwidth as a function of $K$ for $d=1.0$: $w$ grows linearly with $K$ in the small $K$ regime and shows a weaker, sublinear growth at large $K$. Error bars are included in panels (b) and (c), but in many cases they are comparable to the plotting symbols. \textbf{d}. AC quasiholes bandwidth dependence of the twist angle $\theta$ in tMoTe${}_2$ with gate screened Coulomb interaction in the moir\'e Brillouin Zone (mBZ) for $N_e=40,\ d=10 \ \ell_B\approx 20\ \text{nm}$. The twist angle dependence is obtained from a single MC dataset with $\theta$ controlling the length scale and thus the corresponding energy scale. Shades around the curve represent error bars. }
    \label{fig:dispersion-dependence}
\end{figure*}

Applying this method, we compute quasihole dispersion in AC bands as a function of quantum-geometry inhomogeneity and interaction screening length for system sizes up to $N_\Phi = 169$ (13 $\times$ 13), as shown in FIG.~\ref{fig:dispersion-dependence} panels (a)-(c). We find that the quasihole bandwidth increases with both the Berry curvature inhomogeneity and the screening length. The energies are measured in units of the $V_1$ component of the interaction which controls the gap to other excitations in the system. We see that, for the whole range of interaction and quantum-geometry inhomogeneity, the dispersion is a small fraction of this gap which justifies our approach for realistic interactions (without the need to introduce an interaction with the special form $\alpha \hat V_{\rm TK} + \hat V$ with $\alpha \gg 1$). Using realistic parameters relevant to the experimentally observed $\nu = 2/3$ FQAH state in MoTe${}_2$, we obtain a quasihole bandwidth on the order of 1 meV, as shown in FIG.~\ref{fig:dispersion-dependence} panel d. This suggests anyon dispersion may play an important role in the low-energy physics for sufficiently clean samples when the density of doped quasiholes is $\nu_{\rm QH} \gtrsim m_{\rm QH}/\tau_{\rm QH}$ where $m_{\rm QH}$ and $\tau_{\rm QH}$ denote respectively the effective mass and the disorder mean-free time of the quasihole \cite{nosov2025anyonsuperconductivityplateautransitions, shi2025anyondelocalizationtransitionsdisordered}.

Finally, we show that our dispersion formula emerges naturally from a coherent-state path-integral formulation. In this description, a quasihole in a 
$\nu = 1/q$ Laughlin state experiences a periodic potential arising from interactions and quantum-geometry inhomogeneity, as well as an effective magnetic field of average strength $1/q$. This perspective provides a clear physical picture for the origin of quasihole dispersion: the breaking of CMTS by quantum-geometry variations produces a periodic potential, while the many-body quasihole Berry curvature defines a noncommutative quasihole guiding-center coordinate, turning this potential into an effective kinetic energy. At the end, we outline how our approach can enable us to write a microscopic Lagrangian for quasiholes in terms of a \emph{many-body} guiding center coordinate. Such theory provides a controlled minimal description for the quasihole dynamics since it only contains quasihole degrees of freedom. Furthermore, this description contains information about both long distance universal physics as well as short distance physics (binding, dispersion, etc). We briefly discuss how we can use this framework to obtain a projected Hamiltonian description for the space of quasiholes using the formalism of geometric quantization. We leave further analysis of this theory in the multiple quasihole case to follow-up works.

\section{Origin of anyon dispersion: qualitative picture}
\label{sec:origin-quali}
Before discussing the technical details for the computation of anyon dispersion, it is instructive to first consider its physical origin. Anyon dispersion is forbidden in systems with Galilean invariance (or continuous magnetic translation symmetry). Hence, conventional fractional quantum Hall states realized in a uniform magnetic field always have dispersionless anyons. Once Galilean invariance is broken --- either by a periodic potential or by a periodic modulation of the magnetic field --- anyon dispersion becomes symmetry-allowed. Our goal in the following is to identify the physical mechanism that generates such dispersion.

Consider, for example, the case of a uniform-field lowest Landau level (LLL) subject to a periodic potential with period $L_M$ and take a Laughlin quasihole for definiteness. Since the quasihole is a charged object with a spatial extent of order $\ell_B$, its energy depends on its position relative to the potential. For periods much larger than the magnetic length, $L_M \gg \ell_B$, the effective potential felt by the quasihole is essentially that of a point particle with fractional charge. In the opposite limit, $L_M \ll \ell_B$, the quasihole charge is spread over many periods of the potential, and it therefore experiences a much weaker effective potential. The physically relevant regime for flat Chern bands lies between these two limits: the unit cell of the periodic potential encloses one flux quantum, implying $L_M \sim \ell_B$. In this case, the finite size of the quasihole partially smears the periodic potential, suppressing its higher harmonics, yielding a smoother potential with roughly similar magnitude.

A similar situation arises when a periodic modulation of the magnetic field is introduced, which describes what happens in a flat Chern band in the ideal or Aharonov-Casher limit \cite{ledwithFractionalChernInsulator2020a, wang2021exact, ledwith_vortexability_2023, TMDAharonocCasher}. In this case, we get an \emph{interaction-generated} periodic potential for the quasiholes, which can be attributed to a spatially modulated magnetic length. We expect similar physics to the case of a periodic potential: slowly varying modulations are inherited by the quasihole (as shown in Ref.~\cite{schleith2025anyon}), while rapidly oscillating modulations are suppressed. We explicitly compute such modulation for the physically relevant limit in Sec.~\ref{Sec:tMoTe2} (see FIG.~\ref{fig:tMoTe2-nonuniform-field}). The central question then becomes: how does adding a periodic potential to a flat quasihole band generate a dispersion?

To understand the origin of this effect, note that a single quasihole in a $\nu = 1/q$ Laughlin state experiences a background magnetic field equal to $1/q$ of that felt by the electrons. We can therefore define quasihole guiding-center coordinates $\hat{\xi} = \hat{X} + i \hat{Y}$, which satisfy $[\hat{\xi}, \hat{\xi}^\dagger] = 2 q \ell_B^2$, or equivalently $[\hat{X}, \hat{Y}] = i q \ell_B^2$. The noncommutation of $\hat{X}$ and $\hat{Y}$ mixes real and momentum space, since $\hat{P}_x \propto \hat{Y}$ and $\hat{P}_y \propto \hat{X}$. As a result, a purely potential term can generate kinetic energy and give rise to dispersion.

To see this explicitly, consider a quasihole at $\v{\xi}$ subject to a periodic potential $V(\v{\xi}) := \langle \psi_{\v{\xi}} | \hat{H} | \psi_{\v{\xi}} \rangle$, obtained by projecting a microscopic Hamiltonian $\hat{H}$ onto the space of quasihole wave functions. This potential can be expanded in reciprocal lattice vectors as
\begin{equation}
    V(\v{\xi}) = \sum_{\v{G}} V_{\v{G}} e^{i \v{G} \cdot \v{\xi}} .
    \label{QHPotential}
\end{equation}
We will discuss the structure of this reciprocal lattice in detail later. Our earlier analysis suggests that the Fourier coefficients $V_{\v{G}}$ decay for $|\v{G}| \ell_B \gtrsim 1$, since the quasihole charge is smeared over a region of size $\ell_B$. The quantum Hamiltonian acting on the space of quasiholes is obtained from Eq.~(\ref{QHPotential}) by replacing $\v{\xi}$ with the corresponding guiding-center operators. This is done by normal ordering the expression such that $\bar{\xi}$ appears to the left of $\xi$, and then replacing them by the corresponding operators $\bar{\xi} \mapsto \hat{\xi}^\dagger$ and $\xi \mapsto \hat{\xi}$ with $[\hat{\xi}, \hat{\xi}^\dagger] = 2 q \ell_B^2$. This procedure yields
\begin{equation}
    \hat{V} = \sum_{\v{G}} V_{\v{G}} e^{\frac{i}{2} G \hat{\xi}^\dagger} e^{\frac{i}{2} \bar{G} \hat{\xi}}
    = \sum_{\v{G}} V_{\v{G}} e^{\frac{q \ell_B^2}{4} \v{G}^2} e^{\frac{i}{2} (G \hat{\xi}^\dagger + \bar{G} \hat{\xi})} .
    \label{q-QHPotential}
\end{equation}
The operators $e^{\frac{i}{2}(G \hat{\xi}^\dagger + \bar{G} \hat{\xi})}$ correspond to mutually commuting lattice magnetic translations, whose common eigenstates are quasihole Bloch states with eigenvalues $\eta_{\v{G}} e^{i \v{G} \wedge \v{k}}$, where $\eta_{\v{G}} = +1$ if $\v{G}/2$ is a reciprocal lattice vector and $-1$ otherwise, leading to the dispersion
\[
    \epsilon_{\v{k}} = \sum_{\v{G}} \eta_{\v{G}} V_{\v{G}} e^{\frac{q \ell_B^2}{4} \v{G}^2} e^{i \v{G} \wedge \v{k}} .
\]
The growing Gaussian factor in the expression above partially cancels the decay of $V_{\v{G}}$ at large $|\v{G}|$, as we will show in detail. Consequently, the dispersion can be sensitive to higher harmonics of the periodic potential despite the smearing from Landau-level projection. 

The analysis above identifies the microscopic origin of anyon dispersion in flat Chern bands as arising from two distinct effects. First, the nonuniformity of the \emph{single-particle} quantum geometry generates a periodic potential for the quasihole. This effect can also be generated by any perturbation that breaks CMTS such as a periodic potential. Second, the \emph{many-body} quantum geometry of the quasihole states enforces the noncommutation of the projected position operators, which converts the periodic potential into a kinetic (dispersive) term. Although the quasihole quantum geometry (i.e. the real space Berry curvature felt by the quasihole) is also expected to be nonuniform, this is not required for the mechanism discussed here. In the analysis above, we assumed a uniform magnetic field for the quasiholes when specifying their guiding-center commutation relation. In Sec.~\ref{Sec:qhberry}, we will explicitly compute the real space Berry curvature experienced by the quasihole and show that it inherits the nonuniformity of the single-particle quantum geometry.

While this analysis provides a clear physical picture for the origin of quasihole dispersion, it also suggests that computing the dispersion accurately from the quasihole potential $V(\v{\xi})$ can be challenging. The reason is that the Fourier components $V_{\bG}$ of the real-space potential are strongly suppressed at large $|\bG|$, reflecting the finite spatial extent of the quasihole. However, the dispersion is obtained from these coefficients after multiplication by the growing Gaussian factor $e^{q \ell_B^2 \bG^2/4}$ in Eq.~(\ref{q-QHPotential}), which partially compensates this suppression. As a result, extracting the harmonics relevant for the dispersion from a real-space evaluation of $V(\xi)$ is numerically difficult, as it requires inverting this suppression. A similar behavior arises in the calculation of the LLL guiding-center structure factor.
Although the guiding-center structure factor itself does not decay at large momenta, Monte Carlo naturally accesses the physical density structure factor, which is related to the guiding-center structure factor by a Gaussian. Recovering the guiding-center structure factor therefore requires undoing the Gaussian suppression, which amplifies numerical noise at large momenta \cite{WangLatticeMC}. To overcome this challenge, we evaluate the dispersion directly in momentum-space by Monte Carlo sampling. This approach includes an additional cost of $\sim N_\Phi$ per Monte Carlo step. Furthermore, since anyon dispersion is an $O(1)$ number on top of the extensive $O(N_e)$ total energy, we need to run the Monte Carlo for $O(N_e^2)$ extra steps as explained in Sec.~\ref{Sec:numerics}. This makes it harder to go to the very large system sizes typically employed in Monte Carlo for Laughlin correlators ($\sim 500 $ electrons) but, as we will show, it is still possible to achieve system sizes $N_e \sim 60$ ($N_\Phi \sim 170$) which are significantly larger than what can be achieved in ED. Importantly, due to the polynomial scaling of the method, going to large systems does not face the same limitations as in methods with exponential scaling.

\section{Magnetic translation and momentum space quasihole wavefunctions}

\subsection{Quasiholes and magnetic translation symmetry: general discussion}
\label{sec:MagneticTranslation}
We first review the magnetic translation algebra on the torus, as well as its action on quasihole wavefunctions. Throughout the manuscript, we will use boldface letters to indicate 2D vectors and the corresponding unbolded letters to indicate the corresponding complex variables. For instance, for the vector $\v{z} = (z_x, z_y)$, $z = z_x + i z_y$. We consider a torus spanned by $\v{L}_1, \v{L}_2$ and denote its area by $A = |\v{L}_1 \times \v{L}_2|$ with $N_\Phi$ flux quanta such that $A = 2\pi \ell_B^2 N_\Phi$. In the following, we will set $\ell_B = 1$.

The single particle magnetic translation algebra on the torus is given by $t_\bu t_\bv = e^{i \bu \wedge \bv} t_\bv t_\bu$, where $\v{u} \wedge \v{v} = u_x v_y - u_y v_x$ and $\wedge \v{v} = (v_y, -v_x)$. For $N_e$ electrons, we can also define the many-body translation operator $T_\bu := \prod_{i=1}^{N_e} t^i_\bu$ which satisfies $T_{\v{u}} T_{\v{v}} = e^{i N_e \v{u} \wedge \v{v}} T_{\v{v}} T_{\v{u}}$. We define the flux sector of the torus by the action of magnetic translations $t_{\bL_j}$ on the single particle states: $t_{\v{L}_j} \ket{\Psi} = e^{i\Phi_j} \ket{\Psi}$. The magnetic translation algebra implies that infinitesimal translations alter the flux sector. Flux sector-preserving translations are therefore discrete, and they are generated by $\v{d}_i = \frac{1}{N_\Phi} \v{L}_i$. There are $N_\Phi^2$ distinct magnetic translations.

Let us define the Bravais lattice vectors $\v{a}_1 = \v{L_1}/N_1 = N_2 \v{d}_1, \v{a}_2 = \v{L}_2 /N_2 = \v{d}_2 N_1$ with $N_1, N_2 \in \mathbb{Z}$. We take $N_1 N_2 = N_\Phi$ such that the unit cell encloses a unit flux i.e. $|\v{a}_1 \times \v{a}_2| = 2\pi$. We will denote such lattice by $\Lambda = \{n \v{a}_1 + m \v{a}_2\}$. As such, these two translations commute: $t_{\v{a}_1} t_{\v{a}_2} = t_{\v{a}_2} t_{\v{a}_1}$ and generate a set of $N_\Phi$ commuting magnetic translations.
We can thus define the momentum via the eigenvalue of these translation operators $t_{\v{a}_j}\ket{\Psi} = - e^{i\v{k} \cdot \v{a}_j} \ket{\Psi}$. Let $\v{b}_1 = \wedge \v{a}_2,\ \v{b}_2 = -\wedge \v{a}_1$  denote the reciprocal lattice vectors. We will denote the reciprocal lattice by $\Lambda^* = \{n \v{b}_1 + m \v{b}_2\}$. The Brillouin zone is then given by $ \v{k} \in \{k_1 \v{b}_1 + k_2 \v{b}_2 | k_1, k_2 \in [-\frac{1}{2}, \frac{1}{2}) \}$. Specifically, the allowed values of momenta are given by $k_i =   \frac{m_i}{N_i}+\frac{\Phi_i}{2\pi N_i} \in [-\frac{1}{2}, \frac{1}{2})$ with $m_i \in \mathbb{Z}$.

The Laughlin state at $N_\Phi = q N_e$ is $q$-fold degenerate. At this filling, many-body magnetic translations satisfy $T_{\v{d}_1} T_{\v{d}_2} = e^{iN_e \v{d}_1 \wedge \v{d}_2} T_{\v{d}_2} T_{\v{d}_1} = e^{\frac{2\pi i}{q}} T_{\v{d}_2} T_{\v{d}_1}$. This algebra requires a $q$-dimensional representation, corresponding to different topological sectors. For example, for $q = 3$, we can choose a basis of the degenerate eigenstates such that the magnetic translations act as
\begin{equation}
    T_{\v{d}_1} = \begin{pmatrix}
        1 & 0 & 0 \\
        0 & e^{i\frac{2\pi}{3}} & 0 \\
        0 & 0 & e^{i\frac{4\pi}{3}}
    \end{pmatrix}, \quad T_{\v{d}_2} = \begin{pmatrix}
        0 & 1 & 0 \\
        1 & 0 & 0 \\
        0 & 0 & 1 \\
    \end{pmatrix}.
    \label{eq:Laughlin_permutation}
\end{equation}
We see that the eigenvalue of $T_{\v{d}_1}$ characterizes topological sectors, and $T_{\v{d}_2}$ toggles between different topological sectors. For later convenience, we define $\v{\delta}_1 = \v{d}_1$ and $ \v{\delta}_2 = q \v{d}_2$, such that $T_{ \v{\delta}_1} T_{\v{\delta}_2} = T_{ \v{\delta}_2} T_{\v{\delta}_1}$. This generates a set of $N_\Phi^2/q = N_\Phi N_e$ commuting magnetic translations, such that we can take Laughlin states to be simultaneous eigenstates of these translations.

For a Laughlin quasihole at $N_\Phi = q N_e + 1$, the minimal translations obey the algebra $T_{\v{d}_1} T_{\v{d}_2} = e^{2\pi i \frac{N_e}{qN_e + 1}} T_{\v{d}_2} T_{\v{d}_1} =  e^{\frac{2\pi i}{q}} e^{-\frac{2\pi i}{q N_\Phi}} T_{\v{d}_2} T_{\v{d}_1}$. Since $N_e$ and $qN_e + 1$ are coprime, the degeneracy of the quasihole states is a multiple of $N_\Phi$.
Let us first consider the ``Landau gauge'' realization of this algebra, in analogy with Eq.~(\ref{eq:Laughlin_permutation}). The action of the magnetic translation can then be written as $T_{\v{d}_1} \ket{\Psi_j} = e^{i\frac{2\pi j N_e}{qN_e +1 }}\ket{\Psi_j}, T_{\v{d}_2}\ket{\Psi_j} = \ket{\Psi_{j+1}}$. $T_{\v{d}_2}$ changes the eigenvalue of $T_{\v{d}_1}$ by $e^{i\frac{2\pi N_e}{qN_e +1}} \approx e^{i\frac{2\pi}{q}}$ while $T_{q\v{d}_2}$ changes the eigenvalue of $T_{\v{d}_1}$ by $e^{i\frac{2\pi qN_e}{qN_e +1}} \approx 1$ in the limit of $N_e \to \infty$. Heuristically, we see $T_{\v{d}_2}$ permutes between topological sectors while $T_{q\v{d}_2}$ approximately keeps it invariant.

In terms of $T_{\v{\delta}_1}$ and $T_{\v{\delta}_2}$ defined earlier, we have $T_{\v{\delta}_1} T_{\v{\delta}_2} =  e^{-\frac{2\pi i}{N_\Phi}} T_{\v{\delta}_2} T_{\v{\delta}_1}$. This motivates us to define $\v{\alpha}_1 = \v{a}_1$ and $ \v{\alpha}_2 = q \v{a}_2$, such that $T_{ \v{\alpha}_1}$ and $T_{\v{\alpha}_2}$ leave the topological sector invariant. These define a new lattice of commuting translations, which we will denote by $\Lambda_q$ with the corresponding reciprocal lattice denoted by $\Lambda^*_q = \{n \v{\beta}_1 + m \v{\beta}_2\}$ where $\v{\beta}_1 = \wedge \v{\alpha}_2/q = \v{b}_1$ and $\v{\beta}_2 = -\wedge \v{\alpha}_1/q = \v{b}_2/q$.
The Brillouin zone is now smaller: $\v{\kappa} \in \{\kappa_1 \v{b}_1 + \kappa_2 \v{b}_2 | \kappa_1 \in [-\frac{1}{2}, \frac{1}{2}), \kappa_2 \in [-\frac{1}{2q}, \frac{1}{2q}) \}$.
Since $N_1$ and $q$ are coprime, $\v{\alpha}_1$ and $\v{\alpha}_2$ generate a set of $N_\Phi$ translations, just like $\v{a}_1$ and $\v{a}_2$, which means that the reduced BZ has \emph{the same number of points} despite being a $q$-folded version of the original BZ. The reason is that no two folded points coincide for any finite system. The allowed discrete values of momenta are given explicitly by $\kappa_1 =   \frac{m}{N_1}+\frac{N_e\Phi_1}{2\pi N_1} + \frac 12, m \in \mathbb{Z}, \kappa_1 \in [-\frac{1}{2}, \frac{1}{2}), \kappa_2 =   \frac{m}{qN_2}+\frac{N_e\Phi_2}{2\pi N_2} + \frac {1}{2q}, m \in \mathbb{Z}, \kappa_2 \in [-\frac{1}{2q}, \frac{1}{2q})$. 

 The previous discussion implies there are two equivalent ways of labeling the states: one as $q$ flavors in a reduced BZ with the same density of points as the original BZ, i.e. with a total of $\sim N_\Phi/q$ points, and the other is as one flavor in a reduced BZ with a density of points that is $q$ times higher due to the enlargement of the system, i.e. with a total $N_\Phi$ points. In either case, the total number of quasihole states is $N_\Phi$ which is consistent with the count from the thin torus limit \cite{regnaultFractionalChernInsulator2011,bernevigEmergentManybodyTranslational2012}.

\begin{figure}[h]
    \centering
    \includegraphics[width=0.9\linewidth]{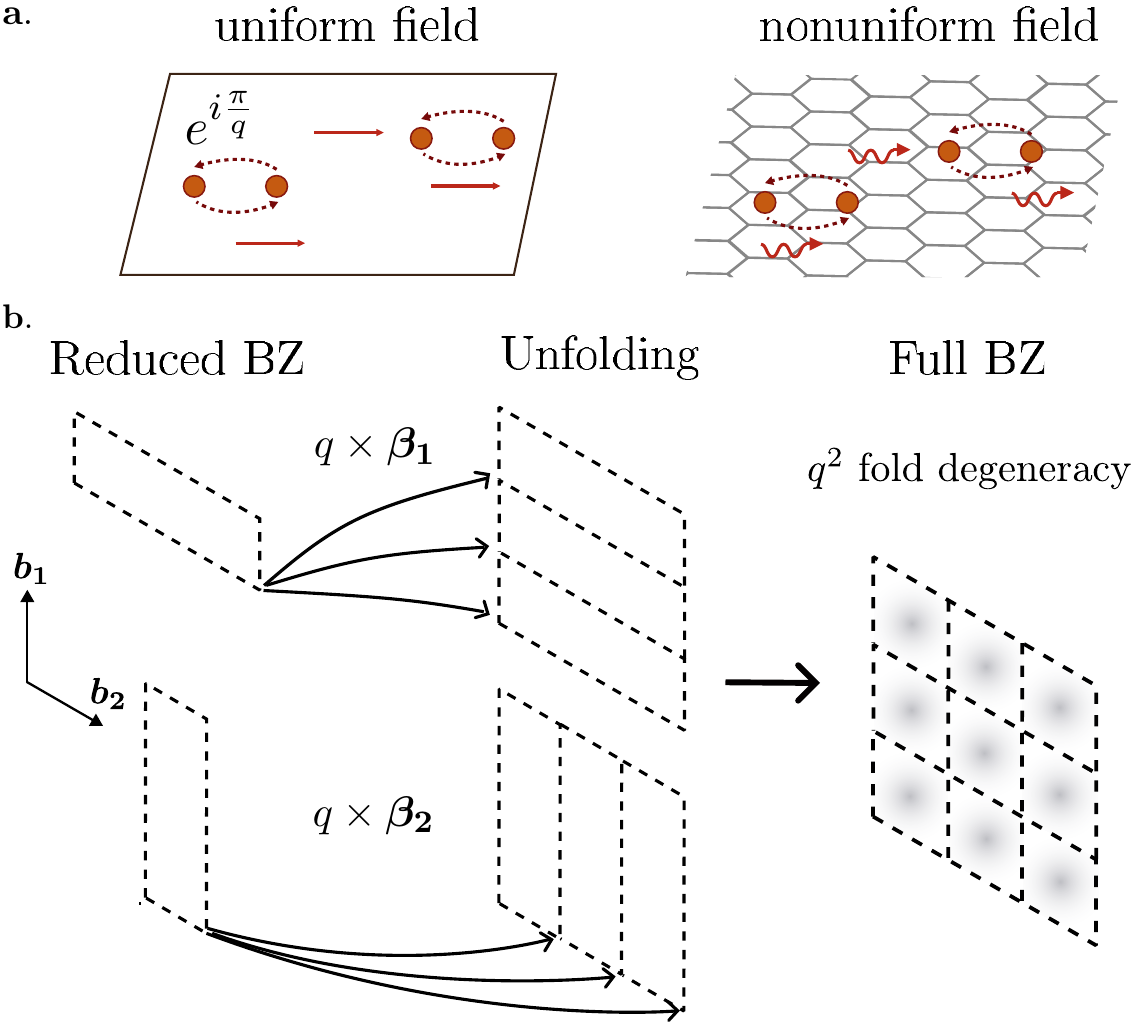}
    \caption{Illustration of quasiholes obtaining dispersion in nonuniform field. \textbf{a}. Quasiholes with exchange phase $e^{i\frac{\pi}{q}}$ moving in uniform field with continuous magnetic translation symmetry and moving in non-uniform magnetic field with discrete magnetic translation symmetry. \textbf{b}. Emergence of the $q^2$ fold degeneracy of quasiholes, taking $q=3$ as an example: the reduced BZ has folding direction either $\bb_1$ or $\bb_2$, unfolding procedure maps adjacent states in the reduced BZ to $q$ sectors in that direction and thus gives rise to $q$ fold degeneracy along the folding direction. Since the folding direction is chosen to be either $\bb_1$ or $\bb_2$, the physical system has to be $q^2$ fold degenerate to be consistent with the unfolding procedure. 
    }
    \label{fig::dispersion-dependence}
\end{figure}

If the Hamiltonian of the system is invariant under the full magnetic translation group, then the energy of a quasihole cannot depend on its momentum since we can map quasiholes with different momenta to each other using magnetic translation $T_{\v{\delta}_j} \ket{\Psi_{\v{\kappa}}} \propto \ket{\Psi_{\v{\kappa}- \wedge \frac{1}{q}\v{\delta}_j} }$. On the other hand, if the magnetic translation group is broken down to its discrete subgroup generated by $\v{a}_j$, then the energy could be a function of $\v{\kappa}$. Assuming the energy is continuous in the folded BZ in the limit $N_e \rightarrow \infty$ implies a $q$-fold degeneracy along the folding direction since $q$ nearby points in the reduced BZ map to equally spaced points in the full BZ. Since the folding direction can be chosen as either of the lattice directions, this implies $q^2$-fold degeneracy in the full BZ \cite{shi2025doping,bernevig2025ED}. This is illustrated schematically in FIG.~\ref{fig::dispersion-dependence}.

\subsection{Landau levels on the torus and Aharonov-Casher bands: a brief review}

In this section, we will review recent developments in understanding bands with ideal quantum geometry, where analytical pseudopotentials remain applicable. There are three similar but distinct conditions discussed in the literature for bands which share the same holomorphic structure as the LLL and thus allow for writing exact Laughlin states: (i) \emph{ideal bands} which are bands satisfying the trace condition equating the trace of quantum metric  to the absolute value of the Berry curvature at every momentum \cite{roy2014band, ledwithFractionalChernInsulator2020a, wang2021exact, meraKahlerGeometryChern2021, meraRelatingTopologyDirac2021, meraEngineeringGeometricallyFlat2021}. This condition is also equivalent to the statement that a wavefunction in the band remains in the band upon multiplying by the vortex function $z = x \pm i y$. (ii) \emph{vortexable bands} are bands whose wavefunctions remain in the band after multiplication by a vortex function $z(x,y)$ which has a more general form than $x \pm i y$ \cite{ledwith_vortexability_2023}. Thus, every ideal band is vortexable but the opposite is not true. (iii) \emph{Aharonov-Casher} (AC) bands are bands that can be written as the zero modes (or zeroth Landau level) of an Aharonov-Casher Hamiltonian \cite{aharonov1979ground} or a Dirac particle in a periodic magnetic field, up to an overall momentum-independent spinor. As shown in Refs.~\cite{ledwithFractionalChernInsulator2020a, wang2021exact, ledwith2021strong, meraEngineeringGeometricallyFlat2021,  meraKahlerGeometryChern2021}, every ideal Chern 1 band admits an AC Hamiltonian with a periodic magnetic field, with $2\pi$ flux per unit cell, that can be constructed from the parent band's wavefunctions. For higher Chern numbers, ideal bands have more complicated structure \cite{dong2023manybody, wang2023origin}. Since our main experimental motivation is MoTe${}_2$, whose bands were shown to satisfy the ideal condition to a good approximation \cite{morales-duran2024magic, TMDAharonocCasher, dong2023composite} and where an explicit mapping to the AC problem was derived, we will restrict ourselves to AC bands \cite{morales-duran2024magic, TMDAharonocCasher}.

\subsubsection{Uniform magnetic field: Landau levels on the torus}
We begin by briefly reviewing the construction of LLL wavefunctions on the torus in uniform magnetic field. While the analysis below is mainly concerned with the parent Bloch band associated with the lattice whose unit vectors are $\v{a}_{1,2}$ defined in the previous section, we will consider below the more general case of any lattice $\Pi$ with lattice vectors $\v{R}_1$ and $\v{R}_2$ and define $\ell_\Pi$ such that $|\v{R}_1 \times \v{R}_2| = 2\pi \ell_\Pi^2$. This will be useful later when we discuss the modified unit cell for quasiholes.

Consider a set of commuting magnetic translations $t_\v{R}$ where $\v{R} = n \v{R}_1 + m \v{R}_2$. Define the magnetic Bloch states as common eigenstates of all $t_\v{R}$ with
\begin{equation}
    t_\v{R} |\phi_\v{k} \rangle = \eta_\v{R} e^{i \v{k} \cdot \v{R}} |\phi_\v{k} \rangle, \quad \eta_{n \v{R}_1 + m \v{R}_2} = (-1)^{n + m + n m}
\end{equation}
Here, the factor $\eta_\v{R}$ which is $+1$ if $\v{R}/2$ is a lattice vector and $-1$ otherwise is introduced for later convenience. Clearly, it depends on the lattice $\Pi$ but we will leave this dependence implicit.

We can choose the action of single particle magnetic translation on position eigenstates to be $t_\v{u} |\br \rangle = e^{-\frac{i}{2 \ell_\Pi^2} \v{u} \wedge \br} |\br - \v{u} \rangle$ such that $t_{\bu} \phi(\br) :=  \langle \br|t_{\bu}|\phi \rangle = e^{-\frac{i}{2 \ell_\Pi^2} \v{u} \wedge \br} \phi(\br + \v{u})$. With this choice, we can write the magnetic Bloch states as
\begin{align}
    \phi^\Pi_\v{k}(\v{z}) &:= \langle \v{z}| \phi \rangle \nonumber \\ &= C_\Pi \sigma(z + i k \ell_\Pi^2|R_1, R_2) e^{\frac{i}{2} z \bar k} e^{-\frac{1}{4 \ell_\Pi^2}|z|^2} e^{-\frac{\ell_\Pi^2}{4 }|k|^2}
    \label{MagneticBloch}
\end{align}
 Here, $\sigma(z|R_1, R_2)$ is the modified Weierstrass sigma function \cite{HaldaneModifiedWeierstrass, wang2021exact} satisfying $\sigma(z + R|R_1, R_2) = \eta_{\v{R}} e^{\frac{\bar R}{2}(z + R/2)} \sigma(z|R_1, R_2)$ and $C_\Pi$ is a normalization constant chosen such that $\int_{\rm UC} d^2 \v{z} |\phi^\Pi_\v{k}(\v{z})|^2 = 1$, which we will compute explicitly later. We will usually drop the superscript/subscript $\Pi$ and leave it implicit and will also drop the lattice indices $R_{1,2}$ from the Weierstrass function.
 
 The wavefunctions $\phi_\bk(\v{z})$ are clearly orthonormal $\langle \phi_\bk|\phi_{\bk'} \rangle = \int d^2 \v{z} \phi^*_\bk(\v{z}) \phi_{\bk'}(\v{z}) = N_\Phi \delta_{\bk,\bk'}$, where the integral here is over the entire area $A$ not just the unit cell. They also satisfy
\begin{equation}
    \frac{1}{N_\Phi} \sum_{\bk \in {\rm BZ}} \phi_\v{k}(\v{z}) \bar \phi_\v{k}(\v{w}) = \frac{1}{2\pi \ell^2} e^{-\frac{1}{4 \ell^2}(|z|^2 + |w|^2 - 2 w^* z)}
\end{equation}

\subsubsection{Periodically-modulated magnetic field: Aharonov-Casher bands}
For AC bands, we consider the LLL in a field which has the form
\begin{equation}
    B^{\rm AC}(\v{z}) = \ell^{-2} + \delta B(\v{z})
\end{equation}
where $\delta B(\v{z})$ is a periodic field over the lattice $\Pi$, i.e. $\delta B(\v{z} + \v{R}) = \delta B(\v{z})$, whose average over the unit cell is zero. We define the magnetic potential $Q(\v{z})$ such that 
\begin{equation}
    \frac{1}{2} \Delta Q(\v{z}) = 2 \partial_z \partial_{\bar z} Q(z,\bar z) = -\delta B(\v{z})
    \label{eq:mag_potential}
\end{equation}
The single particle Bloch states now have the form 
\begin{multline}
    \phi^{\rm AC}_\bk(\v{z}) = \phi_\bk(\v{z}) e^{-\frac{1}{2} Q(\v{z})} e^{-\frac{1}{2} R(\bk)} \\ =  C e^{\frac{i}{2} \bar k z} \sigma(z + i k \ell^2) e^{-\frac{1}{4 \ell^2} (|z|^2 + \ell^4 |k|^2)} e^{-\frac{1}{2} Q(\v{z})} e^{-\frac{1}{2} R(\bk)}
    \label{MagneticBlochAC}
\end{multline}
The factor $R(\bk)$ is a momentum space K\"ahler potential defined by the normalization condition
\begin{align}
    e^{R(\bk)} = \int_{\rm UC} d^2 \v{z} \, e^{-Q(\v{z})} |\phi_{\bk}(\v{z})|^2 
    \label{eq:Rk_magpot}
\end{align}
where the integral is over the unit cell. We can simplify the expression by introducing the Fourier transform $e^{-Q(\v{z})} = \sum_{\v{G} = n \v{G}_1 + m \v{G}_2} K_\v{G} e^{i \v{G} \cdot \v{z}}$, where $\v{G}_{1,2}$ are the reciprocal lattice vectors dual to $\v{R}_{1,2}$ defined such that $\v{G}_i \cdot \v{R}_j = 2\pi \delta_{i,j}$. We now use the identity
\begin{equation}
    |C|^2 \int_{\rm UC} d^2 \v{z} |\sigma(z)|^2 e^{-\frac{1}{2 \ell^2} |z|^2} e^{i \v{G} \cdot \v{z}} = \eta_\v{G} e^{-\frac{\ell^2}{4} \v{G}^2}
    \label{SigmaIdentity}
\end{equation}
where $\eta_{\v{G} = n \v{G}_1 + m \v{G}_2} = (-1)^{n+m+nm}$ and $C$ is fixed by evaluating the integral on the LHS for $\v{G} = 0$. This implies the relation
\begin{align}
    |\phi_\v{k}(\v{z})|^2 &= |C|^2 \sigma(z + i k \ell^2) e^{-\frac{1}{2\ell^2}|z + i k \ell^2|^2} \nonumber \\
    &= \frac{1}{2\pi \ell^2} \sum_\v{G} \eta_{\v{G}} e^{-\frac{\ell^2}{4} \v{G}^2} e^{-i \v{G} \cdot (\v{z} - \wedge \v{k} \ell^2)}
\end{align}

Substituting in (\ref{eq:Rk_magpot}) leads to
\begin{align}
    e^{R(\bk)} = \sum_\v{G} \eta_\v{G} K_\v{G} e^{i \ell^2 \v{G} \wedge \bk} e^{-\frac{\ell^2}{4} \v{G}^2}
    \label{RkExpansion}
\end{align}
The momentum space potential $R(\bk)$ is related to the Berry curvature through
\begin{equation}
    \Omega^{\rm AC}(\bk) = \ell^2 + 2\partial_k \partial_{\bar k} R(\bk)
\end{equation}

We can understand the mapping between the real space and momentum space K\"ahler potentials $Q(\v{z})$ and $R(\v{k})$ as a twisted version of the Weierstrass transform. The Weierstrass transform is defined as $\hat W_\Pi f(\v{z}) = e^{\frac{\ell^2}{4}\Delta} f(\v{z})$ where $\Delta = 4\partial_z \partial_{\bar z}$ is the Laplace operator. Noting that $e^{\frac{\ell^2}{4}\Delta} e^{i \v{k} \cdot \v{z}} = e^{-\frac{\ell^2}{4} \v{k}^2} e^{i \v{k} \cdot \v{z}}$, we see that the Weierstrass transform acts by multiplying the Fourier components by a Gaussian suppression factor, i.e. it smoothens the wavefunction. For any lattice $\Pi$, we can define a twisted Weierstrass transform, which we will call $\eta$-Weierstrass transform, as
\begin{equation}
    \hat W^\eta_\Pi f(\v{\v{z}}) := e^{\frac{\ell^2}{4} \Delta_{\v{z}}} \hat \eta_\Pi f(\v{z})
    \label{TwistedWeierstrass}
\end{equation}
where $\hat \eta_\Pi f(\v{z}) = \frac{1}{2}[-f(\v{z}) + f(\v{z} - \v{R}_1/2) + f(\v{z} - \v{R}_2/2) + f(\v{z} - \v{R}_1/2 - \v{R}_2/2)]$. The action of $\hat \eta$ in Fourier space is simple since for $\v{G} = n \v{G}_1 + m \v{G}_2$, we have $\hat \eta_\Pi e^{i \v{G} \cdot \v{z}} = \frac{1}{2}[-1 + (-1)^n + (-1)^m + (-1)^{n+m}] e^{i \v{G} \cdot \v{z}} = (-1)^{n+m+nm} e^{i \v{G} \cdot \v{z}} = \eta_{\v{G}} e^{i \v{G} \cdot \v{z}}$. That is, the operator $\hat \eta$ acts in Fourier space by multiplying by $\eta_{\v{G}}$. We can formally define the inverse $\eta$-Weierstrass transform via $[\hat W^\eta_\Pi]^{-1} f(\v{\v{z}}) = e^{-\frac{\ell^2}{4} \Delta_{\v{z}}} \hat \eta_\Pi f(\v{z})$. This acts in Fourier space by multiplication by $\eta_\v{G} e^{+\frac{\ell^2}{4} \v{G}^2}$. In an infinite system, this will only yield a well-defined function when acting on a function that is sufficiently smooth such that its Fourier coefficients decay faster than $e^{-\frac{\ell^2}{4} \v{G}^2}$. On a finite system, this inverse is always well-defined but can be numerically very unstable leading to strongly oscillatory functions.

Comparing with Eq.~(\ref{RkExpansion}), we see that
\begin{equation}
    e^{R(\v{k})} = \hat W_{\Lambda}^\eta e^{-Q(\v{z})}|_{\v{z} = \wedge \v{k} \ell^2}
    \label{RkWeierstrass}
\end{equation}
This implies that the Berry curvature of AC bands, derived from $R(\v{k})$ is generally expected to be a lot smoother than the magnetic field, derived from $Q(\v{z})$. Eq.~(\ref{RkWeierstrass}) also implies that it is easy to construct the Berry curvature from the magnetic field but hard to do the opposite. 
If the Berry curvature is given numerically, inverting the relation (\ref{RkWeierstrass}) to obtain the magnetic field would require the knowledge of higher Fourier components to an accuracy that increases as a Gaussian in the wave vector.

\subsection{Projection onto the space of Laughlin quasiholes}
\label{sec:Quasiholes}
To develop an analytically controlled theory of anyon dispersion, we consider hole doping the Laughlin state at filling $\nu = 1/q$, with $q$ being an odd integer. Both in the LLL and in AC bands, it is possible to write a (positive-semidefinite) Trugman-Kivelson interaction $\hat V_{\rm TK}$ by expanding a short-range interaction such that the Laughlin state is an exact zero mode \cite{trugmanExactResultsFractional1985a,ledwithFractionalChernInsulator2020a, wang2021exact, ledwith_vortexability_2023}. For the simplest Laughlin state with $q = 3$, this is simply the first pseudopotential $\hat V_1$ \cite{haldaneFractionalQuantizationHall1983a}. The Laughlin zero mode is separated by a gap of the order of the interaction scale from the rest of the spectrum. Upon electron doping, we generally do not have any zero modes for $\hat V_{\rm TK}$ and thus there is no sharp way to analytically define quasi-electron excitations (though several variational states exist in the literature \cite{laughlin1983anomalous,jeon2003NatureQuasiparticleExcitations,hansson2007CompositefermionWaveFunctions,rodriguez2012QuasiparticlesExcitonsPfaffian,bernevig2009ClusteringPropertiesModel}). On the other hand, for hole doping, we can construct a family of zero modes of $\hat V_{\rm TK}$ for any number of quasiholes labeled by the quasihole positions $\bxi_l$. In the following, we will focus on the case of one quasihole.

We now consider an interaction with the form $\alpha \hat V_{\rm TK} + \hat V$, where $\alpha \gg 1$ such that the first term is large compared to the second one \footnote{More precisely, we require the matrix elements of $\hat V$ in the eigenbasis of $\hat V_{\rm TK}$ to be small compared to the gap between the zero modes of $\hat V_{\rm TK}$ and the excited states}. Then, the low energy states are determined by the zero modes of $\hat V_{\rm TK}$ and their energy will be obtained by projecting $\hat V$ on the space of these zero modes. If we have an orthonormal basis $|\psi_n \rangle$ of such zero modes, then we can construct an effective Hamiltonian $\H_{nm} := \langle \psi_n|\hat V|\psi_m \rangle$ whose eigenvalues provide the low energy excitation spectrum. In particular, for any number of doped quasiholes, this corresponds to a quasihole Hamiltonian which lives fully in the Hilbert space of quasihole states. For the case of a single quasihole, as shown earlier, there is one quasihole momentum eigenstate per momentum. This means that we can immediately read off the quasihole dispersion as $\epsilon_\bk = \langle \psi_\bk|\hat V|\psi_\bk \rangle$ where $|\psi_\bk \rangle$ are the normalized quasihole momentum eigenstates. Our goal in the following is to construct such states explicitly. We will first construct such momentum eigenstates in the LLL in uniform field and then show that by mapping the uniform field LLL to AC bands, we can construct the corresponding states in the latter.

\subsubsection{Laughlin quasiholes in uniform magnetic field}

The Laughlin state for a uniform magnetic field on the torus is given by \cite{WangLatticeMC}
\begin{equation}
    \psi^l(\{\v{z}_i\}) = \sigma_l \prod_{i<j} f(z_i - z_j)^q \prod_{k=1}^q f(z_{\rm CM} - \gamma_k^l) e^{-\frac{i}{2N_\Phi} \bz_{\rm CM} \wedge \v{\gamma}^l_k}
    \label{eq:laughlin}
\end{equation}
where we define $\sigma_{l} = e^{-\frac{\pi i (q-1)}{2q}l}$ which is an overall gauge choice for later convenience.
Here, $z_{\rm CM} = \sum_{i=1}^{N_e} z_i$ and $f(z)$ is defined in terms of the modified Weierstrass sigma function $\sigma(z|L_1, L_2)$ introduced earlier as
\begin{equation}
    f(z) = \sigma(z|L_1, L_2) e^{-\frac{1}{4N_\Phi} |z|^2}
\end{equation}
This function satisfies $f(z + L_j) = - e^{\frac{i}{2N_\Phi} \bL_j \wedge \bz} f(z)$ and has a single zero in the parallelogram spanned by $\bL_1$ and $\bL_2$. Thus, the factor $f(z_i - z_j)^q$ is the proper generalization of the factor $(z_i - z_j)^q$ for the Laughlin state on the plane. The remaining $q$ factors of $f$ ensure that the wavefunction, viewed as a function for one of the particle coordinates with all other coordinates fixed, has $N_\Phi$ zeros in the parallelogram spanned by $\bL_1$ and $\bL_2$.

The factors $\v{\gamma}_k^l$ specify the positions of the extra zeros with $l = 0,\dots,q-1$ labeling the $q$ topologically distinct ground states on the torus. These factors can be chosen to be
\begin{align}
    &\v{\gamma}^{l}_{k} = \frac{k - 1}{q}\bL_1+\frac{l}{q}\bL_2 = \v{\gamma}^{l\mod q}_{k\mod q}, \notag\\ &\v{\gamma}^l_{\rm CM} = \frac{q-1}{2} \bL_1 + l \bL_2
\end{align}
which implies that for $\bL = n \bL_1 + m \bL_2$, $\bL \wedge \v{\gamma}_{\rm CM}^l = 2\pi N_\Phi (n l - \frac{q-1}{2} m)$ which is an integer multiple of $2\pi N_\Phi$ since $q$ is odd. Thus, we find the action of translation for the coordinate of the $i$-th particle by $\bL_j$, $j = 1,2$ gives $t_{i, \v{L}_j} \ket{\Psi^l} = e^{i\Phi_j} \ket{\Psi^l}$ with $e^{i \Phi_j} = (-1)^{N_\Phi}$. 

The action of the minimal translations $T_{\v{d}_1}$ and $T_{\v{d}_2}$ on the Laughlin state are
\begin{equation}
    T_{\v{d}_1} \psi^l = - e^{-\frac{2\pi i l}{q}} \psi^l, \quad T_{\v{d}_2} \psi^l =  \psi^{l-1}
\end{equation}

The construction of the wavefunctions for the Laughlin quasiholes proceeds similarly. Now we have $N_\Phi = q N_e + 1$, and the wavefunction takes the form
\begin{multline}
    \psi^l_{\bxi}(\{\bz_i\}) = \sigma_{l,\v{\xi}} \prod_{i<j} f(z_i - z_j)^q \prod_{i} f(z_i - \xi)  \\ \times \prod_{k=1}^q f(z_{\rm CM} - \gamma^l_k + \frac{\xi}{q}) e^{\frac{i}{2N_\Phi} \v{\gamma}^l_k \wedge (\bz_{\rm CM} + \v{\xi}/q)}
    \label{PsixhiQuasihole}
\end{multline}

Here, we define $\sigma_{l,\v{\xi}} = e^{-\frac{\pi i (q-3)}{2q}l} e^{\frac{i N_\Phi}{2q} \v{\xi} \wedge (\v{\delta}_1 + \v{\delta}_2)} $ which is an overall gauge choice for later convenience. Under the action of the minimal translation, this transforms as
\begin{gather}
    T_{\v{d}_1} \psi_{\bxi}^l = e^{\frac{i}{2q}\v{d}_1\wedge\bxi} e^{-\frac{2\pi l}{q}i} \psi_{\bxi-\v{d}_1}^l, \\
     T_{\v{d}_2} \psi_{\bxi}^l = e^{\frac{i}{2q}\v{d}_2\wedge\bxi} \psi_{\bxi-\v{d}_2}^{l-1}
\end{gather}

Here, the range of $l$ is extended from $0,\dots,q-1$ to $\mathbb{Z}$, and the periodicity for $l$ and $l-q$ is given by $\psi_{\bxi}^{l-q}(\{\bz_i\})= \psi_{\bxi}^{l}(\{\bz_i\})$.

As discussed earlier (cf.~Sec.~\ref{sec:MagneticTranslation}), we can instead consider translations generated by $\v{\delta}_1 = \v{d}_1$ and $\v{\delta}_2 = q\v{d}_2$ and remain within the same topological sector $l$, which we fix to be $l = 0$ and drop the $l$ index in what follows. We then find
\begin{equation}
    T_{\v{\delta}_1} \psi_{\bxi} = e^{\frac{i}{2q}\v{\delta}_1\wedge\bxi} \psi_{\bxi-\v{\delta}_1}, \quad
     T_{\v{\delta}_2} \psi_{\bxi} = e^{\frac{i}{2q}\v{\delta}_2\wedge\bxi} \psi_{\bxi-\v{\delta}_2}
\end{equation}
Thus, translation by a vector of the form $\v{\delta} = n \v{\delta}_1 + m \v{\delta}_2$ that preserve the boundary condition and the topological sector acts as
\begin{equation}
     T_{\v{\delta}} \psi_{\bxi} =  e^{\frac{i}{2q}\v{\delta}\wedge\bxi} \psi_{\bxi-\v{\delta}}, 
     \label{TdsigmaAction}
\end{equation}
and the magnetic algebra for such translations simplifies to
\begin{equation}
    T_{\v{\delta}} T_{\v{\delta}'} = e^{-\frac{i}{q} \v{\delta} \wedge \v{\delta}'} T_{\v{\delta}'} T_{\v{\delta}} = e^{-\frac{i}{2q} \v{\delta} \wedge \v{\delta}'} T_{\v{\delta} + \v{\delta}'}
    \label{QHMagneticAlgebra}
\end{equation}

\subsubsection{Momentum eigenstates}

The commuting set of translations that leave the topological sectors invariant is generated by $\v{\alpha}_1 = \v{a}_1 = \bL_1/N_1 = N_2 \v{\delta}_1$ and $\v{\alpha}_2 = q \v{a}_2 = q\bL_2/N_2 = N_1 \v{\delta}_2$ as discussed earlier. Now we will show how to construct quasihole momentum eigenstates which are eigenstates of all magnetic translations $T_\v{\alpha}$ where $\v{\alpha} = n \v{\alpha}_1 + m \v{\alpha}_2$ satisfying
\begin{equation} \label{magnetic-translation-eq}
    T_{\v{\alpha}} |\psi_{\v{\kappa}} \rangle = \eta_\v{\alpha} e^{i \v{\kappa} \cdot \v{\alpha}} |\psi_{\v{\kappa}} \rangle, \qquad \eta_{n \v{\alpha}_1 + m \v{\alpha}_2} = (-1)^{n+m+nm}
\end{equation}

The quasihole wavefunctions $|\Psi_\v{\xi} \rangle$ are labeled by a continuous parameter and can be viewed as coherent states spanning the space of one quasihole states whereas $|\Psi_\v{\kappa} \rangle$ provides an orthonormal basis of such space. Thus, we can expand
\begin{equation}
    |\psi_\v{\xi} \rangle =  \sum_{\v{\kappa}}  F_{\v{\kappa}}(\v{\xi}) |\psi_\v{\kappa} \rangle, \qquad F_{\v{\kappa}}(\v{\xi}) =  \langle \psi_\v{\kappa}| \psi_\v{\xi} \rangle
    \label{PsixiPsik}
\end{equation}

Writing $\langle \psi_\v{\kappa}|T_{-\v{\alpha}_j}| \psi_\v{\xi} \rangle = -e^{-i \v{\kappa} \cdot \v{\alpha}_j} \langle \psi_\v{\kappa}| \psi_\v{\xi} \rangle = e^{-\frac{i}{2q} \v{\alpha}_j \wedge \v{\xi}} \langle \psi_\v{\kappa}| \psi_\v{\xi + \v{\alpha}_j} \rangle$ which implies $F_{\v{\kappa}}(\v{\xi} + \v{\alpha}_j) = -e^{-i \v{\kappa} \cdot \v{\alpha}_j} e^{\frac{i}{2q} \v{\alpha}_j \wedge \v{\xi}} F_{\v{\kappa}}(\v{\xi})$. 
Furthermore, we can see from Eq.~(\ref{PsixhiQuasihole}) that $\psi_\v{\xi}$ has the form of a holomorphic function in $\xi$ times $e^{-\frac{1}{4q} (\v{\xi} - \v{\xi}_0)^2}$ where $\v{\xi}_0 = N_\Phi (\v{\delta}_1 + \v{\delta}_2) = N_1 \v{\alpha}_1 + N_2 \v{\alpha}_2$ which implies the same is true for $F_{\v{\kappa}}(\v{\xi})$. As a result, we have
\begin{equation}
    F_{\v{\kappa}}(\v{\xi}) \propto  \phi^{{\rm QH}}_{-(\v{\kappa} - \v{\kappa}_0)}(\v{\xi} - \v{\xi}_0) 
\end{equation}
where $\v{\kappa}_0 := \frac{1}{2}(N_2 \v{\beta}_1 - N_1 \v{\beta}_2)$ 
and $\phi^{\rm QH}_{-\v{\kappa}}(\v{\xi})$ are magnetic Bloch states defined in Eq.~(\ref{MagneticBloch}) for the lattice $\Lambda_q$ associated with the enlarged unit cell defined by $\v{\alpha}_1$, $\v{\alpha}_2$ with $\ell^2 = q$. 
The requirement that both $|\psi_{\v{\xi}} \rangle$ and $|\psi_{\v{\kappa}} \rangle$ are normalized fixes the proportionality constant leading to the expression 
\begin{equation}
   F_{\v{\kappa}}(\v{\xi}) = \sqrt{\frac{2\pi q}{N_\Phi}} \phi^{\rm QH}_{-(\v{\kappa-\kappa_0})}(\v{\xi} - \v{\xi}_0)
   \label{Fxik}
\end{equation}

The minus sign here comes from the fact that translation acts on the electron coordinate and its effect shifts the quasihole coordinate in the opposite direction.

To construct quasihole momentum eigenstates, we define
\begin{equation}
    P_{\v{\kappa}} = \frac{1}{N_\Phi} \sum_{\v{\alpha} = n \v{\alpha}_1 + m \v{\alpha}_2} \eta_{\v{\alpha}} e^{-i \v{\kappa} \cdot \v{\alpha}} T_{\v{\alpha}}
\end{equation}
where the sum goes over $n = 0, \dots, N_1 - 1$ and $m = 0, \dots, N_2 - 1$. Clearly, $P_{\v{\kappa}}$ is a projector onto the momentum $\v{\kappa}$ eigensector since its action on a momentum eigenstate is $P_{\v{\kappa}} |\psi_{\v{\kappa'}} \rangle = \delta_{\v{\kappa},\v{\kappa}'} |\psi_{\v{\kappa}} \rangle $. This means that we can construct momentum eigenstates by acting with $P_{\v{\kappa}}$ on any quasihole state $|\psi_{\v{\xi}} \rangle$
\begin{equation}
    |\psi_{\v{\kappa}} \rangle = \frac{1}{F_\v{\kappa}(\v{\xi})} P_\v{\kappa}|\psi_{\v{\xi}} \rangle  \label{PsikProjector}
\end{equation}

Note that the LHS is independent of $\v{\xi}$, so we can pick any $\v{\xi}$ on the RHS to compute the momentum eigenstates. However, note that $\phi^{\rm QH}_{-(\v{\kappa-\kappa_0})}(\v{\xi} - \v{\xi}_0)$ vanishes for $\xi - \xi_0 = i q (\kappa-\kappa_0)$, so we need to avoid to hit a zero in the denominator. We can simply do that by choosing $\v{\xi}$ to avoid any point on the momentum space grid. This however, could make the denominator very small for fine grids (large systems). A better option is to pick $\v{\xi}$ to depend on $\v{k}$ in such a way that we are always far away from the zeros of the denominator. This ensures any numerical expressions using (\ref{PsikProjector}) do not encounter any issues associated with the denominator. Writing back in first quantization, we see that the first quantized momentum space quasihole wavefunctions in uniform magnetic field can thus be expressed in terms of the quasihole wavefunction (\ref{PsixhiQuasihole}) (with $l = 0$) as
\begin{equation}
    \psi_\v{\kappa}(\{\bz_i\}) = \frac{1}{N_\Phi F_\v{\kappa}(\v{\xi})} \sum_{\v{\alpha}} \eta_{\v{\alpha}} e^{-i \v{\kappa} \cdot \v{\alpha}} e^{\frac{i}{2q}\v{\alpha} \wedge \v{\xi}} \psi_\v{\xi - \alpha}(\{\bz_i\})
    \label{eq:momentum-state}
\end{equation}

\subsubsection{Aharonov-Casher bands}
Given any wavefunction that is a zero mode of a Trugman-Kivelson pseudopotential in the LLL, we can construct the corresponding (unnormalized) wavefunctions in AC band as
\begin{equation}
    \psi^{\rm AC}(\{\bz_i\}) = \psi(\{\bz_i\}) \Gamma(\{\bz_i\}), \qquad \Gamma(\{\bz_i\}) = \prod_{i=1}^{N_e} e^{-\frac{1}{2} Q(\v{z}_i)}
    \label{PsiAC}
\end{equation}
This enables us to define quasihole momentum eigenstates in an AC band using Eqs.~(\ref{PsiAC}) and (\ref{eq:momentum-state}) and allows us to write an explicit formula for the dispersion as
\begin{equation}
    \epsilon_{\v{\kappa}} = \frac{N_e(N_e-1)}{2} \frac{\int \prod_{i=1}^{N_e} d^2 \v{z}_i V(\v{z}_1, \v{z}_2) |\psi^{\rm AC}_\v{\kappa}(\{\bz_i\})|^2}{\int \prod_{i=1}^{N_e} d^2 \v{z}_i |\psi^{\rm AC}_\v{\kappa}(\{\bz_i\})|^2}
    \label{eq:AC-MC}
\end{equation}
which can be evaluated directly in Monte Carlo as we will see in the next section.

Note that the non-uniform magnetic field breaks the full magnetic translation group to its discrete subgroup consisting of translations by a lattice vector $\v{a} = n \v{a}_1 + m \v{a}_2$. As discussed earlier, requiring the topological sector to be unchanged limits us to lattice vectors $\v{\alpha} = n \v{\alpha}_1 + m \v{\alpha}_2$. One way to understand the breaking of continuous magnetic translation is to define the deformed magnetic translations
\begin{equation}
    t^{\rm AC}_{\v{v}} = \Gamma t_{\v{v}} \Gamma^{-1}, 
\end{equation}
Here, $\Gamma$ is an invertible but non-unitary transformation. This means that $t^{\rm AC}_{\v{v}}$ satisfies the same magnetic algebra as before but they are in general not unitary operators. However, for lattice vectors $\v{a}$, $[t_{\v{a}}, \Gamma] = 0$ which implies that $t^{\rm AC}_{\v{a}} = t_{\v{a}}$ is unitary and the group of \emph{discrete} magnetic translation remains unchanged.

\section{Numerical results for anyon dispersion}
\label{Sec:numerics}
The explicit construction of the first-quantized  momentum-space quasihole wavefunction given in Eq.~(\ref{eq:momentum-state}) allows us to use the Monte Carlo (MC) method to evaluate anyon dispersion, reaching large system sizes beyond other methods such as ED. In this section, we will first begin by computing dispersion of Laughlin quasiholes for a simple model of an AC band where the periodic magnetic field modulation is restricted to the leading harmonic, before moving to the more realistic case relevant to twisted MoTe${}_2$ bands.

For convenience, we recall the geometry: a torus spanned by $\bL_1,\bL_2$ whose area with $N_\Phi$ flux quanta is denoted as $A=|\bL_1\times \bL_2|=2\pi N_{\Phi}$. Real space lattice $\Lambda$ has unit cell defined by $\ba_1=\bL_1/N_1, \ba_2=\bL_2/N_2$ where $N_1N_2=N_{\Phi}$ such that the unit cell encloses a unit flux. The topological sector preserving lattice $\Lambda_q$ has unit cell 
$\balpha_1=\ba_1, \balpha_2=q\ba_2$.
We choose $\v{a}_1, \v{a}_2$ to form a triangular lattice: $|\v{a}_1| = |\v{a}_2|$ with the relative angle $\pi/3$.
We consider quasiholes for $1/3$ Laughlin state, such that $q=3$ throughout this section.

To evaluate expectation values in the form of\begin{equation}
    \langle O\rangle= \frac{\int \prod_{l=1}^{N_e} d^2 \v{z}_i O(\{z_i\}) |\psi(\{z_i\})|^2}{\int \prod_{l=1}^{N_e} d^2 \v{z}_i |\psi(\{z_i\})|^2}
\end{equation}
we use Metropolis-Hastings algorithm \cite{metropolis1953equation,hastings1970monte} where $|\psi(\{z_i\})|^2$ can be chosen as our quasihole wavefunction of interest and used as the Metropolis weight, specifically, we shall use the momentum-space wavefunction to evaluate the quasihole dispersion.

As a sanity check, we verify the validity of our Monte Carlo algorithm by benchmarking physical quantities in the $1/3$ Laughlin state, i.e. structure factor and Coulomb energy. We found good consistency with previous works \cite{levesque1984CrystallizationIncompressibleQuantumfluid,WangLatticeMC}: benchmark tests and finite-size checks are presented in SM Appendix~\ref{consistencycheck}.

Before proceeding to the results, there are several subtleties about the Monte Carlo we use: 
\begin{enumerate}[(i)]
    \item The variations in the anyon energy as a function of its position, which gives rise to the dispersion, are $O(1)$ on top of an extensive $O(N_e)$ energy of the many-body state. On the other hand, the error on the energy scales as $\sigma_{\bk}\sim O(N_e/\sqrt{N_{\text{step}}})$. In order to resolve the $O(1)$ dispersion on top of the $O(N_e)$ energy, the number of Monte Carlo steps $N_{\text{step}}$ needs to increase as $O(N_e^2)$ to reach fixed precision.  
    \item For each single Monte Carlo evaluation, we are using the momentum-space wavefunctions composed of $N_{\Phi}$ terms that require being updated individually. Thus we obtain an additional complexity $O(N_{\Phi})$ for each single update.
\end{enumerate}
Compared to the classical Monte Carlo in fractional quantum Hall \cite{morfMonteCarloEvaluation1987,zhu1993WignerCrystallizationFractional,ciftja2011FinitesizeMonteCarlo,tserkovnyak2003MonteCarloEvaluation,meirVariationalGroundState1996,biddleVariationalMonteCarlo2013}, we have the extra complexity $O(N_e^2)$ from the smallness of dispersion and $O(N_\Phi)$ from the wavefunctions, which are in principle still polynomial but in practice limit the system size we can reach.

\subsection{Dispersion of Laughlin quasiholes in AC band with first harmonic approximation}
\label{Sec:numerics-AC}
We will first study the dispersion for a Laughlin quasihole in a general AC band where we will parametrize the non-uniformity of the quantum geometry by a single parameter. In the next section, we will consider the concrete setup and parameters relevant to t-MoTe${}_2$. Throughout, we will take the interaction to be double-gate screened Coulomb interaction. This choice is motivated by the experimental setup where metallic gates from both sides together control displacement and carrier density. It also enables interpolation between the short-range pseudopotential limit and the Coulomb interaction. We consider two gates sitting symmetrically at distance $d$ from the sample separated by dielectric with dielectric constant $\epsilon$, leading to modified Coulomb interaction:
\begin{equation}
    V(\bq)=\frac{2\pi e^2}{\epsilon |\bq|}\tanh(|\bq|d)
    \label{eq:sgC-momen}
\end{equation}
where $e$ is the elementary charge and $\bq$ is the momentum. 
In this section, no physical parameters for real systems enter yet, thus it is natural to measure energies in units of the first pseudopotential coefficient: $V_1 = \int \frac{d^2 \bq}{(2\pi)^2} V(\bq) e^{-\bq^2}L_1(\bq^2)$, which provides the scale of the gap to other excitations in the system.

The momentum space interaction Eq.~(\ref{eq:sgC-momen}) is nonlocal in real space, thus a single trial move alters all Fourier modes and complicates our Metropolis–Hastings sampling. To avoid this issue, we rewrite the interaction in real space:
\begin{equation}
    V(\bz_1,\bz_2)=\frac{2}{d}\sum_{m=0}^{\infty}K_0(\frac{(2m+1)\pi|\bz_1-\bz_2|}{2d})
    \label{V:sCoul-real}
\end{equation}
where $K_0$ is the zeroth Bessel K function. The expansion in $K_0$ here is used for faster convergence.

With the gate screened Coulomb interaction, we can evaluate the anyon dispersion in the AC band through relation Eq.~(\ref{eq:AC-MC}). Specifically, we take the quasihole momentum eigenstate wavefunction $\psi_{\v{\kappa}}(\{\bz_i\})$ and real space K\"ahler potential \cite{Abouelkomsan2025}
\begin{equation}
    Q(\v{z})=-2 K \sum_{\bb_i}\cos(\bb_{i}\cdot\bz)
\end{equation}
where $\bb_i,\ i=1,2,3$ are the reciprocal lattice vectors of $\Lambda^*$ with $\bb_3=-\bb_1-\bb_2$ where $K$ is a dimensionless real parameter that controls the strength of the non-uniformity.
We can then evaluate the energy $\epsilon_{\v{\kappa}}$ at each given momentum $\v{\kappa}$ to obtain the dispersion. Note that $\v{\kappa}$ is defined on $\Lambda^{*}_q$ and forms the reduced BZ. In order to get dispersion in the original BZ, we need to unfold the reduced direction.
In our case, since we chose our lattice $\v{\alpha}_2 = q \v{a}_2$, we need to unfold the $\v{b}_2$ direction. The unfolding procedure can be described as follows: consider a momentum eigenstate $|\psi_{\v{k}_{nm}}\rangle$ in the original BZ labeled by momenta $\v{k}_{nm}=\frac{n}{N_1}\bb_1+\frac{m}{N_2}\bb_2$ and a momentum eigenstate $|\psi_{\v{\kappa}_{n'm'}}\rangle$ in the reduced BZ labeled by momenta $\v{\kappa}_{n'm'}=\frac{n'}{N_1}\v{\beta}_1+\frac{m'}{N_2}\v{\beta}_2$. 
We have $n=n'$ from $\v{\beta}_1=\v{b}_1$. For the $\v{b}_2$ direction,
the momentum eigenstate transforms as: 
\begin{equation}
    \begin{split}
        T_{q\ba_2}|\psi_{\bk_m}\rangle&=\eta_{q\ba_2}e^{i\bk_m\cdot q\ba_2}|\psi_{\bk_m}\rangle=\eta_{q\ba_2}e^{i2\pi q m/N_2}|\psi_{\bk_m}\rangle\\
    T_{\balpha_2}|\psi_{\v{\kappa}_{m'}}\rangle&=\eta_{\balpha_2}e^{i\v{\kappa}_{m'}\cdot\balpha_{2}}|\psi_{\v{\kappa}_{m'}}\rangle=\eta_{\balpha_2}e^{i2\pi m'/N_2}|\psi_{\v{\kappa}_{m'}}\rangle
    \end{split}
\end{equation}

Since $\eta_{\balpha_2}=\eta_{q\ba_2}=-1$, the mapping condition is then given by the modular equation:
\begin{equation}
    qm=m' \mod N_2,\ \  m,m'=0,...,N_2-1
    \label{eq:momentum-map}
\end{equation}
By using Eq.~(\ref{eq:momentum-map}), one can perform the momentum mapping and restore the dispersion $\epsilon_{\bk}$ in the original BZ.

Using MC simulations, we systematically vary the interaction range $d$, non-uniformity $K$ and system size $N_e$, obtaining the result shown as FIG. ~\ref{fig:dispersion-dependence}.
We can see that
\begin{enumerate}[(i)]
    \item  the quasihole AC band indeed has approximate $q^2=9$-fold degeneracy, as a result from the folding of BZ as we discussed in Sec.~\ref{sec:MagneticTranslation}.
    \item The energy vanishes in the small $d$ limit as $d^4$ when measured in units of $V_1$ (see SM FIG.~\ref{fig:bandwidthscaling} panel b). This can be understood by expanding the interaction for small $d$ in terms of pseudopotentials $V(\v{r}) = \sum_n c_{2n+1} d^{2(2n+1)+1} \hat V_{2n+1}$ and noting that the quasiholes are zero modes of $\hat V_1$. Since the $V_1$ component scales as $d^3$, we see that the energy of the quasiholes goes as $d^7/d^3 \sim d^4$. Since in this regime the bandwidth is extracted from the maximum and minimum energies over a discrete Brillouin-zone grid, and several candidate momenta for the energy extrema remain close in energy, features in $w(d)$ at small $d$ are numerically more delicate and should not be overinterpreted.
    \item The bandwidth of the quasihole AC bands grows linearly in the small $K$ limit and then shows a weaker, sublinear growth at larger $K$. 
    This can be understood from the fact that, as $K$ increases, the modulation in the normalized wavefunction approaches an extreme limit in which the weight becomes increasingly concentrated near the maxima of the wavefunction amplitude and thus sets an upper bound on the bandwidth.
    The more physical picture of quasihole dispersion from guiding center will be discussed in Sec.~\ref{Sec:qhGC}.
    \item The quasihole bandwidth has very weak system size dependence (see SM FIG.~\ref{fig:bandwidthscaling} panel a).  Since quasiholes are local excitations in real space the size of which remains the same across system size, they only feel the field modulation locally within the same spread, thus the resulting energy variation is $O(1)$ and does not scale with system size. The weak system size dependence originates mainly from finite-size effects and can be extracted by scaling. 
    \item The overall scale of the dispersion is relatively small when measured in units of $V_1$. This provides justification for our projected treatment and suggests the gap to other excitations remains much larger than the dispersion for realistic interactions.
\end{enumerate}

\subsection{Application to Twisted MoTe${}_2$}
\label{Sec:tMoTe2}
Homobilayer transition metal dichalcogenides moir\'e superlattices have been experimentally verified to host a wide range of strongly correlated phases, including the fractional quantum anomalous Hall (FQAH) or zero field fractional Chern insulators (FCI) \cite{cai2023signatures, park2023observation, zeng2023thermodynamic, xu2023observation}. The nearly ideal band geometry therein is believed to be crucial for understanding these novel phases \cite{dong2023composite, morales-duran2023pressureenhanced, TMDAharonocCasher}. Previous studies showed that with the adiabatic approximation, the continuum model describing the moir\'e superlattice can be described by an Aharonov-Casher band with non-uniform field arising from the layer pseudospin \cite{morales-duran2024magic, TMDAharonocCasher}. In this section, we apply our MC algorithm to evaluate the AC band quasiholes and estimate the corresponding bandwidth.

In the continuum model of TMD homobilayers, valence band holes with effective mass $m$ experience a spatially periodic potential $\Delta_0(\br),\bDelta(\br)$ that couples to layer pseudospin:
\begin{equation}
    H_{\mathrm{cont}}=\frac{\bp^2}{2m}+\sigma_0\cdot\Delta_0(\br)+\bsigma\cdot\bDelta(\br)
    \label{eq::H-tMoTe${}_2$}
\end{equation}
where $\sigma_0=I,\ \bsigma=(\sigma_x,\sigma_y,\sigma_z)$. The scalar potential $\Delta_0(\br)$ and the vector field $\bDelta(\br)$ satisfy the relation:

\begin{align}
     \Delta_0(\br)\pm\Delta_{z}(\br)&=2V\sum_{i=1}^{3}\cos(\bb_i\cdot\br\mp\psi)\\
    \Delta_{x}(\br)\pm i \Delta_{y}(\br)&=t_{M}\sum_{i=1}^{3}e^{\pm i\bq_i\cdot\br}
\end{align}
$\bb_i,i=1,2,3$ are the reciprocal lattice vectors in $\Lambda^*$ with $\bb_3=-\bb_1-\bb_2$. $\bq_i,i=1,2,3$ satisfy the relation: $\bq_i=-\frac{1}{\sqrt{3}}\wedge\bb_i$. $\bb_i$ and $\bq_i$  describe intralayer and interlayer moi\'re momentum transfer respectively. Parameters $t_{M},V,\psi$ characterize the strength of interlayer tunneling, the strength of potential energy within the layer and a phase angle for moir\'e potential maxima \cite{wu2019topological,morales-duran2024magic,TMDAharonocCasher, devakul2021magic}.

Let $\bn(\br)$ describe direction of the layer pseudospin at location $\br$, the vector field $\bDelta(\br)$ from moir\'e sets the lowest energy configuration for the layer pseudospin $\bn_0(\br)=\bDelta(\br)/|\bDelta(\br)|$.
The adiabatic approximation assumes $\bn_0$ varies slowly in space and electrons are assumed to occupy the locally aligned band such that the layer pseudospin texture tracks $\bn_0$ at every point $\bn(\br)=\bn_0(\br)$.
Within the adiabatic approximation, the continuum model can be mapped to an AC Hamiltonian with single particle potential $U(\br)$:
\begin{align}
    H_{\mathrm{cont}}&=H_{\mathrm{AC}}+U(\br)\\
    U(\br)&=\Delta_{+}(\br)-\omega_c\chi(\br)
\end{align}
$\Delta_{+}(\br)=\Delta_0(\br)+|\bDelta(\br)|$ is the strength of the layer pseudospin potential from aligning to the moi\'re. The moir\'e layer pseudospin texture creates an AC effective field $B^{\rm AC}(\br)$, uniform part of which determines the cyclotron frequency $\omega_c=|B_{0}|/m$. $\chi(\br)=\frac{1}{2}[G(\br)-B^{\rm AC}(\br)]$ describes the energy associated with creating the layer pseudospin texture in real space where $G(\br)$ comes from the energy cost of twisting the layer pseudospin texture:
\begin{align}
G(\br)&=\frac{|\partial_x\bn(\br)|^2+|\partial_y\bn(\br)|^2}{4}\\
    B^{\rm AC}(\br)&=\frac{\bn(\br)\cdot(\partial_{x}\bn(\br)\times \partial_{y}\bn(\br))}{2}
\end{align}
The spatially non-uniform field $B^{\rm AC}(\br)$ corresponds to the real space K\"ahler potential $Q(\br)$ that satisfies $\frac{1}{2}\Delta Q(\br)=B_0-B^{\rm AC}(\br)$, which is responsible for the AC band mapping. Before proceeding with numerical results, note that we have set $\ell_B=1$ throughout. In order to convert energy into physical units, we use the relation between moir\'e unit cell vector and magnetic length: $a_M=\sqrt{4\pi/\sqrt{3}}\ell_B=a_0/[2\sin(\theta/2)]$. $a_0=3.52 \text{\AA}$ is the physical unit cell length of MoTe${}_2$ and $\theta$ is the twist angle between the two MoTe${}_2$ layers. 
\begin{figure}[h]
    \centering
    \includegraphics[width=\linewidth]{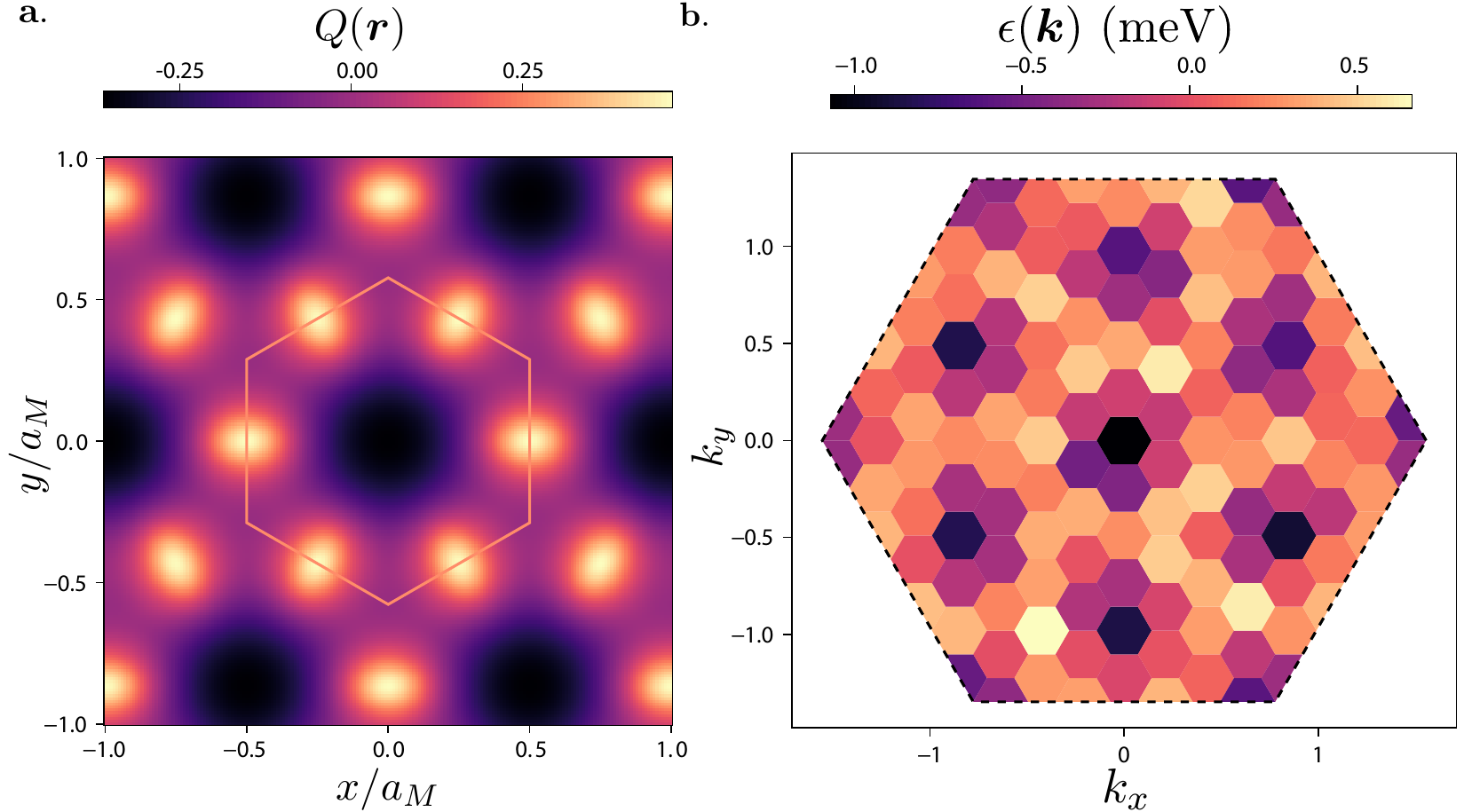}
    \caption{\textbf{a.} Single particle quantum geometry, i.e. real space K\"ahler potential $Q(\br)$, for twisted MoTe${}_2$. The hexagon outlines Wigner-Seitz moir\'e unit cell. \textbf{b}. Dispersion of quasiholes in tMoTe${}_2$ with gate screened Coulomb interaction in the mBZ for $N_e=40,\ d=10 \ \ell_B,\ \theta=3^\circ$.}
    \label{fig:tMoTe2-nonuniform-field}
\end{figure}
Here we take realistic parameters for twisted MoTe${}_2$ $(t_M,V,\psi)=(23.8\  \mathrm{meV},20.8\  \mathrm{meV},107.7^\circ)$ and $\theta=3.0^\circ$ \cite{wu2019topological,morales-duran2024magic, TMDAharonocCasher,liangfu2023tMoTe2} to study dispersion of the corresponding AC band quasiholes using Monte Carlo.
We first obtain the effective magnetic field $B^{\rm AC}(\br)$ and thus the real space K\"ahler potential $Q(\br)$ from the periodic vector field $\bDelta(\br)$, as shown in FIG.~\ref{fig:tMoTe2-nonuniform-field} (a): we see that the nonuniform part of the effective magnetic field is concentrated around the M point where charge also concentrates.
Comparing the magnitude of the real space K\"ahler potential $Q(\br)$ with the leading harmonic approximated potential in previous section, we see that it corresponds to $K \approx 0.1 \ll 1$ \footnote{This is seen by comparing the bandwidth of the cosine potential $9K$ to the results in FIG.~\ref{fig:tMoTe2-nonuniform-field}(a)}, placing it firmly in the weak field regime where a linear approximation might apply.

\begin{figure}
    \centering
    \includegraphics[width=\linewidth]{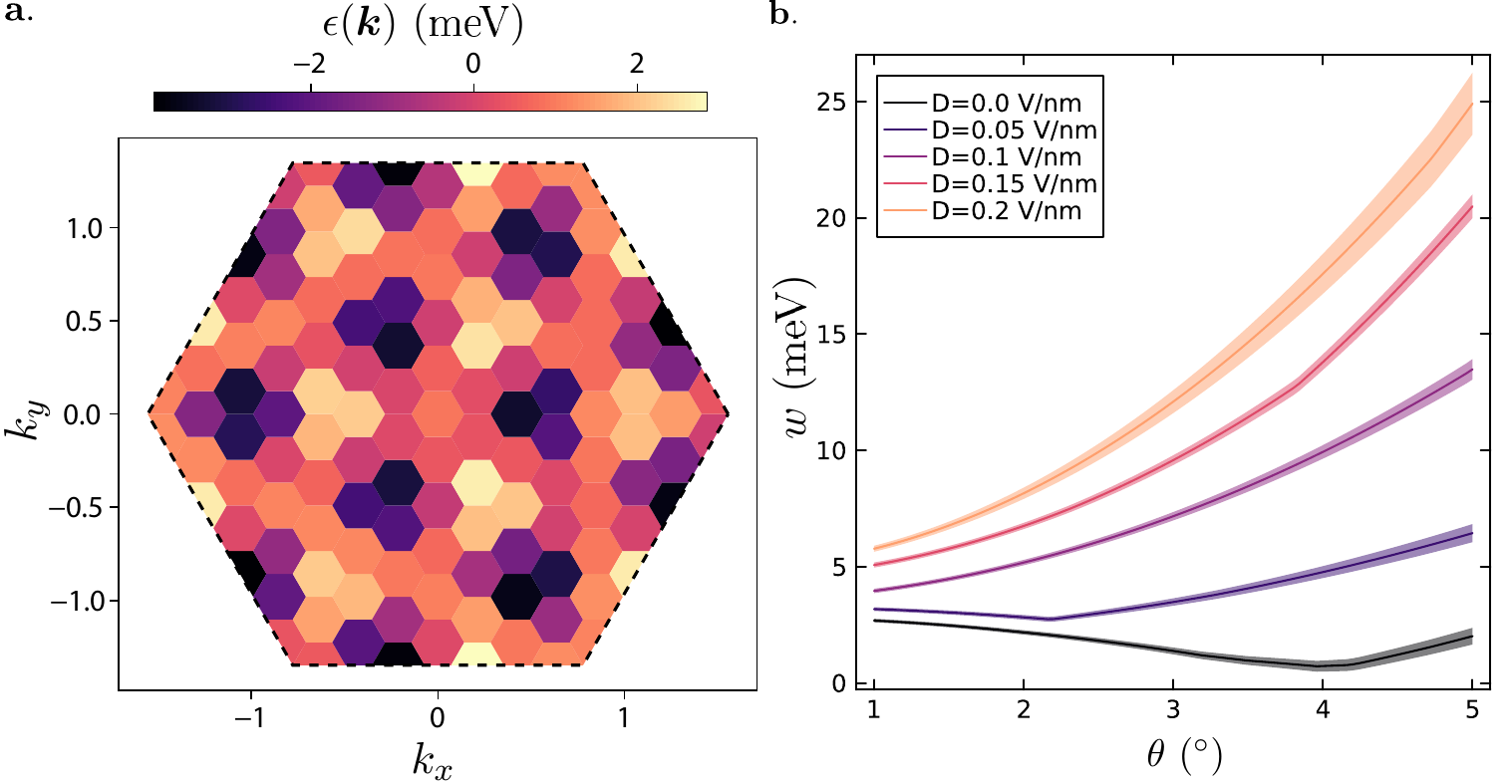}
    \caption{Dependence of AC quasihole dispersion in tMoTe${}_2$ on displacement field $D$ and twist angle $\theta$ under gate screened Coulomb interaction, with screening length $d=10 \ \ell_B $. \textbf{a.} Quasihole dispersion in the mBZ for $N_e=40, \ D=0.1 \ \text{V/nm}, \ \theta=3^\circ$. \textbf{b.} Dependence of the quasihole bandwidth on twist angle and displacement field for $N_e=16$. Each twist angle dependence is obtained from a single MC dataset with $\theta$ controlling the length scale and thus the corresponding energy scale. Shades around the curve represent error bars.}
    \label{fig:tMoTe2-Dfield}
\end{figure}

We then calculate the AC quasiholes in tMoTe${}_2$ to obtain  the dispersion shown as FIG.~\ref{fig:dispersion-dependence} (d) and FIG.~\ref{fig:tMoTe2-nonuniform-field} (b).
We see that (i) the dispersion still has $q^2=9$-fold degeneracy similar to the case of the cosine modulation discussed in the previous section. (ii) the bandwidth changes non-monotonically with the twist angle $\theta$ where a minimum is reached around $\theta\approx 3.5^\circ\sim 4^\circ$ as shown in FIG.~\ref{fig:dispersion-dependence} (d). This can be understood from the near cancellation between $\Delta_+(\br)$ and $\omega_c\chi(\br)$ as twist angle varies. The same mechanism explains why this nonmonotonic twist-angle dependence remains visible at small displacement field in FIG.~\ref{fig:tMoTe2-Dfield} (b). (iii) The dispersion we obtained for quasiholes in tMoTe${}_2$ is of order 1 meV and specifically is  $1.1\pm 0.3 \ \text{meV}$ when $\theta\approx3.7^\circ$, which is consistent with previous ED work on the quasiholes in tMoTe${}_2$ \cite{bernevig2025ED, WuAnyon}. We discuss potential implications of this value in Sec.~\ref{Sec::discussion}.

Let us now consider the effect of displacement field on the anyon dispersion. The displacement field $D$ adds an interlayer electrostatic potential difference $V_D$ which contributes as an additional term in the $z$ component of the vector field $\Delta_z(\br)$:
\begin{equation}
    \Delta_{z}'(\br)=\Delta_{z}(\br)+\frac{1}{2}V_{D}
\end{equation}
The modified vector field $\bDelta'(\br)$ changes the layer pseudospin texture and the resulting dependence of AC quasiholes dispersion on displacement field is shown in FIG.~\ref{fig:tMoTe2-Dfield}.
We see from FIG.~\ref{fig:tMoTe2-Dfield} that (i) The ninefold degeneracy in AC quasihole dispersion is preserved in the presence of displacement field as expected.
(ii) The quasihole bandwidth increases with increasing displacement field, indicating a decrease in the quasihole mass. As anyons become lighter with displacement field, itinerant anyon physics should be enhanced. This is not surprising since a vertical displacement field pushes the electrons to one layer, which deforms the layer skyrmion textures leading to stronger field inhomogeneity. The experimental implications of this fact will be discussed in Sec.~\ref{Sec::discussion}.

\section{Origin of quasihole dispersion: quasihole guiding center and Quantum Geometry}
\label{Sec:Origin}
In this section, we will present an alternative way to derive anyon dispersion. This way will provide a more transparent physical picture for its origin along the lines of what we discussed in Sec.~\ref{sec:origin-quali}.

\subsection{Quasihole guiding center and coherent state path integral}
\label{Sec:qhGC}
Let us begin by rewriting the formula for anyon dispersion as
\begin{equation}
    \epsilon_{\v{\kappa}} = \frac{\langle \psi_{\v{\kappa}}^{\rm AC}|\hat V|\psi_{\v{\kappa}}^{\rm AC} \rangle}{\langle \psi_{\v{\kappa}}^{\rm AC}|\psi_{\v{\kappa}}^{\rm AC} \rangle} = \frac{\langle \psi_{\v{\kappa}}|\Gamma \hat V \Gamma |\psi_{\v{\kappa}} \rangle}{\langle \psi_{\v{\kappa}}|\Gamma^2|\psi_{\v{\kappa}} \rangle} 
    \label{epskND}
\end{equation}
This expression suggests the non-uniform magnetic field affects anyon dispersion in two ways. First, it alters the interactions $\hat V \mapsto \Gamma \hat V \Gamma$ in a spatially dependent way such that it breaks continuous magnetic translation to a discrete subgroup i.e. interaction acquires a center of mass dependence \cite{wang2021exact, schleith2025anyon}. Second, it alters the Hilbert space of quasiholes, essentially introducing a new ``metric" on that space that redefines wavefunction overlaps. Earlier discussions for mapping the interacting problem in non-uniform field to that of a uniform field either ignored the denominator \cite{wang2021exact} or used a special form of the magnetic field on the sphere to factor out the dependence on the denominator \cite{schleith2025anyon}. However, in the general case, we need to account for the non-uniform magnetic field in both numerator and denominator. Each of the numerator and the denominator has the form $\langle \psi_{\v{\kappa}}|\O|\psi_{\v{\kappa}} \rangle$  where $\O$ breaks continuous magnetic translation but commutes with lattice translations $T_\v{\alpha}$. The object $\langle \psi_{\v{\kappa}}|\O|\psi_{\v{\kappa}} \rangle$ can be understood as the effective operator acting on the space of quasiholes in a given momentum sector. Since there is only one quasihole state per momentum, this is just a number. 

 An alternative way to build an effective theory in the projected space of quasihole is to note that the quasihole states in the LLL have the form of coherent states which can be used to derive a coherent state path integral \cite{schleith2025anyon,flavin2011AbelianNonAbelianStatistics}. For simplicity, we will use a different gauge from Sec.~\ref{sec:Quasiholes} where the term $\sigma_{l,\v{\xi}}$ will be dropped. It will be restored later when comparing to earlier results. We can show that for this gauge choice, we have
\begin{gather}
    \langle \psi_\v{\xi}|\psi_{\v{\xi}'} \rangle = e^{-\frac{1}{4q}(\xi \bar \xi + \xi' \bar \xi' - 2 \xi' \bar \xi)} \\
    \int\frac{d^2\bxi}{2\pi q}|\psi_\bxi \rangle \langle \psi_{\bxi}| = \sum_{\v{\kappa}}|\psi_\v{\kappa} \rangle \langle \psi_\v{\kappa}| = \mathbbm{1}
\end{gather}
where $\mathbbm{1}$ denotes the projector on the space of quasiholes in the LLL.
 Following the standard procedure for deriving coherent state path integral, we obtain 
\begin{gather}
    \langle \psi_{\bxi_f} | e^{-\beta \O} | \psi_{\bxi_i}\rangle = \int_{\substack{\bxi(0)=\bxi_i\\\bxi(\beta)=\bxi_f}} \mathcal{D} \v{\xi} \,\, e^{-\int_0^\beta d \tau \mathcal L(\v{\xi}, \v{\dot \xi})}
    \label{PathIntegral} \\
     \mathcal L(\v{\xi}, \v{\dot \xi}) = \frac{\bar \xi\partial_\tau \xi}{2q} + \O_\v{\xi}, \qquad  O_\v{\xi} := \langle \psi_\v{\xi}| \O| \psi_\v{\xi} \rangle
\end{gather}
This path integral treats the quasihole position as a dynamical variable. The first term encodes the non-trivial Berry phase felt by a quasihole which is equivalent to a uniform magnetic field of $1/q$ per unit cell. 

An effective Hamiltonian for the space of quasiholes can be derived from the path integral (\ref{PathIntegral}) as follows. The first term in the Lagrangian implies that the momentum conjugate to the complex variable $\xi$ is $\frac{1}{2q} \bar \xi$. The standard quantization procedure promotes $\xi$ and $\bar \xi$ to operators satisfying
\begin{equation}
    [\hat \xi, \hat{\xi}^\dagger] = 2 q
\end{equation}
where the operator $\hat \xi$ satisfies $\hat \xi |\psi_\v{\xi} \rangle = \xi |\psi_\v{\xi} \rangle$. The operator $\hat \xi$ can be interpreted as a quasihole guiding center. The quantum Hamiltonian is obtained by first normal ordering the expression $\O_{\xi, \bar \xi}$ such that all $\bar \xi$ is to the left of $\xi$ and replacing them with the respective operators. To do this, we note that $\O_{\v{\xi}} = \O_{\v{\xi} + \v{\alpha}}$ (note that translations by the original lattice vectors $\v{a}$ that are not part of the new lattice $\v{\alpha}$ commute with $\O$ but switch the topological sector). Now we can introduce the Fourier transform $\O_\v{\xi} = \sum_\v{\alpha} \O_\v{\alpha} e^{\frac{i}{q} \v{\alpha} \wedge \v{\xi}}$. Here, we use the correspondence between the reciprocal lattice vector and real space which enables us to write any reciprocal lattice vector $\v{\beta}$ as $\frac{1}{q} \wedge \v{\alpha}$ for some lattice vector $\v{\alpha}$. Using the quantization procedure described above yields 
\begin{multline}
    : \O_\v{\xi} : =  \sum_\v{\alpha} \O_\v{\alpha} :e^{\frac{i}{q} \v{\alpha} \wedge \v{\xi}}: = \sum_\v{\alpha} \O_\v{\alpha} e^{-\frac{1}{2q} \alpha \xi^*} e^{\frac{1}{2q} \alpha^* \xi} \\
    \mapsto \hat \O = \sum_\v{\alpha} \O_\v{\alpha} e^{-\frac{1}{2q} \alpha \hat \xi^\dagger} e^{\frac{1}{2q} \alpha^* \hat \xi} = \sum_\v{\alpha} \O_\v{\alpha} e^{\frac{\v{\alpha}^2}{4q}} e^{\frac{1}{2q} (\alpha^* \hat \xi - \alpha \hat \xi^\dagger)}
\end{multline}
The operator $e^{\frac{1}{2q} (\alpha^* \hat \xi - \alpha \hat \xi^\dagger)}$ satisfies the magnetic algebra (\ref{QHMagneticAlgebra}) and can be identified with the magnetic translation operator $T_\v{\alpha}$, which implies that the path integral (\ref{PathIntegral}) describes the quantum Hamiltonian
\begin{equation}
    \hat \O = \sum_\v{\alpha} \O_\v{\alpha} e^{\frac{\v{\alpha}^2}{4q}} T_\v{\alpha}
\end{equation}
To compare with the results of Sec.~\ref{sec:Quasiholes}, we gauge transform to the gauge used in that section by taking $T_\v{\alpha} \mapsto \sigma_{0,\v{\xi}}^* T_\v{\alpha} \sigma_{0,\v{\xi}} = e^{i \kappa_0 \cdot \v{\alpha}} T_\v{\alpha}$.
Taking the eigenstates of $T_\v{\alpha}$ to be momentum eigenstates $|\psi_\v{\kappa} \rangle$ satisfying $T_\v{\alpha} |\psi_\v{\kappa} \rangle = \eta_\v{\alpha} e^{i \v{\kappa} \cdot \v{\alpha}} |\psi_\v{\kappa} \rangle$, we get the eigenvalues
\begin{equation}
    \lambda_\v{\kappa} = \sum_\v{\alpha} \O_\v{\alpha} \eta_\v{\alpha} e^{\frac{\v{\alpha}^2}{4q}} e^{i (\v{\kappa} - \v{\kappa_0}) \cdot \v{\alpha}}
    \label{lambdak}
\end{equation}
We now want to show that $\lambda_\v{\kappa} = \langle \psi_{\v{\kappa}}|\O|\psi_{\v{\kappa}} \rangle$ with $|\psi_\v{\kappa} \rangle$ related to $|\psi_\v{\xi} \rangle$ through Eqs.~(\ref{PsixiPsik}) and (\ref{Fxik}). First, use  Eq.~(\ref{PsixiPsik}) to write
\begin{equation}
    \O_\v{\xi} = \sum_{\v{\kappa}} |F_{\v{\kappa}}(\v{\xi})|^2 \langle \psi_\v{\kappa}|\O|\psi_\v{\kappa} \rangle
    \label{Oxi}
\end{equation}

Now using Eq.~(\ref{MagneticBloch}) and (\ref{SigmaIdentity}), we can write 
\begin{equation}
    |F_{\v{\kappa}}(\v{\xi})|^2 = \frac{1}{N_\Phi} \sum_{\v{\alpha}} \eta_{\v{\alpha}} e^{-\frac{1}{4q} \v{\alpha}^2} e^{\frac{i}{q} \v{\alpha} \wedge \v{\xi}} 
    e^{-i\v{\alpha} \cdot (\v{\kappa} - \v{\kappa}_0)}
    \label{Fsquared}
\end{equation}
which implies
\begin{equation}
    \frac{1}{N_\Phi}\sum_{\v{\kappa}} e^{-i\v{\alpha} \cdot (\v{\kappa} - \v{\kappa}_0)} \langle \psi_\v{\kappa}|\O|\psi_\v{\kappa} \rangle = \eta_\alpha e^{\frac{1}{4q} \v{\alpha}^2}  \O_{\v{\alpha}}
\end{equation}
Comparing with Eq.~(\ref{lambdak}), we see that this implies $\lambda_\v{\kappa} = \langle \psi_\v{\kappa}|\O|\psi_\v{\kappa} \rangle$.

We can understand the mapping between $\O_\v{\xi}$ and $\O_{\v{\kappa}}$ in terms of the twisted Weierstrass transform defined in Eq.~(\ref{TwistedWeierstrass}) for the lattice $\Lambda^*_q$ which acts as $\hat W_{\Lambda^*_q}^\eta e^{i \v{\alpha} \cdot (\v{\kappa} - \v{\kappa}_0)} = \eta_{\Lambda^*_q} e^{\frac{1}{4q} \Delta_{\v{\kappa}}} e^{i \v{\alpha} \cdot (\v{\kappa} - \v{\kappa}_0)} =  \eta_{\v{\alpha}} e^{-\frac{\v{\alpha}^2}{4 q}} e^{i \v{\alpha} \cdot (\v{\kappa} - \v{\kappa}_0)}$. Comparing with Eq.~(\ref{Fsquared}), we see that we can write 
\begin{align}
    |F_{\v{\kappa}}(\v{\xi})|^2 &= \frac{1}{N_\Phi}\hat W_{\Lambda^*_q}^\eta\sum_{\v{\alpha}} e^{-i \v{\alpha} \cdot (\v{\kappa} - \v{\kappa}_0)} e^{\frac{i}{q} \v{\alpha} \wedge \v{\xi}} \nonumber \\
    & = \frac{1}{N_\Phi} \hat W_{\Lambda_q^*}^\eta \delta_{\v{\kappa},\v{\kappa}_0 +\wedge\frac{\v{\xi}}{q}}
\end{align}
This leads to the simple relation
\begin{equation}
    \O_\v{\xi} = \langle \psi_\v{\xi}|\hat O|\psi_{\v{\xi}} \rangle = \hat W_{\Lambda_q^*}^\eta  \langle \psi_\v{\kappa}|\O|\psi_{\v{\kappa}} \rangle |_{\v{\kappa} = \v{\kappa}_0 + \wedge \frac{\v{\xi}}{q} }
\end{equation}

This implies that, in principle, one can invert this relation to obtain the dispersion from the coherent state potential
$\O_\v{\xi} = \langle \psi_\v{\xi} | \O | \psi_\v{\xi} \rangle$,
which corresponds to a plasma correlation function. In practice, however, this inversion is numerically unstable due to the growing Gaussian factor in Eq.~(\ref{lambdak}), which makes it difficult to reliably extract the dispersion from real-space plasma Monte Carlo. Under additional assumptions---such as imposing \emph{a priori} that the Fourier components decay sufficiently rapidly---the inversion can still be performed for the lowest Fourier components, for example in the limit of a weak field where only the leading harmonic is retained. We present some numerical results of this procedure in the SM. We further note that the difficulty in extracting the dispersion from plasma Monte Carlo is closely related to the difficulty of determining the guiding center structure factor $\langle \rho^{\rm GC}_\v{q} \rho^{\rm GC}_{-\v{q}} \rangle$ at large $\v{q}$, since Monte Carlo techniques naturally access only correlators of the unprojected physical densities $\rho_\v{q} = e^{-\v{q}^2/4} \rho^{\rm GC}_\v{q}$ \cite{WangLatticeMC}.

The path integral formulation provides a transparent physical picture of the origin of quasihole dispersion. Any perturbation that reduces continuous magnetic translation symmetry to a discrete subgroup is sufficient to generate a periodic potential for the quasihole. In the experimentally relevant case of MoTe${}_2$, this periodicity primarily arises from the non-uniform quantum geometry encoded in the K\"ahler potential $Q(\v{z})$. However, this mechanism alone does not generate a dispersion. A second ingredient is required: the anyon Berry phase, which is equivalent to a background magnetic field of strength $1/q$ per unit cell. The quasihole dispersion therefore emerges from the combined effects of the anyon Berry phase (anyon quantum geometry) and the non-uniform potential generated by the underlying electron quantum geometry.

We highlight one important subtlety regarding operator interpretation. Although the identification of $\hat{\xi}$ as a quasihole guiding center may suggest an analogy with LLL projection, the procedures are not equivalent. In the LLL projection, one places all $\bar z$ operators to the left of $z$ and replaces each $\bar z$ by $2 \frac{\partial}{\partial z}$. Because acting with $z$ on LLL wavefunctions raises angular momentum (analogous to a creation operator) while $\frac{\partial}{\partial z}$ lowers it (analogous to an annihilation operator), the resulting projection corresponds to \emph{anti-normal} ordering. By contrast, the coherent state path integral approach maps a quantum operator to its classical potential $\O_\v{\xi} = \langle \psi_\v{\xi} | \O | \psi_\v{\xi} \rangle$, which is the Husimi symbol of the operator and is obtained by \emph{normal ordering}, i.e., placing all annihilation operators---of which coherent states are eigenstates---to the left. In the SM, we review the mapping between quantum operators and their classical symbols to clarify this important distinction in operator ordering.

\subsection{Quasihole Hamiltonian and Berry phase}
\label{Sec:qhberry}
In the previous section, we developed a path integral formulation to compute the numerator and denominator of the energy dispersion expression, Eq.~(\ref{epskND}), separately. However, this makes it not possible to associate the path integral and the associated quantum operator with the physical Hamiltonian of the system. In this section, we will show how we can derive a path integral formulation whose quantization captures the full projected Hamiltonian. This means it will simultaneously capture the changes in the Hilbert space and the periodic potential generated by the non-uniform field. Here, we will use the same gauge as in the previous section which differs from the gauge of Sec.~\ref{sec:Quasiholes} by dropping the term $\sigma_{l,\v{\xi}}$.

To do this, we consider the normalized quasihole states for an AC band given by
\begin{equation}
    |\psi^{\rm AC}_{\v{\xi}} \rangle = e^{-\frac{1}{2} \Q(\bxi)} \Gamma |\psi_{\v{\xi}} \rangle, \qquad e^{\Q(\bxi)} = \langle \psi_\bxi|\Gamma^2|\psi_\bxi \rangle
\end{equation}
Note that $\Gamma$ can be written in second quantization as $\Gamma = e^{-\frac{1}{2}\sum_\v{k} R_\v{k} c_{\v{k}}^\dagger c_{\v{k}}}$. Using Eq.~(\ref{PsixiPsik}), we can write
\begin{align}
    |\psi^{\rm AC}_\bxi \rangle &= \sum_\bk  e^{-\frac{1}{2} \Q(\bxi)} F_{\v{\kappa}}(\bxi) \Gamma |\psi_\v{\kappa} \rangle \nonumber \\ 
    &= \sum_\v{\kappa}  e^{-\frac{1}{2} \Q(\bxi)} F_{\v{\kappa}}(\bxi) e^{\frac{1}{2}\R(\v{\kappa})} |\psi^{\rm AC}_\v{\kappa} \rangle
\end{align}
 Here, we defined $\R(\v{\kappa})$ such that $|\psi^{\rm AC}_\v{\kappa} \rangle$ is normalized. It is given explicitly by
\begin{equation}
     e^{\R(\v{\kappa})} = \langle \psi_\v{\kappa}|\Gamma^2| \psi_\v{\kappa} \rangle 
\end{equation}
$\Q(\bxi)$ and $\R(\v{\kappa})$ are related via
\begin{equation}
    e^{\Q(\bxi)} = \sum_\v{\kappa} |F_{\v{\kappa}}(\bxi)|^2 e^{\R(\v{\kappa})} = \hat W_{ \Lambda_q^*}^\eta e^{\R(\v{\kappa})} |_{\v{\kappa} = \v{\kappa}_0 + \wedge \frac{\v{\xi}}{q} }
    \label{eq:Qxi-Rk-mapping}
\end{equation}

That is, $e^{\Q(\v{\xi})}$ is a smoothened version of $e^{\R(\v{\kappa})}$.
We can now derive a path integral using $|\psi^{\rm AC}_\bxi \rangle$ as coherent states which satisfy the resolution of unity
\begin{equation}
    \int \frac{d^2 \bxi}{2\pi q} \F(\v{\xi}) | \psi^{\rm AC}_\bxi \rangle \langle \psi^{\rm AC}_\bxi| = \sum_{\bk} | \psi^{\rm AC}_\v{\kappa} \rangle \langle \psi^{\rm AC}_{\v{\kappa}}| 
\end{equation}
where $\F(\bxi)$ is an integration measure defined through the reproducing Kernel condition
\begin{equation}
    \int \frac{d^2 \bxi}{2\pi q} \F(\v{\xi}) 
    \langle \psi^{\rm AC}_\v{\omega}| \psi^{\rm AC}_\v{\xi} \rangle \langle \psi^{\rm AC}_\v{\xi}|\psi^{\rm AC}_\v{\zeta} \rangle = \langle \psi^{\rm AC}_\v{\omega}| \psi^{\rm AC}_\v{\zeta} \rangle
\end{equation}
which should hold for all 2D vectors $\v{\omega}$ and $\v{\zeta}$. Note that the integration measure $\F$ is fully fixed by the \emph{diagonal} overlaps $e^{\Q(\v{\xi})} = \langle \psi_\v{\xi}|\Gamma^2|\psi_\v{\xi} \rangle$. To see this, note that we can define $|\xi \rangle = e^{\frac{1}{4q} \v{\xi}^2} \Gamma |\psi_{\v{\xi}} \rangle$ which depends holomorphically on $\xi$. This allows us to fix the overlaps $\langle \psi^{\rm AC}_\v{\omega}| \psi^{\rm AC}_\v{\zeta} \rangle$ from the knowledge of $\Q(\v{\xi})$ by analytic continuation. In SM, we provide a systematic discussion of this procedure using the formalism of K\"ahler or geometric quantization.

We can express $\F$ explicitly in terms of $\R(\v{\kappa})$ as $\F(\bxi) = e^{\Q(\bxi)} [\hat W^\eta_{\Lambda_q^*}]^{-1} e^{-\R(\v{\kappa})}|_{\v{\kappa} =\v{\kappa}_0 +\wedge\v{\xi}/q}$ and note that $\R(\v{\kappa})$ can be obtained from $\Q(\v{\xi})$ by inverting Eq.~(\ref{eq:Qxi-Rk-mapping}), which enables us to construct the measure $\F$ for a given $\Q$.
It may seem worrisome that we need the inverse $\eta$-Weierstrass transform to define $\F(\v{\xi})$. However, as we noted earlier, the inverse $\eta$-Weierstrass transform is well-defined on any finite system, even if it produces highly oscillatory functions, so we can formally do the inversion before taking the thermodynamic limit. Furthermore, the function $\F(\v{\xi})$ does not explicitly appear in any of the final answers and is only a part of the path integral measure. 

The resulting path integral will have the same form as Eq.~(\ref{PathIntegral}) with the path integral measure $\D \v{\xi}$ replaced by $\D \v{\xi} \F(\v{\xi})$ and the Lagrangian replaced by
\begin{equation}
    {\mathcal L} = i \A_\mu[\v{\xi}] \dot \xi_\mu + \O_\v{\xi}^{\rm AC}
\end{equation}
where $\O_\v{\xi}^{\rm AC} = \langle \psi_\v{\xi}^{\rm AC}|\O|\psi_\v{\xi}^{\rm AC} \rangle$ and $\A_\mu$ given by 
\begin{equation}
    \A = \frac{1}{2}(\A_x -iA_y) = -i \langle \psi_\v{\xi}^{\rm AC}|\partial_{\xi} \psi_\v{\xi}^{\rm AC} \rangle= -\frac{i}{4q}\bar\xi - \frac{i}2 \partial_{\xi} \Q(\xi,\bar \xi)
\end{equation}

This corresponds to the effective magnetic field
\begin{equation}
    \mathcal B(\xi,\bar \xi) = \frac{1}{q} + 2 \partial_\xi \partial_{\bar \xi} \Q(\xi,\bar \xi)
    \label{BQH}
\end{equation}
Since $\Q$ is a periodic function in the quasihole UC, the second term is a periodic field modulation whose average value in a UC vanishes. The first term corresponds to a reduced average magnetic field felt by the quasihole which averages to $1/q$ over the original UC associated with the lattice $\Lambda$ and to 1 over quasihole (enlarged) UC associated with the lattice $\Lambda_q$.

In SM, we show how this action reproduces the full quantum Hamiltonian in the quasihole Hilbert space whose spectrum provides the quasihole dispersion, Eq.~(\ref{epskND}), using the formalism of geometric quantization. However, beyond reproducing that, this Hamiltonian also provides information about the quasihole Berry phase or alternatively, the effective magnetic field felt by the quasihole. As we can see from Eq.~(\ref{BQH}), in addition to the uniform piece $1/q$, the quasihole also feels periodic modulations related to the quasihole K\"ahler potential $\Q$. 

\begin{figure*}
    \centering
    \includegraphics[width=0.9\linewidth]{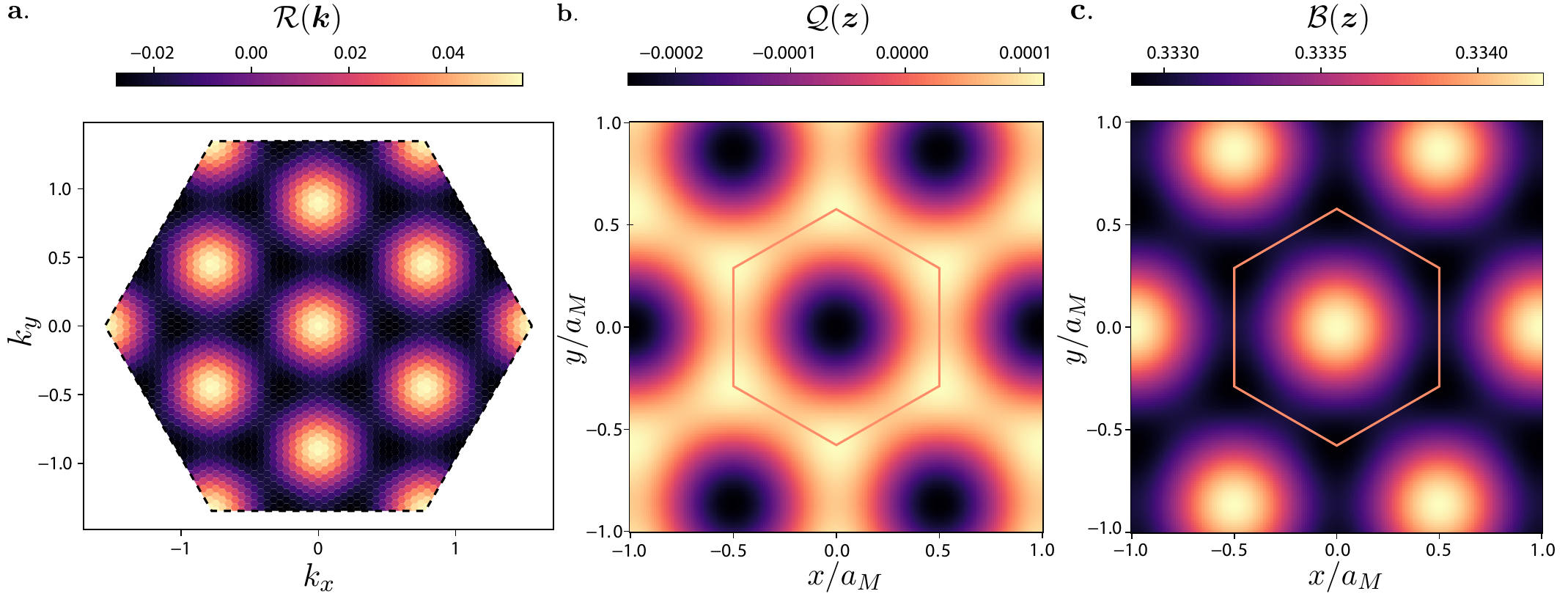}
    \caption{Quasihole quantum geometry in AC band with first harmonic approximation, $K=0.25$. Reconstructed from weak field approximation and finite size scaling, $N_{\Phi}=2500$. \textbf{a}. Momentum space K\"ahler potential $\R(\bk)$. \textbf{b}. Real space K\"ahler potential $\Q(\bz)$. \textbf{c}. Effective magnetic field $\B(\bz)$. In b-c, the hexagon outlines Wigner-Seitz moir\'e unit cell.} 
    \label{fig:omegak}
\end{figure*}

Taking the limit of weak field for simplicity, where we can expand the expression for $\Q$ to linear order in K, we can compute the non-uniform part of the quasihole K\"ahler potential $\Q$ by extrapolating to the thermodynamic limit. The results are shown in FIG.~\ref{fig:omegak} (b) for $K = 0.25$. We see that the modulations are extremely small (about $0.03\%$) and that the quasihole feels an effectively uniform magnetic field. The smallness of this modulation may suggest that it has no effect on the dispersion. However, note that the magnetic field affects the dispersion through the momentum space K\"ahler potential $\R(\v{\kappa})$ which is related to $\Q(\v{\xi})$ by the inverse Weierstrass transform, Eq.~(\ref{eq:Qxi-Rk-mapping}). Indeed, since $e^{\Q(\v{\xi})} \approx 1 + \Q(\v{\xi})$ contains only the leading Fourier harmonics in the weak field limit, we can perform the inverse Weierstrass transform explicitly to obtain $\R(\v{\kappa})$. The result is shown in FIG.~\ref{fig:omegak} (a) and exhibits more substantial variations (about $7\%$), which is non-negligible and can be of the same order as the variations coming from the Hamiltonian matrix elements. This contribution to the dispersion comes purely from the non-uniform quantum geometry of quasiholes. 

\subsection{Toward a multi-anyon theory: microscopic Lagrangian and long distance limit}
In this section, we outline how our theory can be generalized to the case of multiple anyons. This can address questions of quasihole binding and can serve as a starting point for a controlled theory for anyon phases which: (i) fully contains the microscopic information, and (ii) only accounts for the quasihole degrees of freedom. We also anticipate the theory to reproduce the Chern-Simons effective theory at long distances but we leave such discussion to follow-up work. We consider $m$ quasiholes parametrized by $m$ complex numbers $\xi_1,\dots, \xi_m$. For convenience, we introduce the $2m$ complex vector $\xi = (\xi_1,\dots,\xi_m)$. A state describing $m$ Laughlin quasiholes can be easily constructed by including the appropriate factors of $f(z - \xi_l)$ (or $z - \xi_l$ on the plane) in the Laughlin wavefunction. We can always choose the dependence on the quasihole coordinates to be holomorphic. Here, we do not separate the part of the K\"ahler potential that gives rise to a uniform Berry curvature, so our $\K$ is not the same as $\Q$ in the previous two sections. As a result, we can define a K\"ahler potential $\K(\bar \xi, \xi):= \ln\langle \xi|\xi \rangle$ where $|\xi \rangle$ denotes the multi-quasihole state which depends holomorphically on quasihole coordinates. In this case, the space of quasihole states defines a K\"ahler manifold with $m$ complex dimensions. We can define a so-called reproducing Kernel $\F(\bar \xi, \xi)$ by requiring
\begin{equation}
    \int \D \xi \F(\bar \xi, \xi) e^{-\K(\bar \xi, \xi)} e^{\K(\bar \omega,\xi)} f(\bar \xi) = f(\bar \omega)
\end{equation}
for any anti-holomorphic function $f(\bar z)$, where $D \xi = \prod_{l=1}^m d^2 \v{\xi}_l$. This is equivalent to the existence of a resolution of unity of the form
\begin{equation}
    \int \D \xi \F(\bar \xi, \xi) e^{-\K(\bar \xi, \xi)} |\xi \rangle \langle \xi| = \mathbbm{1}
\end{equation}
and gives rise to a path integral with the integration measure $D \xi \F(\bar \xi, \xi) e^{-\K(\bar \xi, \xi)}$ and the Lagrangian 
\begin{equation}
    \mathcal L = i \sum_l \A_{l}[\bar\xi,\xi] \dot \xi_{l} + V[\bar\xi,\xi]
\end{equation}
where $\A_{l}[\bar\xi,\xi] = -i \partial_{\xi_l} \K(\bar \xi, \xi)$ and $V[\bar\xi,\xi] = \frac{\langle \xi|\hat V|\xi \rangle}{\langle \xi|\xi \rangle}$.
The Berry phase term encodes the Hilbert space structure and reduces at long distances to the expected mutual Berry phase between different quasiholes. To see this, separate the long-distance part by writing 
 $\langle \xi|\xi\rangle = e^{\Q(\bar \xi, \xi)} e^{\frac{1}{4q} \sum_l \bar \xi_l \xi_l} \prod_{i<j} |f(\xi_i-\xi_j)|^{-\frac{2}{q}}$, the Berry phase term is
\begin{align}
    \A_{l} = -\frac{i\bar\xi_l}{2q} + \frac{i}{q}\sum_{i\neq l}\zeta(\xi_l-\xi_i) -i\partial_{\xi_l}\Q(\bar\xi,\xi)
\end{align}
where the modified Weierstrass zeta function $\zeta(z) = \frac{\sigma'(z)}{\sigma(z)}$ which behaves as $1/z$ in the thermodynamic limit.

The potential $V[\xi]$ encodes both short distance correlations responsible for dispersion and binding as well as long distance interactions. If the particle interaction is short-ranged, this reduces to a sum of single particle terms due to plasma screening. This theory generalizes the single quasihole guiding center coordinate defined in this work to a \emph{many-body} guiding center coordinate which does not decompose into a sum of individual guiding centers due to the mutual Berry phases between different quasiholes. The Hamiltonian corresponding to this theory is obtained using the formalism of geometric quantization reviewed in supplemental material. We leave a more detailed analysis of this theory for multiple quasiholes and how it can be used to study quasihole binding and collective phases to follow-up works.

\section{Discussion}
\label{Sec::discussion}
We close by critically examining the assumptions underlying our analysis and by discussing several implications of our results for experiments.

In this work, we study how anyon dispersion arises from non-uniform quantum geometry in ideal (Aharonov–Casher) bands. Our analysis is fully controlled when considering an interaction of the form $\alpha \hat V_{\rm TK} + \hat V$ with large $\alpha$, for the case of doping quasiholes on top of Laughlin $\nu = 1/q$ states. In this regime, we can show that the space of quasiholes forms a manifold of zero modes of $\hat V_{\rm TK}$, separated by a large gap $\alpha$ from the remaining many-body states. As a result, we can project $\hat V$ onto this zero-mode subspace and obtain an effective quasihole Hamiltonian. The form $\alpha \hat V_{\rm TK} + \hat V$ is natural for short-range interactions which admit an expansion in terms of Trugman-Kivelson pseudopotentials. Furthermore, our results indicate that, for most realistic interaction, the gap to other excitations far exceeds the energy scale associated with the projected Hamiltonian justifying our approach away from this idealized limit. Focusing on the case of a single quasihole, we show that one can construct quasihole momentum eigenstates whose energies form a quasihole band, with a bandwidth set by the details of the interaction and by fluctuations in the quantum geometry of the parent band.
Beyond the analytically controlled limit where the space of low-energy quasihole excitations is well-separated from other excitations, our approach can be viewed as a variational scheme that constructs momentum-resolved quasihole wavefunctions and evaluates their energies. Although we focus on quasiholes--because they can be sharply defined as zero modes of pseudopotentials, enabling a controlled analysis--we anticipate similar results for the dispersion of quasielectrons once analogous momentum eigenstates are constructed in the quasielectron sector.

Beyond our specific results for quasiholes in ideal bands, we identify a generic mechanism for anyon dispersion that depends only on the effective periodic potential and the effective magnetic field experienced by the anyons. The potential can arise either from non-uniform quantum geometry or from an actual periodic potential, and it is expected for any charged quasiparticle whose size is not too large compared to the period. The effective magnetic field is generically expected for a charged quasiparticle and encodes the nontrivial \emph{many-body} quantum geometry of the anyons. Thus, we anticipate that this mechanism applies to other charged quasiparticles, such as charged bound states of quasiholes and quasielectrons.

Our results are directly applicable to twisted MoTe${}_2$, where flat, isolated, topological bands with nearly ideal quantum geometry host FQAH states at several filling fractions—for example, at $\nu = 2/3$ as seen in recent experiments \cite{xu2023observation,cai2023signatures, park2023observation, zeng2023thermodynamic, MoTe2SC}. We also expect the qualitative aspects of our analysis to be relevant in rhombohedral graphene, where FQAH states have been observed despite the absence of isolated bands. The existence of finite anyon dispersion implies that the physics of FQAH states for small doping differs qualitatively from that of FQH states in a uniform magnetic field in the following way.

When a finite density of anyons—say quasiholes with density $\nu_{\rm QH}$—is doped into a FQAH state, the anyons feel an average statistical Chern–Simons magnetic field proportional to their density, in addition to the background $1/q$ field experienced in the $\nu = 1/q$ Laughlin state. At small densities, the primary effect of this excess statistical magnetic field is the formation of Landau levels at the minima of the single-anyon dispersion. These Landau levels have a spacing given by the cyclotron gap $\omega_{c,\rm QH} \sim B_{\rm CS}/m_{\rm QH} \sim \nu_{\rm QH}/m_{\rm QH}$ \cite{LaughlinAnyonSC, FetterHannaLaughlin, 
laughlin1988, WilczekWittenHalperinAnyonSC,shi2025doping, nosov2025anyonsuperconductivityplateautransitions, shi2025anyondelocalizationtransitionsdisordered}. In the presence of disorder, which induces a mean-free time $\tau_{\rm QH}$ for the quasiholes, the system exhibits two qualitatively distinct regimes depending on the value of the dimensionless parameter $\omega_{c,\rm QH}\tau_{\rm QH} \sim \nu_{\rm QH}\tau_{\rm QH}/m_{\rm QH}$. When this parameter is small, disorder dominates pushing the delocalized states within each Landau level to higher energies so that they do not contribute to transport \cite{nosov2025anyonsuperconductivityplateautransitions, shi2025anyondelocalizationtransitionsdisordered}. This is the physics of quantum Hall plateaus, where doped quasiparticles are localized by disorder and the response is governed by the parent FQH state. Such behavior is expected in disordered samples (small $\tau_{\rm QH}$), at very small anyon doping (small $\nu_{\rm QH}$), or for large quasihole mass. In conventional FQH systems, $m_{\rm QH} = \infty$, placing the system squarely in this disorder-dominated regime for arbitrarily weak disorder.

By contrast, for anyons with finite mass—realized in FQAH systems as we have shown—the system can undergo a transition, as a function of doping $\nu_{\rm QH}$ and for sufficiently clean samples, into a regime where Landau quantization of the quasiholes modifies the response and stabilizes a different phase. The nature of this phase depends on the lowest-energy anyon species present. For instance, for the $1/3$ Laughlin state, a re-entrant integer quantum anomalous Hall (RIQAH) state arises if the $1/3$ anyons are the lowest-energy excitations, whereas a superconductor emerges if the $2/3$ anyons are lowest in energy (i.e., if $1/3$ anyons pair to form $2/3$ anyons) \cite{shi2025doping}. 
Within this scenario, our results also allow us to determine the doping at which a transition from a FQAH to a RIQAH phase is expected for a given disorder strength. Conversely, they enable estimates of disorder strength in samples where such a transition is observed. For example, in the sample of Ref.~\cite{MoTe2SC}, a transition is observed at electron or hole doping of roughly $\sim 0.02$. Given the anyon bandwidth of $\sim 1$ meV we find, this gives an estimate of $\tau_{\rm QH,c} \sim 10$ ps.

In this work, we have focused on the energetics of single $1/3$ quasiholes and have not computed the anyon-anyon interaction, even though it is important in determining the collective behavior of anyons. At the level of a rough estimate, however, one may compare the single-quasihole kinetic scale to the interaction at the typical inter-anyon spacing.
 In particular, the effective mass extracted from the local curvature of the quasihole band helps distinguish regimes in which the doped anyons behave approximately as a weakly interacting gas from regimes in which residual interactions may become important. 
From the local quadratic curvature of the realistic tMoTe$_2$ quasihole dispersion near a band minimum, we estimate an effective quasihole mass of order $m_{\mathrm{QH}}\sim 0.5\,m_e$. At quasihole filling $\nu_{\mathrm{QH}}$, the characteristic inter-anyon spacing is $a_{\mathrm{QH}}\sim 1/\sqrt{\pi n_{\mathrm{QH}}}\sim \sqrt{2}\,\ell_B/\sqrt{\nu_{\mathrm{QH}}}$. For unscreened Coulomb, one can define a standard $r_s$-like parameter $r_s^{\mathrm{QH}}\sim a_{\mathrm{QH}}/a_B^{\mathrm{QH}}$ where $a_B^{\mathrm{QH}}=\epsilon\hbar^2/m_{\mathrm{QH}}(e/3)^2$ so interaction effects become increasingly important upon lowering the density, with a crystal phase expected at sufficiently low anyon density. By contrast, the gate-screened Coulomb interaction relevant to experiments decays exponentially in the long-distance limit, a more appropriate dimensionless measure is 
$r_{\mathrm{scr}}\sim V_{\mathrm{scr}}(a_{\mathrm{QH}})/\frac{\hbar^2}{(m_{\mathrm{QH}}a_{\mathrm{QH}}^2)}$. 
 Its qualitative behavior is controlled by the ratio $a_{\mathrm{QH}}/d_{\mathrm{scr}}$: when $a_{\mathrm{QH}}\gg d_{\mathrm{scr}}$, the interaction is strongly suppressed and the system approaches a dilute anyon gas; when $a_{\mathrm{QH}}\ll d_{\mathrm{scr}}$, the interaction is effectively unscreened on the scale of the mean spacing and the behavior crosses over to the usual Coulomb regime; and when $a_{\mathrm{QH}}\sim d_{\mathrm{scr}}$, the interaction and kinetic scales can be comparable.
 Determining whether the regime supports interaction-driven phases such as anyon liquids or anyon superconductivity \cite{shi2025doping, divic2024anyon, nosov2025anyonsuperconductivityplateautransitions, shi2025anyondelocalizationtransitionsdisordered,PICHLERanyonSC,han2025anyon,Pichlerspectral} requires a microscopic calculation of the residual anyon--anyon interaction and binding energies, which we leave for future work.

Our results in FIG.~\ref{fig:tMoTe2-Dfield} also indicate that anyon dispersion can be tuned by applying vertical displacement field with charge $e/3$ anyons getting lighter as the displacement field increases. The decrease in the $1/3$ anyon mass with displacement field suggests the phase boundary separating the $\nu = 2/3$ FQAH from the RIQAH should move \emph{towards} $2/3$ with increasing displacement field. While our calculation was done only for quasiholes (doping $\nu = 2/3 + \delta$), we expect displacement field to increase anyon dispersion also for quasi-electrons (doping $\nu = 2/3 - \delta$) since it increases the inhomogeneity of the AC magnetic field. On the quasi-electron side, this is consistent with the experimental observation \cite{MoTe2SC} that the phase boundary to the RIQAH state at doping $\nu = 2/3 - \delta$ ($\delta > 0$) moves to smaller $\delta$ with displacement field. For doping $\nu = 2/3 + \delta$, there is no RIQAH at zero displacement field, but when it appears at finite field, the phase boundary also moves towards $2/3$ with field consistent with our picture. We note also that binding of $e/3$ anyons into $2e/3$ anyons, if it exists, is expected to be destabilized as $e/3$ anyon dispersion increases (since the bigger object $2e/3$ is heavier and would be disfavored by kinetic energy for small anyon mass). Thus, our results are consistent with a picture where $e/3$ anyons bind to form $2e/3$ anyons for doping $\nu = 2/3 + \delta$ at zero field, which leads to the anyon SC phase, but the binding is lost at some finite displacement field, leading to a RIQAH \cite{nosov2025anyonsuperconductivityplateautransitions}.

Finally, we emphasize that while our approach has been applied here to anyon dispersion, it should also enable the systematic study of more general problems involving anyon phases and interactions by working \emph{entirely within the Hilbert space of anyon degrees of freedom}. This strategy has the advantage of incorporating only the minimal ingredients required to describe anyon physics, while retaining full information about the short-distance processes responsible for anyon dispersion and binding. Thus, our approach paves the way to a systematic and fully controlled framework that bridges long-wavelength effective theories and microscopic models capturing material-specific details.

\section*{Acknowledgments}
We thank Jie Wang, Tobias Wolf, Nicolás Morales-Durán, Amir Yacoby, Ashvin Vishwanath, Pavel Nosov, Zhaoyu Han, and Patrick Ledwith for helpful discussion. Eslam Khalaf and Qingchen Li were supported by NSF CAREER grant DMR no. 2441781. Zihan Yan was supported by the Army Research Office under Grant number: W911NF-23-1-0110.
This research is funded in part by the Gordon and Betty Moore Foundation’s EPiQS Initiative, Grant GBMF8683 to T.S.
The authors thank the Harvard FAS Research
Computing (FASRC) for computational support.

\let \temp \v
\let \v \oldv
\bibliography{bibliography}

@article{schleith2025anyon,
  title={Anyon dispersion from non-uniform magnetic field on the sphere},
  author={Schleith, Mina-Lou and Soejima, Tomohiro and Khalaf, Eslam},
  journal={arXiv preprint arXiv:2506.11211},
  year={2025}
}

@article{WangLatticeMC,
  title = {Lattice Monte Carlo for quantum Hall states on a torus},
  author = {Wang, Jie and Geraedts, Scott D. and Rezayi, E. H. and Haldane, F. D. M.},
  journal = {Phys. Rev. B},
  volume = {99},
  issue = {12},
  pages = {125123},
  numpages = {13},
  year = {2019},
  month = {Mar},
  publisher = {American Physical Society},
  doi = {10.1103/PhysRevB.99.125123},
  url = {https://link.aps.org/doi/10.1103/PhysRevB.99.125123}
}

@article{HaldaneModifiedWeierstrass,
  title={A modular-invariant modified Weierstrass sigma-function as a building block for lowest-Landau-level wavefunctions on the torus},
  author={Haldane, FDM},
  journal={Journal of Mathematical Physics},
  volume={59},
  number={7},
  year={2018},
  publisher={AIP Publishing}
}

@article{Wang2021exact,
  title = {Exact {{Landau Level Description}} of {{Geometry}} and {{Interaction}} in a {{Flatband}}},
  author = {Wang, Jie and Cano, Jennifer and Millis, Andrew J. and Liu, Zhao and Yang, Bo},
  year = {2021},
  month = dec,
  journal = {Phys. Rev. Lett.},
  volume = {127},
  number = {24},
  pages = {246403},
  issn = {0031-9007, 1079-7114},
  doi = {10.1103/PhysRevLett.127.246403},
  urldate = {2023-10-21},
  langid = {english}
}

@article{ledwithFractionalChernInsulator2020a,
	author = {Ledwith, Patrick J. and Tarnopolsky, Grigory and Khalaf, Eslam and Vishwanath, Ashvin},
	doi = {10.1103/PhysRevResearch.2.023237},
	issn = {2643-1564},
	journal = {Phys. Rev. Research},
	journaltitle = {Phys. Rev. Research},
	month = may,
	number = 2,
	pages = {023237},
	publisher = {{American Physical Society}},
	shorttitle = {Fractional {Chern} insulator states in twisted bilayer graphene},
	title = {Fractional {{Chern}} Insulator States in Twisted Bilayer Graphene: {{An}} Analytical Approach},
	url = {https://link.aps.org/doi/10.1103/PhysRevResearch.2.023237},
	urldate = {2022-04-22},
	volume = 2,
	year = 2020}

@article{ledwith_vortexability_2023,
	title = {Vortexability: {A} {Unifying} {Criterion} for {Ideal} {Fractional} {Chern} {Insulators}},
	volume = {108},
	issn = {2469-9950, 2469-9969},
	shorttitle = {Vortexability},
	url = {http://arxiv.org/abs/2209.15023},
	doi = {10.1103/PhysRevB.108.205144},
	number = {20},
	urldate = {2025-01-29},
	journal = {Physical Review B},
	author = {Ledwith, Patrick J. and Vishwanath, Ashvin and Parker, Daniel E.},
	month = nov,
	year = {2023},
	note = {arXiv:2209.15023 [cond-mat]},
	pages = {205144},
}

@article{Lu2024,
  author  = {Zhengguang Lu and Tonghang Han and Yuxuan Yao and Aidan P. Reddy and 
             Jixiang Yang and Junseok Seo and Kenji Watanabe and Takashi Taniguchi 
             and Liang Fu and Long Ju},
  title   = {Fractional quantum anomalous Hall effect in multilayer graphene},
  journal = {Nature},
  volume  = {626},
  number  = {8000},
  pages   = {759--764},
  year    = {2024},
  doi     = {10.1038/s41586-023-07010-7}
}

@article{TMDAharonocCasher,
  title = {Adiabatic approximation and Aharonov-Casher bands in twisted homobilayer transition metal dichalcogenides},
  author = {Shi, Jingtian and Morales-Dur\'an, Nicol\'as and Khalaf, Eslam and MacDonald, A. H.},
  journal = {Phys. Rev. B},
  volume = {110},
  issue = {3},
  pages = {035130},
  numpages = {17},
  year = {2024},
  month = {Jul},
  publisher = {American Physical Society},
  doi = {10.1103/PhysRevB.110.035130},
  url = {https://link.aps.org/doi/10.1103/PhysRevB.110.035130}
}

@article{roy2014band,
  title = {Band Geometry of Fractional Topological Insulators},
  author = {Roy, Rahul},
  year = {2014},
  month = oct,
  journal = {Phys. Rev. B},
  volume = {90},
  number = {16},
  pages = {165139},
  issn = {1098-0121, 1550-235X},
  doi = {10.1103/PhysRevB.90.165139},
  urldate = {2023-10-28},
  langid = {english}
}

@article{meraEngineeringGeometricallyFlat2021,
	author = {Mera, Bruno and Ozawa, Tomoki},
	date = {2021-09},
	year=2021,
	doi = {10.1103/PhysRevB.104.115160},
	journal = {Phys. Rev. B},
	number = 11,
	pages = 115160,
	publisher = {{American Physical Society}},
	title = {Engineering Geometrically Flat {{Chern}} Bands with {{Fubini-Study K{\"a}hler}} Structure},
	url = {https://link.aps.org/doi/10.1103/PhysRevB.104.115160},
	volume = 104}

@article{meraKahlerGeometryChern2021,
	author = {Mera, Bruno and Ozawa, Tomoki},
	date = {2021-07},
	doi = {10.1103/PhysRevB.104.045104},
	issn = {2469-9950, 2469-9969},
	journal = {Phys. Rev. B},
	journaltitle = {Phys. Rev. B},
	month = jul,
	number = 4,
	pages = {045104},
	publisher = {{American Physical Society}},
	shorttitle = {K{\"a}hler geometry and {Chern} insulators},
	title = {K{\"a}hler Geometry and {{Chern}} Insulators: {{Relations}} between Topology and the Quantum Metric},
	url = {https://link.aps.org/doi/10.1103/PhysRevB.104.045104},
	urldate = {2022-04-16},
	volume = 104,
	year = 2021
	}

@unpublished{meraRelatingTopologyDirac2021,
	archiveprefix = {arXiv},
	author = {Mera, Bruno and Zhang, Anwei and Goldman, Nathan},
	date = 2021,
	eprint = {2106.00800},
	eprinttype = {arxiv},
	title = {Relating the Topology of {{Dirac Hamiltonians}} to Quantum Geometry: {{When}} the Quantum Metric Dictates {{Chern}} Numbers and Winding Numbers}}

@article{dong2023composite,
  title = {Composite {{Fermi Liquid}} at {{Zero Magnetic Field}} in {{Twisted MoTe}} 2},
  author = {Dong, Junkai and Wang, Jie and Ledwith, Patrick J. and Vishwanath, Ashvin and Parker, Daniel E.},
  year = {2023},
  month = sep,
  journal = {Phys. Rev. Lett.},
  volume = {131},
  number = {13},
  pages = {136502},
  issn = {0031-9007, 1079-7114},
  doi = {10.1103/PhysRevLett.131.136502},
  urldate = {2023-10-17},
  langid = {english}
}

@article{dong2023manybody,
  title = {Many-Body Ground States from Decomposition of Ideal Higher {{Chern}} Bands: {{Applications}} to Chirally Twisted Graphene Multilayers},
  shorttitle = {Many-Body Ground States from Decomposition of Ideal Higher {{Chern}} Bands},
  author = {Dong, Junkai and Ledwith, Patrick J. and Khalaf, Eslam and Lee, Jong Yeon and Vishwanath, Ashvin},
  year = {2023},
  month = jun,
  journal = {Phys. Rev. Research},
  volume = {5},
  number = {2},
  pages = {023166},
  issn = {2643-1564},
  doi = {10.1103/PhysRevResearch.5.023166},
  urldate = {2024-02-12},
  langid = {english}
}

@article{wang2023origin,
  title = {Origin of Model Fractional {{Chern}} Insulators in All Topological Ideal Flatbands: {{Explicit}} Color-Entangled Wave Function and Exact Density Algebra},
  shorttitle = {Origin of Model Fractional {{Chern}} Insulators in All Topological Ideal Flatbands},
  author = {Wang, Jie and Klevtsov, Semyon and Liu, Zhao},
  year = {2023},
  month = jun,
  journal = {Phys. Rev. Research},
  volume = {5},
  number = {2},
  pages = {023167},
  issn = {2643-1564},
  doi = {10.1103/PhysRevResearch.5.023167},
  urldate = {2024-02-12},
  langid = {english}
}

@article{zeng2023thermodynamic,
  title = {Thermodynamic Evidence of Fractional {{Chern}} Insulator in Moir{\'e} {{MoTe2}}},
  author = {Zeng, Yihang and Xia, Zhengchao and Kang, Kaifei and Zhu, Jiacheng and Kn{\"u}ppel, Patrick and Vaswani, Chirag and Watanabe, Kenji and Taniguchi, Takashi and Mak, Kin Fai and Shan, Jie},
  year = {2023},
  month = oct,
  journal = {Nature},
  volume = {622},
  number = {7981},
  pages = {69--73},
  issn = {0028-0836, 1476-4687},
  doi = {10.1038/s41586-023-06452-3},
  urldate = {2024-02-07},
  langid = {english}
}

@article{xu2023observation,
  title = {Observation of {{Integer}} and {{Fractional Quantum Anomalous Hall Effects}} in {{Twisted Bilayer MoTe}} 2},
  author = {Xu, Fan and Sun, Zheng and Jia, Tongtong and Liu, Chang and Xu, Cheng and Li, Chushan and Gu, Yu and Watanabe, Kenji and Taniguchi, Takashi and Tong, Bingbing and Jia, Jinfeng and Shi, Zhiwen and Jiang, Shengwei and Zhang, Yang and Liu, Xiaoxue and Li, Tingxin},
  year = {2023},
  month = sep,
  journal = {Phys. Rev. X},
  volume = {13},
  number = {3},
  pages = {031037},
  issn = {2160-3308},
  doi = {10.1103/PhysRevX.13.031037},
  urldate = {2023-10-17},
  langid = {english}
}

@article{liangfu2023tMoTe2,
  title = {Fractional quantum anomalous Hall states in twisted bilayer ${\mathrm{MoTe}}_{2}$ and ${\mathrm{WSe}}_{2}$},
  author = {Reddy, Aidan P. and Alsallom, Faisal and Zhang, Yang and Devakul, Trithep and Fu, Liang},
  journal = {Phys. Rev. B},
  volume = {108},
  issue = {8},
  pages = {085117},
  numpages = {10},
  year = {2023},
  month = {Aug},
  publisher = {American Physical Society},
  doi = {10.1103/PhysRevB.108.085117},
  url = {https://link.aps.org/doi/10.1103/PhysRevB.108.085117}
}

@misc{bernevig2025ED,
      title={Spinless and spinful charge excitations in moir\'e Fractional Chern Insulators}, 
      author={Miguel Gonçalves and Juan Felipe Mendez-Valderrama and Jonah Herzog-Arbeitman and Jiabin Yu and Xiaodong Xu and Di Xiao and B. Andrei Bernevig and Nicolas Regnault},
      year={2025},
      eprint={2506.05330},
      archivePrefix={arXiv},
      primaryClass={cond-mat.str-el},
      url={https://arxiv.org/abs/2506.05330}, 
}

@article{bernevigEmergentManybodyTranslational2012,
  title = {Emergent Many-Body Translational Symmetries of {{Abelian}} and Non-{{Abelian}} Fractionally Filled Topological Insulators},
  author = {Bernevig, B. Andrei and Regnault, N.},
  year = {2012},
  month = feb,
  journal = {Physical Review B},
  volume = {85},
  number = {7},
  pages = {075128},
  issn = {1098-0121, 1550-235X},
  doi = {10.1103/PhysRevB.85.075128},
  copyright = {http://link.aps.org/licenses/aps-default-license},
  langid = {english}
}

@article{metropolis1953equation,
  title={Equation of state calculations by fast computing machines},
  author={Metropolis, Nicholas and Rosenbluth, Arianna W and Rosenbluth, Marshall N and Teller, Augusta H and Teller, Edward},
  journal={The journal of chemical physics},
  volume={21},
  number={6},
  pages={1087--1092},
  year={1953},
  publisher={American Institute of Physics}
}

@article{hastings1970monte,
  title={Monte Carlo sampling methods using Markov chains and their applications},
  author={Hastings, W. Keith},
  journal={Biometrika},
  volume={57},
  number={1},
  pages={97--109},
  year={1970},
  publisher={Oxford University Press}
}

@article{levesque1984CrystallizationIncompressibleQuantumfluid,
  title = {Crystallization of the Incompressible Quantum-Fluid State of a Two-Dimensional Electron Gas in a Strong Magnetic Field},
  author = {Levesque, D. and Weis, J. J. and MacDonald, A. H.},
  year = {1984},
  month = jul,
  journal = {Physical Review B},
  volume = {30},
  number = {2},
  pages = {1056--1058},
  issn = {0163-1829},
  doi = {10.1103/PhysRevB.30.1056},
  copyright = {http://link.aps.org/licenses/aps-default-license},
  langid = {english}
}

@article{morfMonteCarloEvaluation1987,
  title = {Monte {{Carlo}} Evaluation of Trial Wavefunctions for the Fractional Quantized {{Hall}} Effect: {{Spherical}} Geometry},
  shorttitle = {Monte {{Carlo}} Evaluation of Trial Wavefunctions for the Fractional Quantized {{Hall}} Effect},
  author = {Morf, R. and Halperin, B. I.},
  year = {1987},
  month = jun,
  journal = {Zeitschrift f{\"u}r Physik B Condensed Matter},
  volume = {68},
  number = {2},
  pages = {391--406},
  issn = {1431-584X},
  doi = {10.1007/BF01304256},
  langid = {english}
}

@article{zhu1993WignerCrystallizationFractional,
  title = {Wigner Crystallization in the Fractional Quantum {{Hall}} Regime: {{A}} Variational Quantum {{Monte Carlo}} Study},
  shorttitle = {Wigner Crystallization in the Fractional Quantum {{Hall}} Regime},
  author = {Zhu, Xuejun and Louie, Steven G.},
  year = {1993},
  month = jan,
  journal = {Physical Review Letters},
  volume = {70},
  number = {3},
  pages = {335--338},
  publisher = {American Physical Society},
  doi = {10.1103/PhysRevLett.70.335}
}

@article{ciftja2011FinitesizeMonteCarlo,
  title = {Finite-Size {{Monte Carlo}} Results for Anisotropic Quantum {{Hall}} Liquids},
  author = {Ciftja, Orion and Cornelius, Brittney and Brown, Kesha and Taylor, Emery},
  year = {2011},
  month = may,
  journal = {Physical Review B},
  volume = {83},
  number = {19},
  pages = {193101},
  publisher = {American Physical Society},
  doi = {10.1103/PhysRevB.83.193101}
}

@article{meirVariationalGroundState1996,
  title = {A {{Variational Ground-State}} for the \${\textbackslash}nu=2/3\$ {{Fractional Quantum Hall Regime}}},
  author = {Meir, Yigal},
  year = {1996},
  month = may,
  journal = {International Journal of Modern Physics B},
  volume = {10},
  number = {12},
  eprint = {cond-mat/9509108},
  pages = {1425--1437},
  issn = {0217-9792, 1793-6578},
  doi = {10.1142/S0217979296000544},
  archiveprefix = {arXiv}
}

@article{tserkovnyak2003MonteCarloEvaluation,
  title = {Monte {{Carlo Evaluation}} of {{Non-Abelian Statistics}}},
  author = {Tserkovnyak, Yaroslav and Simon, Steven H.},
  year = {2003},
  month = jan,
  journal = {Physical Review Letters},
  volume = {90},
  number = {1},
  pages = {016802},
  publisher = {American Physical Society},
  doi = {10.1103/PhysRevLett.90.016802}
}

@article{biddleVariationalMonteCarlo2013,
  title = {Variational {{Monte Carlo}} Study of Spin Polarization Stability of Fractional Quantum {{Hall}} States against Realistic Effects in Half-Filled {{Landau}} Levels},
  author = {Biddle, J. and Peterson, Michael R. and Sarma, S. Das},
  year = {2013},
  month = jun,
  journal = {Physical Review B},
  volume = {87},
  number = {23},
  eprint = {1304.1174},
  primaryclass = {cond-mat},
  pages = {235134},
  issn = {1098-0121, 1550-235X},
  doi = {10.1103/PhysRevB.87.235134},
  archiveprefix = {arXiv}
}

@article{kalinay2000sixthMomentSumRule,
  title   = {The sixth-moment sum rule for the pair correlations of the two-dimensional one-component plasma: Exact result},
  author = {Kalinay, P. and Markoš, P. and Šamaj, L. and Travěnec, I.},
  journal = {Journal of Statistical Physics},
  year    = {2000},
  month   = feb,
  volume  = {98},
  number  = {3},
  pages   = {639--666},
  issn    = {1572-9613},
  doi     = {10.1023/A:1018667207145},
}

@article{shi2025doping,
  title={Doping a fractional quantum anomalous Hall insulator},
  author={Shi, Zhengyan Darius and Senthil, T},
  journal={Physical Review X},
  volume={15},
  number={3},
  pages={031069},
  year={2025},
  publisher={APS}
}

@article{laughlin1983anomalous,
  title={Anomalous quantum Hall effect: An incompressible quantum fluid with fractionally charged excitations},
  author={Laughlin, Robert B},
  journal={Physical Review Letters},
  volume={50},
  number={18},
  pages={1395},
  year={1983},
  publisher={APS}
}

@article{jeon2003NatureQuasiparticleExcitations,
  title = {Nature of Quasiparticle Excitations in the Fractional Quantum {{Hall}} Effect},
  author = {Jeon, Gun Sang and Jain, Jainendra K.},
  year = {2003},
  month = oct,
  journal = {Physical Review B},
  volume = {68},
  number = {16},
  pages = {165346},
  issn = {0163-1829, 1095-3795},
  doi = {10.1103/PhysRevB.68.165346},
  copyright = {http://link.aps.org/licenses/aps-default-license},
  langid = {english}
}

@article{bernevig2009ClusteringPropertiesModel,
  title = {Clustering {{Properties}} and {{Model Wave Functions}} for {{Non-Abelian Fractional Quantum Hall Quasielectrons}}},
  author = {Bernevig, B. Andrei and Haldane, F. D. M.},
  year = {2009},
  month = feb,
  journal = {Physical Review Letters},
  volume = {102},
  number = {6},
  pages = {066802},
  issn = {0031-9007, 1079-7114},
  doi = {10.1103/PhysRevLett.102.066802},
  copyright = {http://link.aps.org/licenses/aps-default-license},
  langid = {english}
}

@article{rodriguez2012QuasiparticlesExcitonsPfaffian,
  title = {Quasiparticles and Excitons for the {{Pfaffian}} Quantum {{Hall}} State},
  author = {Rodriguez, Ivan D. and Sterdyniak, A. and Hermanns, M. and Slingerland, J. K. and Regnault, N.},
  year = {2012},
  month = jan,
  journal = {Physical Review B},
  volume = {85},
  number = {3},
  pages = {035128},
  publisher = {American Physical Society},
  doi = {10.1103/PhysRevB.85.035128}
}

@article{hansson2007CompositefermionWaveFunctions,
  title = {Composite-Fermion Wave Functions as Correlators in Conformal Field Theory},
  author = {Hansson, T. H. and Chang, C.-C. and Jain, J. K. and Viefers, S.},
  year = {2007},
  month = aug,
  journal = {Physical Review B},
  volume = {76},
  number = {7},
  pages = {075347},
  publisher = {American Physical Society},
  doi = {10.1103/PhysRevB.76.075347}
}

@article{flavin2011AbelianNonAbelianStatistics,
  title = {Abelian and {{Non-Abelian Statistics}} in the {{Coherent State Representation}}},
  author = {Flavin, John and Seidel, Alexander},
  year = {2011},
  month = dec,
  journal = {Physical Review X},
  volume = {1},
  number = {2},
  pages = {021015},
  issn = {2160-3308},
  doi = {10.1103/PhysRevX.1.021015},
  copyright = {http://creativecommons.org/licenses/by/3.0/},
  langid = {english}
}

@article{abouelkomsan2023quantum,
  title = {Quantum Metric Induced Phases in {{Moir{\'e}}} Materials},
  author = {Abouelkomsan, Ahmed and Yang, Kang and Bergholtz, Emil J.},
  year = {2023},
  month = feb,
  journal = {Phys. Rev. Research},
  volume = {5},
  number = {1},
  pages = {L012015},
  issn = {2643-1564},
  doi = {10.1103/PhysRevResearch.5.L012015},
  urldate = {2024-03-08},
  langid = {english}
}

@article{aharonov1979ground,
  title = {Ground State of a Spin-{$\frac{1}{2}$} Charged Particle in a Two-Dimensional Magnetic Field},
  author = {Aharonov, Y. and Casher, A.},
  year = {1979},
  month = jun,
  journal = {Phys. Rev. A},
  volume = {19},
  number = {6},
  pages = {2461--2462},
  issn = {0556-2791},
  doi = {10.1103/PhysRevA.19.2461},
  urldate = {2023-10-17},
  langid = {english}
}

@article{laughlin1988,
author = {R. B. Laughlin },
title = {The Relationship Between High-Temperature Superconductivity and the Fractional Quantum Hall Effect},
journal = {Science},
volume = {242},
number = {4878},
pages = {525-533},
year = {1988},
doi = {10.1126/science.242.4878.525},
URL = {https://www.science.org/doi/abs/10.1126/science.242.4878.525},
eprint = {https://www.science.org/doi/pdf/10.1126/science.242.4878.525}}

@article{divic2024anyon,
  title={Anyon Superconductivity from Topological Criticality in a Hofstadter-Hubbard Model},
  author={Divic, Stefan and Cr{\'e}pel, Valentin and Soejima, Tomohiro and Song, Xue-Yang and Millis, Andrew and Zaletel, Michael P and Vishwanath, Ashvin},
  journal={arXiv preprint arXiv:2410.18175},
  year={2024}}

@article{PhysRevB.41.11101,
  title = {Exactness of anyon superconductivity and its finite-temperature behavior},
  author = {Kitazawa, Y. and Murayama, H.},
  journal = {Phys. Rev. B},
  volume = {41},
  issue = {16},
  pages = {11101--11109},
  numpages = {0},
  year = {1990},
  month = {Jun},
  publisher = {American Physical Society},
  doi = {10.1103/PhysRevB.41.11101},
  url = {https://link.aps.org/doi/10.1103/PhysRevB.41.11101}
}

@article{PhysRevLett.63.903,
  title = {Anyon superconductivity and the fractional quantum Hall effect},
  author = {Lee, Dung-Hai and Fisher, Matthew P. A.},
  journal = {Phys. Rev. Lett.},
  volume = {63},
  issue = {8},
  pages = {903--906},
  numpages = {0},
  year = {1989},
  month = {Aug},
  publisher = {American Physical Society},
  doi = {10.1103/PhysRevLett.63.903},
  url = {https://link.aps.org/doi/10.1103/PhysRevLett.63.903}
}

@article{cai2023signatures,
  title = {Signatures of Fractional Quantum Anomalous {{Hall}} States in Twisted {{MoTe2}}},
  author = {Cai, Jiaqi and Anderson, Eric and Wang, Chong and Zhang, Xiaowei and Liu, Xiaoyu and Holtzmann, William and Zhang, Yinong and Fan, Fengren and Taniguchi, Takashi and Watanabe, Kenji and Ran, Ying and Cao, Ting and Fu, Liang and Xiao, Di and Yao, Wang and Xu, Xiaodong},
  year = {2023},
  month = oct,
  journal = {Nature},
  volume = {622},
  number = {7981},
  pages = {63--68},
  issn = {0028-0836, 1476-4687},
  doi = {10.1038/s41586-023-06289-w},
  urldate = {2023-10-17},
  langid = {english}
}

@article{devakul2021magic,
  title = {Magic in Twisted Transition Metal Dichalcogenide Bilayers},
  author = {Devakul, Trithep and Cr{\'e}pel, Valentin and Zhang, Yang and Fu, Liang},
  year = {2021},
  month = nov,
  journal = {Nat Commun},
  volume = {12},
  number = {1},
  pages = {6730},
  issn = {2041-1723},
  doi = {10.1038/s41467-021-27042-9},
  urldate = {2023-06-16},
  langid = {english}
}

@article{ledwith2021strong,
  title = {Strong Coupling Theory of Magic-Angle Graphene: {{A}} Pedagogical Introduction},
  shorttitle = {Strong Coupling Theory of Magic-Angle Graphene},
  author = {Ledwith, Patrick J. and Khalaf, Eslam and Vishwanath, Ashvin},
  year = {2021},
  month = dec,
  journal = {Annals of Physics},
  volume = {435},
  pages = {168646},
  issn = {00034916},
  doi = {10.1016/j.aop.2021.168646},
  urldate = {2023-10-29},
  langid = {english}
}

@article{morales-duran2023pressureenhanced,
  title = {Pressure-Enhanced Fractional {{Chern}} Insulators along a Magic Line in Moir{\'e} Transition Metal Dichalcogenides},
  author = {{Morales-Dur{\'a}n}, Nicol{\'a}s and Wang, Jie and Schleder, Gabriel R. and Angeli, Mattia and Zhu, Ziyan and Kaxiras, Efthimios and Repellin, C{\'e}cile and Cano, Jennifer},
  year = {2023},
  month = aug,
  journal = {Phys. Rev. Research},
  volume = {5},
  number = {3},
  pages = {L032022},
  issn = {2643-1564},
  doi = {10.1103/PhysRevResearch.5.L032022},
  urldate = {2023-10-17},
  langid = {english}
}

@article{morales-duran2024magic,
  title = {Magic {{Angles}} and {{Fractional Chern Insulators}} in {{Twisted Homobilayer Transition Metal Dichalcogenides}}},
  author = {{Morales-Dur{\'a}n}, Nicol{\'a}s and Wei, Nemin and Shi, Jingtian and MacDonald, Allan H.},
  year = {2024},
  month = mar,
  journal = {Phys. Rev. Lett.},
  volume = {132},
  number = {9},
  pages = {096602},
  issn = {0031-9007, 1079-7114},
  doi = {10.1103/PhysRevLett.132.096602},
  urldate = {2024-04-23},
  langid = {english}
}

@article{park2023observation,
  title = {Observation of Fractionally Quantized Anomalous {{Hall}} Effect},
  author = {Park, Heonjoon and Cai, Jiaqi and Anderson, Eric and Zhang, Yinong and Zhu, Jiayi and Liu, Xiaoyu and Wang, Chong and Holtzmann, William and Hu, Chaowei and Liu, Zhaoyu and Taniguchi, Takashi and Watanabe, Kenji and Chu, Jiun-Haw and Cao, Ting and Fu, Liang and Yao, Wang and Chang, Cui-Zu and Cobden, David and Xiao, Di and Xu, Xiaodong},
  year = {2023},
  month = oct,
  journal = {Nature},
  volume = {622},
  number = {7981},
  pages = {74--79},
  issn = {0028-0836, 1476-4687},
  doi = {10.1038/s41586-023-06536-0},
  urldate = {2023-10-17},
  langid = {english}
}

@article{wu2019topological,
  title = {Topological {{Insulators}} in {{Twisted Transition Metal Dichalcogenide Homobilayers}}},
  author = {Wu, Fengcheng and Lovorn, Timothy and Tutuc, Emanuel and Martin, Ivar and MacDonald, A. H.},
  year = {2019},
  month = feb,
  journal = {Phys. Rev. Lett.},
  volume = {122},
  number = {8},
  pages = {086402},
  issn = {0031-9007, 1079-7114},
  doi = {10.1103/PhysRevLett.122.086402},
  urldate = {2023-06-16},
  langid = {english}
}

@misc{liuRecentDevelopmentsFractional2022,
  title = {Recent {{Developments}} in {{Fractional Chern Insulators}}},
  author = {Liu, Zhao and Bergholtz, Emil J.},
  year = {2022},
  month = aug,
  number = {arXiv:2208.08449},
  eprint = {2208.08449},
  eprinttype = {arxiv},
  primaryclass = {cond-mat, physics:math-ph, physics:quant-ph},
  publisher = {{arXiv}},
  doi = {10.48550/arXiv.2208.08449},
  archiveprefix = {arXiv},
  journal = ""
}

@article{MoTe2SC,
  title={Signatures of unconventional superconductivity near reentrant and fractional quantum anomalous Hall insulators},
  author={Xu, Fan and Sun, Zheng and Li, Jiayi and Zheng, Ce and Xu, Cheng and Gao, Jingjing and Jia, Tongtong and Watanabe, Kenji and Taniguchi, Takashi and Tong, Bingbing and others},
  journal={arXiv preprint arXiv:2504.06972},
  year={2025}
}

@article{WilczekWittenHalperinAnyonSC,
  title={On anyon superconductivity},
  author={Chen, Yi-Hong and Wilczek, Frank and Witten, Edward and Halperin, Bertrand I},
  journal={International Journal of Modern Physics B},
  volume={3},
  number={07},
  pages={1001--1067},
  year={1989},
  publisher={World Scientific}
}

@article{PhysRevB.41.240,
  title = {Compressibility and superfluidity in the fractional-statistics liquid},
  author = {Wen, X. G. and Zee, A.},
  journal = {Phys. Rev. B},
  volume = {41},
  issue = {1},
  pages = {240--253},
  numpages = {0},
  year = {1990},
  month = {Jan},
  publisher = {American Physical Society},
  doi = {10.1103/PhysRevB.41.240},
  url = {https://link.aps.org/doi/10.1103/PhysRevB.41.240}
}

@article{doi:10.1142/S0217979291001607,
author = {Lee, Dung-Hai},
title = {Anyon Superconductivity and the Fractional Quantum-Hall Effect},
journal = {International Journal of Modern Physics B},
volume = {05},
number = {10},
pages = {1695-1713},
year = {1991},
doi = {10.1142/S0217979291001607},

URL = { 
    
        https://doi.org/10.1142/S0217979291001607
    
    

},
eprint = { 
    
        https://doi.org/10.1142/S0217979291001607
    
    

}
}

@article{PhysRevB.42.342,
  title = {Superconductivity in the anyon model},
  author = {Hosotani, Yutaka and Chakravarty, Sumantra},
  journal = {Phys. Rev. B},
  volume = {42},
  issue = {1},
  pages = {342--346},
  numpages = {0},
  year = {1990},
  month = {Jul},
  publisher = {American Physical Society},
  doi = {10.1103/PhysRevB.42.342},
  url = {https://link.aps.org/doi/10.1103/PhysRevB.42.342}
}

@article{PhysRevB.39.11413,
  title = {Chiral spin states and superconductivity},
  author = {Wen, X. G. and Wilczek, Frank and Zee, A.},
  journal = {Phys. Rev. B},
  volume = {39},
  issue = {16},
  pages = {11413--11423},
  numpages = {0},
  year = {1989},
  month = {Jun},
  publisher = {American Physical Society},
  doi = {10.1103/PhysRevB.39.11413},
  url = {https://link.aps.org/doi/10.1103/PhysRevB.39.11413}
}

@article{LaughlinAnyonSC,
  title = {Superconducting Ground State of Noninteracting Particles Obeying Fractional Statistics},
  author = {Laughlin, R. B.},
  journal = {Phys. Rev. Lett.},
  volume = {60},
  issue = {25},
  pages = {2677--2680},
  numpages = {0},
  year = {1988},
  month = {Jun},
  publisher = {American Physical Society},
  doi = {10.1103/PhysRevLett.60.2677},
  url = {https://link.aps.org/doi/10.1103/PhysRevLett.60.2677}
}

@article{FetterHannaLaughlin,
  title={Random-phase approximation in the fractional-statistics gas},
  author={Fetter, AL and Hanna, CB and Laughlin, RB},
  journal={Physical Review B},
  volume={39},
  number={13},
  pages={9679},
  year={1989},
  publisher={APS}
}

@preamble{"\ifdefined\DeclarePrefChars\DeclarePrefChars{'’-}\else\fi "}

@preamble{ " \newcommand{\noop}[1]{} " }

@article{kourtisFractionalChernInsulators2014,
  title = {Fractional {{Chern}} Insulators with Strong Interactions Far Exceeding Bandgaps},
  author = {Kourtis, Stefanos and Neupert, Titus and Chamon, Claudio and Mudry, Christopher},
  year = {2014},
  month = mar,
  journal = {Physical Review Letters},
  volume = {112},
  number = {12},
  eprint = {1310.6371},
  eprinttype = {arxiv},
  primaryclass = {cond-mat},
  pages = {126806},
  issn = {0031-9007, 1079-7114},
  doi = {10.1103/PhysRevLett.112.126806},
  archiveprefix = {arXiv}
}

@article{BergholtzReview2013,
	author = {Bergholtz, Emil J. and Liu, Zhao},
	doi = {10.1142/S021797921330017X},
	eprint = {https://doi.org/10.1142/S021797921330017X},
	journal = {Int. J. Mod. Phys. B},
	number = {24},
	pages = {1330017},
	title = {Topological flat band models and fractional Chern insulators},
	url = {https://doi.org/10.1142/S021797921330017X},
	volume = {27},
	year = {2013}}

@article{neupertFractionalQuantumHall2011,
	author = {Neupert, Titus and Santos, Luiz and Chamon, Claudio and Mudry, Christopher},
	date = {2011-06-06},
	doi = {10.1103/PhysRevLett.106.236804},
	issn = {0031-9007, 1079-7114},
	year=2011,
	journal = {Phys. Rev. Lett.},
	langid = {english},
	number = {23},
	pages = {236804},
	shortjournal = {Phys. Rev. Lett.},
	title = {Fractional Quantum Hall States at Zero Magnetic Field},
	url = {https://link.aps.org/doi/10.1103/PhysRevLett.106.236804},
	urldate = {2022-08-12},
	volume = {106}}

@article{shengFractionalQuantumHall2011,
	author = {Sheng, D.N. and Gu, Zheng-Cheng and Sun, Kai and Sheng, L.},
	date = {2011-09},
	doi = {10.1038/ncomms1380},
	year=2011,
	issn = {2041-1723},
	journal = {Nat. Commun.},
	langid = {english},
	number = {1},
	pages = {389},
	shortjournal = {Nat Commun},
	title = {Fractional quantum Hall effect in the absence of Landau levels},
	url = {http://www.nature.com/articles/ncomms1380},
	urldate = {2022-08-12},
	volume = {2}}

@article{regnaultFractionalChernInsulator2011,
	author = {Regnault, N. and Bernevig, B. Andrei},
	date = {2011-12-02},
	doi = {10.1103/PhysRevX.1.021014},
	issn = {2160-3308},
	year=2011,
	journal = {Phys. Rev. X},
	langid = {english},
	number = {2},
	pages = {021014},
	shortjournal = {Phys. Rev. X},
	title = {Fractional Chern Insulator},
	url = {https://link.aps.org/doi/10.1103/PhysRevX.1.021014},
	urldate = {2022-08-12},
	volume = {1}}

@article{ledwithFamilyIdealChern2022,
	author = {Ledwith, Patrick J. and Vishwanath, Ashvin and Khalaf, Eslam},
	date = {2022-04-29},
	year=2022,
	doi = {10.1103/PhysRevLett.128.176404},
	journal = {Phys. Rev. Lett.},
	number = {17},
	pages = {176404},
	shortjournal = {Phys. Rev. Lett.},
	title = {Family of Ideal Chern Flatbands with Arbitrary Chern Number in Chiral Twisted Graphene Multilayers},
	url = {https://link.aps.org/doi/10.1103/PhysRevLett.128.176404},
	urldate = {2022-06-09},
	volume = {128}}

@article{abouelkomsanParticleHoleDualityEmergent2020a,
	author = {Abouelkomsan, Ahmed and Liu, Zhao and Bergholtz, Emil J.},
	date = {2020-03},
	doi = {10.1103/PhysRevLett.124.106803},
	issn = {0031-9007, 1079-7114},
	journal = {Phys. Rev. Lett.},
	month = mar,
	number = 10,
	pages = 106803,
	publisher = {{American Physical Society}},
	title = {Particle-{{Hole Duality}}, {{Emergent Fermi Liquids}}, and {{Fractional Chern Insulators}} in {{Moir{\'e} Flatbands}}},
	url = {https://link.aps.org/doi/10.1103/PhysRevLett.124.106803},
	urldate = {2022-04-22},
	volume = 124,
	year = 2020}

@article{haldaneFractionalQuantizationHall1983a,
	author = {Haldane, F. D. M.},
	date = {1983-08},
	doi = {10.1103/PhysRevLett.51.605},
	journal = {Phys. Rev. Lett.},
	number = 7,
	pages = {605--608},
	publisher = {{American Physical Society}},
	title = {Fractional {{Quantization}} of the {{Hall Effect}}: {{A Hierarchy}} of {{Incompressible Quantum Fluid States}}},
	url = {https://link.aps.org/doi/10.1103/PhysRevLett.51.605},
	volume = 51}

@article{Abouelkomsan2025,
  title = {From Fractionalization to Chiral Topological Superconductivity in a Flat Chern Band},
  author = {Guerci, Daniele and Abouelkomsan, Ahmed and Fu, Liang},
  journal = {Phys. Rev. Lett.},
  volume = {135},
  issue = {18},
  pages = {186601},
  numpages = {8},
  year = {2025},
  month = {Oct},
  publisher = {American Physical Society},
  doi = {10.1103/zm39-dstj},
  url = {https://link.aps.org/doi/10.1103/zm39-dstj}
}

@article{parameswaranFractionalChernInsulators2012,
	author = {Parameswaran, S. A. and Roy, R. and Sondhi, S. L.},
	date = {2012-06},
	doi = {10.1103/PhysRevB.85.241308},
	issn = {1098-0121, 1550-235X},
	journal = {Phys. Rev. B},
	journaltitle = {Phys. Rev. B},
	month = jun,
	number = 24,
	pages = 241308,
	publisher = {{American Physical Society}},
	title = {Fractional {{Chern}} Insulators and the {$W_{\infty}$} Algebra},
	url = {https://link.aps.org/doi/10.1103/PhysRevB.85.241308},
	urldate = {2022-04-22},
	volume = 85,
	year = 2012}

@article{parameswaranFractionalQuantumHall2013,
	author = {Parameswaran, Siddharth A. and Roy, Rahul and Sondhi, Shivaji L.},
	date = 2013,
	doi = {10.1016/j.crhy.2013.04.003},
	issn = {1631-0705},
	journal = {C. R. Phys.},
	journaltitle = {Comptes Rendus Physique},
	month = nov,
	number = 9,
	pages = {816--839},
	title = {Fractional Quantum {{Hall}} Physics in Topological Flat Bands},
	url = {http://www.sciencedirect.com/science/article/pii/S163107051300073X},
	urldate = {2022-04-16},
	volume = 14,
	year = 2013}

@article{repellinFerromagnetismNarrowBands2020,
	author = {Repellin, C{\'e}cile and Dong, Zhihuan and Zhang, Ya-Hui and Senthil, T.},
	date = {2020-05},
	doi = {10.1103/PhysRevLett.124.187601},
	journal = {Phys. Rev. Lett.},
	number = 18,
	pages = 187601,
	publisher = {{American Physical Society}},
	title = {Ferromagnetism in {{Narrow Bands}} of {{Moir{\'e} Superlattices}}},
	url = {https://link.aps.org/doi/10.1103/PhysRevLett.124.187601},
	volume = 124}

@article{tarnopolskyOriginMagicAngles2019,
	author = {Tarnopolsky, Grigory and Kruchkov, Alex Jura and Vishwanath, Ashvin},
	date = {2019-03},
	doi = {10.1103/PhysRevLett.122.106405},
	journal = {Phys. Rev. Lett.},
	year = 2019,
	number = 10,
	pages = 106405,
	publisher = {{American Physical Society}},
	title = {Origin of {{Magic Angles}} in {{Twisted Bilayer Graphene}}},
	url = {https://link.aps.org/doi/10.1103/PhysRevLett.122.106405},
	volume = 122}

@article{trugmanExactResultsFractional1985a,
	author = {Trugman, S. A. and Kivelson, S.},
	year=1985,
	doi = {10.1103/PhysRevB.31.5280},
	journal = {Phys. Rev. B},
	number = 8,
	pages = {5280--5284},
	publisher = {{American Physical Society}},
	title = {Exact Results for the Fractional Quantum {{Hall}} Effect with General Interactions},
	url = {https://link.aps.org/doi/10.1103/PhysRevB.31.5280},
	volume = 31}

@article{wuBlochModelWave2013,
	author = {Wu, Yang-Le and Regnault, N. and Bernevig, B. Andrei},
	year=2013,
	doi = {10.1103/PhysRevLett.110.106802},
	journal = {Phys. Rev. Lett.},
	number = 10,
	pages = 106802,
	publisher = {{American Physical Society}},
	title = {Bloch {{Model Wave Functions}} and {{Pseudopotentials}} for {{All Fractional Chern Insulators}}},
	url = {https://link.aps.org/doi/10.1103/PhysRevLett.110.106802},
	volume = 110}

@article{zhangNearlyFlatChern2019,
	author = {Zhang, Ya-Hui and Mao, Dan and Cao, Yuan and Jarillo-Herrero, Pablo and Senthil, T.},
	date = {2019-02},
	doi = {10.1103/PhysRevB.99.075127},
	journal = {Phys. Rev. B},
	number = 7,
	pages = {075127},
	publisher = {{American Physical Society}},
	title = {Nearly Flat {{Chern}} Bands in Moir{\'e} Superlattices},
	url = {https://link.aps.org/doi/10.1103/PhysRevB.99.075127},
	volume = 99}

@article{qi_generic_2011,
	author = {Qi, Xiao-Liang},
	doi = {10.1103/PhysRevLett.107.126803},
	issn = {0031-9007, 1079-7114},
	journal = {Phys. Rev. Lett.},
	month = sep,
	number = 12,
	pages = 126803,
	title = {Generic {Wave}-{Function} {Description} of {Fractional} {Quantum} {Anomalous} {Hall} {States} and {Fractional} {Topological} {Insulators}},
	url = {https://link.aps.org/doi/10.1103/PhysRevLett.107.126803},
	urldate = {2022-04-16},
	volume = 107,
	year = 2011}

@article{lu2025electromagnetic,
  title={Electromagnetic response and emergent topological orders in transition metal dichalcogenide MoTe $ \_2 $ bilayers},
  author={Lu, Tianhong and Wu, Yi-Ming and Santos, Luiz H},
  journal={arXiv preprint arXiv:2505.04685},
  year={2025}
}

@article{Tianhong2025,
  title = {Fractional Chern Insulators in Twisted Bilayer ${\mathrm{MoTe}}_{2}$: A Composite Fermion Perspective},
  author = {Lu, Tianhong and Santos, Luiz H.},
  journal = {Phys. Rev. Lett.},
  volume = {133},
  issue = {18},
  pages = {186602},
  numpages = {6},
  year = {2024},
  month = {Oct},
  publisher = {American Physical Society},
  doi = {10.1103/PhysRevLett.133.186602},
  url = {https://link.aps.org/doi/10.1103/PhysRevLett.133.186602}
}

@article{shi2025anyondelocalizationtransitionsdisordered,
       title={Anyon delocalization transitions out of a disordered fractional quantum anomalous Hall insulator},
  author={Shi, Zhengyan Darius and Senthil, T.},
  journal={Proc. Natl. Acad. Sci. U.S.A.},
  volume={122},
  pages={e2520608122},
  year={2025} 
}

@misc{nosov2025anyonsuperconductivityplateautransitions,
      title={Anyon superconductivity and plateau transitions in doped fractional quantum anomalous Hall insulators}, 
      author={Pavel A. Nosov and Zhaoyu Han and Eslam Khalaf},
      year={2025},
      eprint={2506.02108},
      archivePrefix={arXiv},
      primaryClass={cond-mat.str-el},
      url={https://arxiv.org/abs/2506.02108}, 
}

@article{han2025anyon,
  title={Anyon superfluidity of excitons in quantum Hall bilayers},
  author={Han, Zhaoyu and Wang, Taige and Dong, Zhihuan and Zaletel, Michael P and Vishwanath, Ashvin},
  journal={arXiv preprint arXiv:2508.14894},
  year={2025}
}

@article{WuAnyon,
  title = {Characterization of fractional Chern insulator quasiparticles in twisted homobilayer ${\mathrm{MoTe}}_{2}$},
  author = {Liu, Zhao and Li, Bohao and Shi, Yuhao and Wu, Fengcheng},
  journal = {Phys. Rev. B},
  volume = {112},
  issue = {24},
  pages = {245104},
  numpages = {18},
  year = {2025},
  month = {Dec},
  publisher = {American Physical Society},
  doi = {10.1103/nddl-729x},
  url = {https://link.aps.org/doi/10.1103/nddl-729x}
}

@article{JainMolecularAnyons,
  title = {Molecular Anyons in the Fractional Quantum Hall Effect},
  author = {Gattu, Mytraya and Jain, J. K.},
  journal = {Phys. Rev. Lett.},
  volume = {135},
  issue = {23},
  pages = {236601},
  numpages = {7},
  year = {2025},
  month = {Dec},
  publisher = {American Physical Society},
  doi = {10.1103/scl5-8pv6},
  url = {https://link.aps.org/doi/10.1103/scl5-8pv6}
}

@article{BoYangAnyonClusters,
  title = {Dynamics of anyon clusters in fractional quantum Hall fluids},
  author = {Xu, Qianhui and Ji, Guangyue and Wang, Yuzhu and Trung, Ha Quang and Yang, Bo},
  journal = {Phys. Rev. B},
  volume = {112},
  issue = {23},
  pages = {235112},
  numpages = {11},
  year = {2025},
  month = {Dec},
  publisher = {American Physical Society},
  doi = {10.1103/vgz6-z98r},
  url = {https://link.aps.org/doi/10.1103/vgz6-z98r}
}

@article{PICHLERanyonSC,
title = {Microscopic mechanism of anyon superconductivity emerging from fractional Chern insulators},
journal = {Newton},
pages = {100340},
year = {2025},
issn = {2950-6360},
doi = {https://doi.org/10.1016/j.newton.2025.100340},
url = {https://www.sciencedirect.com/science/article/pii/S2950636025003329},
author = {Fabian Pichler and Clemens Kuhlenkamp and Michael Knap and Ashvin Vishwanath}
}

@article{Pichlerspectral,
  title = {Single-particle spectral function of fractional quantum anomalous Hall states},
  author = {Pichler, Fabian and Kadow, Wilhelm and Kuhlenkamp, Clemens and Knap, Michael},
  journal = {Phys. Rev. B},
  volume = {111},
  issue = {7},
  pages = {075108},
  numpages = {10},
  year = {2025},
  month = {Feb},
  publisher = {American Physical Society},
  doi = {10.1103/PhysRevB.111.075108},
  url = {https://link.aps.org/doi/10.1103/PhysRevB.111.075108}
}

@article{shi2025effects,
  title={Effects of Berry curvature on ideal fractional Chern insulator many-body gaps},
  author={Shi, Jingtian and Cano, Jennifer and Morales-Dur{\'a}n, Nicol{\'a}s},
  journal={arXiv preprint arXiv:2503.15900},
  year={2025}
}
\let \oldv \v
\let \v \temp 

\clearpage
\newpage

\begin{appendix}
\onecolumngrid
\newpage
\makeatletter 
\begin{center}
\textbf{\large Supplementary material for ``\@title ''}

\vspace{10pt}

\end{center}
\vspace{20pt}

\setcounter{figure}{0}
\setcounter{section}{0}
\setcounter{equation}{0}
\renewcommand{\thefigure}{S\arabic{figure}}
\renewcommand{\theHfigure}{S\arabic{figure}}
\section{Deriving the dispersion formula from path integral using geometric quantization}
In this section, we will present a self-contained derivation of the Hamiltonian operator from the path integral using geometric quantization (Berezin-Toeplitz quantization). This is valid since our path integral is defined on a K\"ahler manifold. A K\"ahler manifold is an even-dimensional manifold with a complex structure, a symplectic structure, and a metric that are compatible with each other. In our problem, the K\"ahler potential is given by $\K(\bar \xi, \xi) = \ln \langle \xi|\xi \rangle$ where $|\xi \rangle$ is a set of states spanning the Hilbert space of interest that depends holomorphically on $\xi$. For uniform field $|\xi \rangle = e^{\frac{1}{4q} \bar \xi \xi} |\psi_\v{\xi} \rangle$ whereas for non-uniform field $|\xi \rangle = e^{\frac{1}{4q} \bar \xi \xi + \frac{1}{2} Q(\v{\xi})} |\psi^{\rm AC}_\v{\xi} \rangle = e^{\frac{1}{4q} \bar \xi \xi} \Gamma |\psi_{\v{\xi}} \rangle$. The K\"ahler potential determines the Berry curvature (effective magnetic field) for the quasihole via
\begin{equation}
    \B(\bar \xi,\xi) = 2\partial_{\bar \xi} \partial_\xi \K(\bar \xi, \xi)
\end{equation}
The Hilbert space can be identified with anti-holomorphic functions of $\xi$ with the appropriate boundary condition on the torus $f(\bar \xi) = \langle \xi|f \rangle$. Boundary conditions are given by
\begin{align*}
    &|\xi+L_1\rangle = (-1)^{N_e+1} e^{\frac{|\bL_1|^2}{4q}} e^{\frac{\bar L_1}{2q}\xi}|\xi\rangle,\qquad|\xi+L_2'\rangle = (-1)^{N_e+1} e^{\frac{|\bL_2'|^2}{4q}} e^{\frac{\bar L_2'}{2q}\xi}|\xi\rangle.
\end{align*}
We will see below, that we can fully construct that space from the knowledge of $\K$ alone. We define the reproducing Kernel $\F(\xi, \bar \xi)$ by requiring that
\begin{equation}
    \int \frac{d^2 \v{\xi}}{2\pi q} \F(\bar \xi, \xi) e^{-\K(\bar \xi, \xi)} e^{\K(\bar \omega,\xi)} f(\bar \xi) = f(\bar \omega)
    \label{RepKernel}
\end{equation}
For any anti-holomorphic function $f(\bar \xi)$. Here, $\K(\bar \xi, \omega)$ is obtained from $\K(\bar \xi, \xi)$ by analytic continuation. This identity is equivalent to the existence of a resolution of unity of the form 
\begin{equation}
    \int \frac{d^2 \v{\xi}}{2\pi q} \F(\bar \xi, \xi) e^{-\K(\bar \xi, \xi)} |\xi \rangle \langle \xi| = 1
\end{equation}
which leads to (\ref{RepKernel}) upon the identification $f(\bar \xi) = \langle \xi|f \rangle$. It can also be viewed as defining an inner product on the space of holomorphic functions given by
\begin{equation}
    \langle f|g \rangle = \int \frac{d^2 \v{\xi}}{2\pi q} \F(\bar \xi, \xi) e^{-\K(\bar \xi, \xi)} \bar f(\bar \xi) g(\bar \xi)
\end{equation}

To construct the quantum operator from a given classical potential $V(\bar \xi, \xi)$, we have to incorporate the choice of operator ordering we made during the map from the quantum operator to a classical Lagrangian (when deriving the path integral). Two standard choices are the Husimi $Q$-symbol defined via
\begin{equation}
    V_Q(\bar \xi, \xi) = \frac{\langle \xi| \hat V|\xi \rangle}{\langle \xi|\xi \rangle} = e^{-\K(\bar \xi, \xi)} \langle \xi| \hat V|\xi \rangle
\end{equation}
This is the choice that naturally emerges in our setup if we try to compute the expectation values from the unprojected operators. The other is the $P$-symbol defined such that
\begin{equation} \label{VP-eq}
    \hat V = \int \frac{d^2 \v{\xi}}{2\pi q} \F(\bar \xi, \xi) e^{-\K(\bar \xi, \xi)} V_P(\bar \xi, \xi) |\xi \rangle \langle \xi|
\end{equation}
The two are related via
\begin{equation}
    V_Q(\bar \omega, \omega) = e^{-\K(\bar \omega, \omega)} \int \frac{d^2 \v{\xi}}{2\pi q} \F(\bar \xi, \xi) e^{-\K(\bar \xi, \xi)} |e^{\K(\bar \xi, \omega)}|^2 V_P(\bar \xi, \xi) 
    \label{PtoQ}
\end{equation}

To clarify this connection on a concrete example, consider the case of a standard harmonic oscillator coherent states which correspond to $q=1$, $\F = 1$ and $[\hat \xi, \hat \xi^\dagger] = 2$ \footnote{The way we define $\hat \xi$ is related to the standard raising/lowering operators by a factor of $\sqrt{2}$}. Then the $Q$-symbol corresponds to \emph{normal-ordering} such that all factors of $\hat \xi^\dagger$ are brought to the left of $
\hat \xi$ followed by the replacement $\hat \xi^\dagger \mapsto \bar \xi$ and $\hat \xi \mapsto \xi$. On the other hand, the $P$-symbol corresponds to \emph{anti-normal ordering} where $\hat \xi$ is placed to the left of $\hat \xi^\dagger$ before the replacement with the classical variables. This is the mapping used to project operators to the LLL where all factors of $\bar z$ are brought to the left of factors $z$, followed by the replacement $\bar z \mapsto 2 \partial_z$. Noting that $z$ acts as a raising operator whereas $\bar z \equiv 2\partial_z$ acts as a lowering operator, we see that this indeed corresponds to anti-normal ordering. The mapping between $V_Q$ and $V_P$ in this case is precisely the standard Weierstrass transform
\begin{align}
    V_Q(\bar \omega, \omega) &= \int \frac{d^2 \v{\xi}}{2\pi} e^{-\frac{|\xi - \omega|^2}{2}} V_P(\bar \xi, \xi) = \int \frac{d^2 \v{\xi}}{2\pi} e^{-\frac{|\xi|^2}{2}} V_P(\bar \xi + \bar \omega, \xi + \omega) \nonumber \\
    &= \int \frac{d^2 \v{\xi}}{2\pi} e^{-\frac{|\xi|^2}{2}} e^{\bar \xi \partial_{\bar \omega} + \xi \partial_{\omega}} V_P(\bar \omega, \omega) = e^{2 \partial_\omega \partial_{\bar \omega}} V_P(\bar \omega, \omega) = e^{\frac{1}{2} \Delta_\omega} V_P(\bar \omega, \omega)
    \label{VQUniform}
\end{align}
We can check this is indeed the case for simple examples. For example, the $Q$ symbol corresponding to the operator $\hat \xi^\dagger \hat \xi$ is $\bar \xi \xi$ while the $P$ symbol is obtained by first anti-normal ordering $\hat \xi^\dagger \hat \xi = \hat \xi \hat \xi^\dagger - 2$ leading to the $P$ symbol $\xi \bar \xi - 2$. Clearly we have $e^{-2 \partial_\xi \partial_{\bar \xi}} \xi \bar \xi = (1 -2 \partial_\xi \partial_{\bar \xi}) \xi \bar \xi = \xi \bar \xi - 2$.

In the Berezin-Toeplitz quantization, given the real function $V_P(\bar \xi, \xi)$, corresponding to the $P$ symbol of an operator, we can define a corresponding operator on the space of anti-holomorphic functions via
\begin{equation}
    [\hat V f](\bar \omega) = \int \frac{d^2 \v{\xi}}{2\pi q} \F(\bar \xi, \xi) e^{-\K(\bar \xi, \xi)} e^{\K(\bar \omega, \xi)} V_P(\bar \xi, \xi) f(\bar \xi) 
    \label{QuantizationMap}
\end{equation}

This provides a concrete recipe for constructing the Hilbert space operators from the K\"ahler form $\K(\bar \xi, \xi)$ and the potential $V(\bar \xi, \xi)$ corresponding to the $Q$ symbol. We first need to solve Eq.~(\ref{RepKernel}) for $\F(\bar \xi, \xi)$, which provides a definition for the Hilbert space structure on the space of (anti-)holomorphic functions. We then need to solve Eq.~(\ref{PtoQ}) for $V_P(\xi, \xi)$ given $V_Q(\bar \xi, \xi) = V(\bar \xi, \xi)$ and finally we use Eq.~(\ref{QuantizationMap}) to define a Hilbert space operator. This procedure requires the use of the inverse Weierstrass transform which is numerically unstable, which explains why we needed to perform Monte Carlo in momentum space. This effectively amounts to directly computing $\F$ and the $P$ symbol instead of obtaining them by inverting Eqs.~(\ref{RepKernel}) and (\ref{PtoQ}).

In the following, we will first discuss how this can be done to quantize operators for uniform anyon Berry curvature, corresponding to what we discussed in Sec.~\ref{Sec:qhGC} when we only considered the effect of non-uniformity in the potential but not the Berry curvature, then we discuss the more general case of non-uniform anyon Berry curvature.

\subsection{Uniform anyon Berry curvature}
The uniform anyon Berry curvature corresponds to the simple expression for the K\"ahler potential $\K(\bar \omega, \xi) = \frac{1}{2q} \bar \omega \xi$. Then we can see that Eq.~(\ref{RepKernel}) is satisfied by simply choosing $\F(\bar \xi, \xi) = 1$ which corresponds to the standard Bergman reproducing Kernel. Below, we will present an elementary derivation of this relation. Consider an anti-holomorphic function $f(\bar \xi)$ and write
\begin{align}
    \int \frac{d^2 \v{\xi}}{2\pi q} e^{-\frac{1}{2q} \bar \xi \xi} e^{\frac{1}{2q} \bar \omega \xi} f(\bar \xi) &= -\int \frac{d^2 \v{\xi}}{2\pi q} \left[\frac{2q}{\bar \xi - \bar \omega} \partial_{\xi} e^{-\frac{1}{2q} \xi (\bar \xi - \bar \omega)}\right]  f(\bar \xi)
\end{align}
Integrating by parts and using $\partial_{\xi} \frac{1}{\bar \xi - \bar \omega} = \pi \delta(z)$, the integral evaluates to $f(\bar  \omega)$. Now consider a periodic function $V_Q(\bar \xi,\xi)$ with Fourier expansion
\begin{equation}
    V_Q(\bar \xi,\xi) = \sum_\v{\alpha} V_{\v{\alpha}} e^{\frac{1}{2q}(\alpha^* \xi - \alpha \bar \xi)}
\end{equation}
The corresponding $P$ symbol can be obtained by noting that
\begin{equation}
    V_Q(\bar \omega,\omega) =\int \frac{d^2 \v{\xi}}{2\pi q} e^{-\frac{|\xi - \omega|^2}{2q}} V_P(\bar \xi, \xi) = e^{2q \partial_\omega \partial_{\bar \omega}} V_P(\bar \omega, \omega)
\end{equation}
which implies
\begin{equation}
    V_P(\bar \xi, \xi) = e^{-2q \partial_\xi \partial_{\bar \xi}} \sum_\v{\alpha} V_{\v{\alpha}} e^{\frac{1}{2q}(\alpha^* \xi - \alpha \bar \xi)} = \sum_\v{\alpha} V_{\v{\alpha}} e^{\frac{\alpha \alpha^*}{2q}} e^{\frac{1}{2q}(\alpha^* \xi - \alpha \bar \xi)}
\end{equation}
To obtain the quantized operator acting on a function $f$, we use Eq.~(\ref{QuantizationMap}):
\begin{align}
    [\hat V f](\bar \omega) &= \sum_\v{\alpha} V_{\v{\alpha}} e^{\frac{\alpha \alpha^*}{2q}} \int \frac{d^2 \v{\xi}}{2\pi q} e^{-\frac{1}{2q} \bar \xi \xi} e^{\frac{1}{2q}  \xi \bar \omega} e^{\frac{1}{2q}(\alpha^* \xi - \alpha \bar \xi)} f(\bar \xi) \nonumber \\
    &=- \sum_\v{\alpha} V_{\v{\alpha}} e^{\frac{\alpha \alpha^*}{2q}} \int \frac{d^2 \v{\xi}}{2\pi q} \left[\frac{2q}{\bar \xi - \bar \omega - \alpha^*} \partial_{\xi} e^{-\frac{1}{2q} \xi (\bar \xi - \bar \omega - \alpha)}\right] e^{-\frac{1}{2q}\alpha \bar \xi} f(\bar \xi) = \sum_\v{\alpha} V_{\v{\alpha}} e^{-\frac{1}{2q}\alpha \bar \omega} f(\bar \omega + \alpha^*)
\end{align} 
We can see that it does lead to the correct dispersion formula by taking $f_{\v{\kappa}}(\bar \xi) = \bar \sigma(\xi - i q \kappa) e^{\frac{i}{2} \kappa \bar \xi}$ leading to
\begin{equation}
    [\hat V f_{\v{\kappa}}](\bar \omega) = \sum_\v{\alpha} V_{\v{\alpha}} \eta_{\v{\alpha}} e^{-\frac{1}{2q}\alpha \bar \omega} e^{\frac{\alpha}{2q}(\bar \omega + i q \kappa^* + \alpha^*/2)} e^{\frac{i}{2} \kappa \alpha^*} f_{\v{\kappa}}(\bar \omega) = \sum_\v{\alpha} V_{\v{\alpha}} \eta_{\v{\alpha}} e^{\frac{\alpha \alpha^*}{4q}} e^{i \v{\kappa} \cdot \v{\alpha}} f_{\v{\kappa}}(\bar \omega)
\end{equation}

consistent with the result of the main text.

\subsection{Non-uniform magnetic field}
Now we will reproduce the result in the main text for the non-uniform magnetic field. Here, the K\"ahler potential is given by
\begin{equation}
    \K(\bar \xi, \xi) = \frac{1}{2q} \bar \xi \xi + \Q(\bar \xi, \xi)
\end{equation}
To solve for $\F$, we introduce write the Fourier transform
\begin{equation}
    e^{\Q(\bar \xi, \xi)} = \sum_{\v{\alpha}} Q_\v{\alpha} e^{\frac{1}{2q}(\alpha^* \xi - \alpha \bar \xi)}, \qquad \F(\bar \xi, \xi) e^{-\Q(\bar \xi, \xi)} = \sum_{\v{\alpha}} F_\v{\alpha} e^{\frac{1}{2q}(\alpha^* \xi - \alpha \bar \xi)}
\end{equation}
Substituing in the LHS of Eq.~(\ref{RepKernel}), we get
\begin{align}
    \int \frac{d^2 \v{\xi}}{2\pi q} \F(\bar \xi, \xi) e^{-\K(\bar \xi, \xi)} e^{\K(\bar \omega,\xi)} f(\bar \xi) &= \sum_{\v{\alpha},\v{\alpha}'} Q_\v{\alpha} F_{\v{\alpha'}} \int \frac{d^2 \v{\xi}}{2\pi q} e^{-\frac{1}{2q} \bar \xi \xi} e^{\frac{1}{2q}  \xi \bar \omega} e^{\frac{1}{2q}(\alpha^* \xi - \alpha \bar \omega)} e^{\frac{1}{2q}(\alpha'^* \xi - \alpha' \bar \xi)} f(\bar \xi) \nonumber \\
    &= \sum_{\v{\alpha},\v{\alpha}'} Q_\v{\alpha} F_{\v{\alpha'}} e^{-\frac{1}{2q}(\alpha + \alpha') \bar \omega} e^{-\frac{1}{2q} \alpha' (\alpha^* + \alpha'^*)} f(\bar \omega + \alpha^* + \alpha'^*) \nonumber \\
    &= \sum_{\v{\alpha}}  e^{-\frac{1}{2q} \alpha \bar \omega} f(\bar \omega + \alpha^*) \sum_{\v{\alpha}'} Q_{\v{\alpha} - \v{\alpha}'} F_{\v{\alpha'}} e^{-\frac{1}{2q} \alpha' \alpha^*}
\end{align}
In order for this to be equal to $f(\bar \omega)$, the summation over $\v{\alpha}'$ should evaluate to $\delta_{\v{\alpha},0}$. To solve this condition, we write
\begin{equation}
    F_\v{\alpha} = \eta_{\v{\alpha}} e^{\frac{\v{\alpha}^2}{4q}} \tilde F_\v{\alpha}, \qquad Q_\v{\alpha} = \eta_{\v{\alpha}} e^{-\frac{\v{\alpha}^2}{4q}} \tilde Q_\v{\alpha}
\end{equation}
leading to the condition
\begin{equation}
    \sum_{\v{\alpha}'} \tilde Q_{\v{\alpha} - \v{\alpha}'} \tilde F_{\v{\alpha'}} = \delta_{\v{\alpha},0}
\end{equation}
where we used $e^{\frac{i}{2q} \v{\alpha} \wedge \v{\alpha}'} = \eta_\v{\alpha} \eta_\v{\alpha'} \eta_{\v{\alpha} - \v{\alpha}'}$. This condition means that the real space functions corresponding to $\tilde Q$ and $\tilde F$ are inverse to each other. This relation can be written in terms of the Weierstrass transform as
\begin{equation} \label{FQ-eq}
    \F(\v{\xi}) = e^{\Q(\v{\xi})}[W_{\Lambda^*_q}^\eta]^{-1} \frac{1}{[W_{\Lambda^*_q}^\eta]^{-1} e^{\Q(\v{\xi})} } 
\end{equation}
in agreement with the expression in the main text.

Now we would like to derive the dispersion formula for the full action derived in Sec.~\ref{Sec:qhberry}. Let $V_P(\bar\xi,\xi)=\sum_{\v{\alpha}} \tilde V_{P,\v{\alpha}} e^{\frac{1}{2q}(\alpha^*\xi - \alpha \xi^*)}$ and $f_{\v{\kappa}}(\bar \xi) = \bar \sigma(\xi - i q \kappa) e^{\frac{i}{2} \kappa \bar \xi}$, Eq.\eqref{QuantizationMap} becomes

\begin{align}
    [\hat V f_{\v{\kappa}}](\bar\omega) &=  \int \frac{d^2 \v{\xi}}{2\pi q} \F(\bar \xi, \xi) e^{-\Q(\bar \xi, \xi)} e^{\Q(\bar \omega, \xi)} e^{-\frac{\bar\xi\xi}{2q}+\frac{\bar\omega\xi}{2q}}V_P(\bar \xi, \xi) f_{\v{\kappa}}(\bar \xi)
    =\epsilon_{\v\kappa} f_{\v\kappa}(\bar\omega),
\end{align}

where the dispersion $\epsilon_{\v\kappa}$ is 
\begin{align}
    \epsilon_{\v\kappa} &= \sum_{\balpha} w_\balpha e^{i\v\kappa\cdot\balpha},\qquad w_\balpha = \eta_\balpha e^{-\frac{\bar\alpha \alpha}{4q}} \sum_{\balpha_1,\balpha_2} F_{\balpha_1}Q_{\balpha_2}\tilde V_{P,\balpha-\balpha_1-\balpha_2} e^{\frac{\bar\alpha\alpha_2}{2q}}.
\end{align}
The component $\tilde V_{P,\balpha}$ satisfies

\begin{align}
    V_Q(\bar\omega,\omega) &= e^{-\Q(\bar\omega,\omega)-\frac{\bar\omega\omega}{2q}} \int \frac{d^2 \v{\xi}}{2\pi q} \F(\bar \xi, \xi) e^{-\Q(\bar \xi, \xi)} e^{-\frac{\bar\xi\xi}{2q}} e^{Q(\bar\omega,\xi)+\frac{\bar\omega\xi}{2q}}e^{Q(\bar\xi,\omega)+\frac{\bar\xi\omega}{2q}}  V_P(\bar\xi,\xi)\\
    &=e^{-\Q(\bar\omega,\omega)}\sum_{\balpha,\balpha_3} \eta_\balpha e^{-\frac{\bar\alpha\alpha}{4q}}e^{-\frac{\bar\alpha \alpha_3}{2q}} w_{\balpha} Q_{\balpha_3} e^{\frac{\omega}{2q}(\bar\alpha+\bar\alpha_3)-\frac{\bar\omega}{2q}(\alpha+\alpha_3)}\\
    &= e^{-\Q(\bar\omega,\omega)}W_{\Lambda^*_q}^\eta\Big[w(\bar\omega,\omega)\tilde Q(\bar\omega,\omega)\Big],
\end{align}
where $w(\bar\omega,\omega) = \sum_{\balpha} w_\balpha e^{\frac{1}{2q}(\bar\alpha\omega - \bar\omega\alpha)}$.
Therefore, the dispersion is 
\begin{align}
    \epsilon_{\v\kappa} = \frac{{W_{\Lambda^*_q}^\eta}^{-1}\Big[ V_Q(\bar\omega,\omega)e^{\Q(\bar\omega,\omega)} \Big]}{{W_{\Lambda^*_q}^\eta}^{-1} \Big[e^{\Q({\bar\omega,\omega})}\Big] }\Bigg|_{\v\omega = -\wedge q\v\kappa}
\end{align}
which agrees with the result in the main text.

\section{Supplementary numerics}
\subsection{Consistency check}
\label{consistencycheck}
In this section, we present some benchmarks for our Monte Carlo algorithm, recall the geometry: our torus is spanned by $\bL_1,\bL_2$ whose area with $N_\Phi$ flux quanta is denoted as $A=|\bL_1\times \bL_2|=2\pi N_{\Phi}$. Real space lattice $\Lambda$ has unit cell defined by $\ba_1=\bL_1/N_1, \ba_2=\bL_2/N_2$ where $N_1N_2=N_{\Phi}$ such that the unit cell encloses a unit flux. For the consistency check,
 we fix the angle between $\ba_1,\ba_2$ to be $\pi/2$ and let $|\ba_1|=|\ba_2|$ such that the unit cell has unit aspect ratio. We still take $\ell_B=1$ throughout this section.
Correlation in quantum hall liquid is captured by the structure factor:
\begin{align}
      S(\bq)&=\frac{1}{N_e}\langle\delta\rho_{\bq}\delta\rho_{-\bq}\rangle\\
   \delta\rho_{\bq} &=\rho_{\bq}-\langle\rho_{\bq}\rangle
\end{align}
where $\rho_{\bq}=\sum_{i}e^{i\bq\cdot\br_i}$ is the physical density that carries momenta $\bq$.
The interaction Hamiltonian is given as:
\begin{equation}
    \H=\frac{1}{2}\int \frac{d^{2}\bq}{(2\pi)^2} V(\bq):\tilde\rho_{\bq}\tilde\rho_{-\bq}:
\end{equation}
$::$ denotes normal ordering, $\tilde\rho_{\bq}$ is the density operator projected to the LLL Hilbert space.
Since we work with Monte Carlo and take already the fractional quantum Hall wavefunction for sampling, we have
\begin{align}
    \langle\rho_{\bq}\rangle_{\rm MC}&=\langle\tilde\rho_{\bq}\rangle\\
    \langle\rho_{\bq}\rho_{-\bq}\rangle_{\rm MC}&=\langle\tilde\rho_{\bq}\tilde\rho_{-\bq}\rangle
\end{align}
Thus the correlation energy per particle is then related to the physical structure factor via:
\begin{equation}
\begin{split}
    \frac{\langle V\rangle}{N}
    &=\frac{1}{2}\int \frac{d^{2}\bq}{(2\pi)^2}  V(\bq)( S(\bq)-1)
\end{split}
\label{E-Sq}
\end{equation}
Another benchmark test is the guiding center structure factor $S^{GC}(\bq)$, where as mentioned in the main text, $\rho_{\bq}=e^{-\bq^2/4}\rho^{\rm GC}_{\bq}$, we have:
\begin{align}
    S^{\rm GC}(\bq)&=\frac{1}{N_e}\langle\delta\rho^{\rm GC}_{\bq}\delta\rho^{\rm GC}_{-\bq}\rangle\\
    S(\bq)-1&=e^{-\bq^2/2}(S^{\rm GC}(\bq)-1)
\end{align}
For $1/3$ Laughlin state, there is no analytic formula for the physical structure factor, but the guiding center structure factor has a known small-$|\bq|$ expansion.
\begin{equation}
    S^{\rm GC}(|\bq|)=c_{2}|\bq|^2+c_{4}|\bq|^4+c_{6}|\bq|^6+...
\end{equation}
Since we are working in the projected Hilbert space, $c_2$ vanishes for a gapped system. In Laughlin state, one can evaluate $c_4$ and $c_6$ exactly from plasma correlations  \cite{kalinay2000SixthMomentSumRule}.

We then perform MC calculations for a $1/3$ Laughlin state and obtain the guiding center structure factor as well as the Coulomb energy as follows:
\begin{figure}[h]
    \centering
    \includegraphics[width=0.9\linewidth]{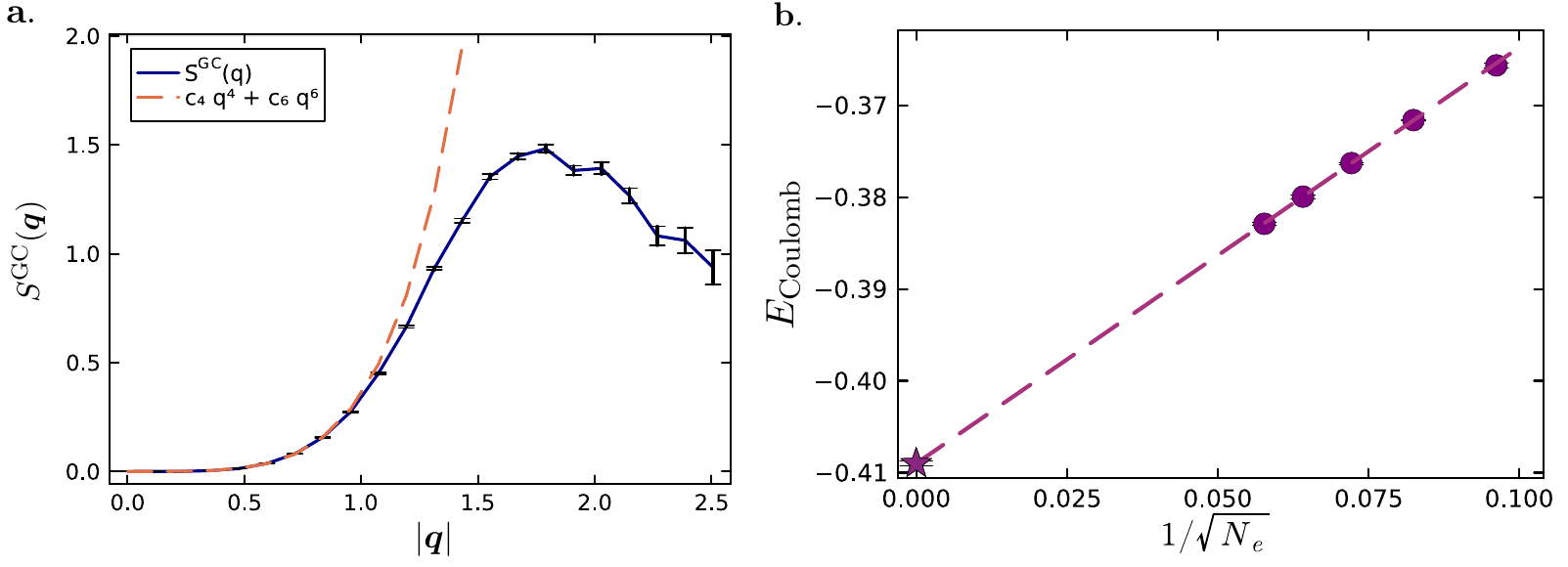}
    \caption{Consistency check with $1/3$ Laughlin state. \textbf{a.} Guiding center structure factor taken for system with $N_e=147$. The dashed line represents analytical small $|\bq|$ expansion. \textbf{b.} Finite size scaling for Coulomb energy calculated from physical structure factor. Thermodynamic value $E_{\rm Coulomb}$ is extracted from scaling.}   
    \label{fig:Sq-E}
\end{figure}

From FIG.~\ref{fig:Sq-E}, we see that the guiding center structure factor agrees nicely with the analytical expansion at small $|\bq|$ region, and from the $1/\sqrt{N_e}$ scaling for Coulomb energy, we get the thermodynamic value: $E_{\rm Coulomb}=-0.4090\pm0.0002 \ \frac{e^2}{l_{B}}$, which is consistent with the thermodynamic value $E_{\rm Coulomb}\approx-0.4100 \ \frac{e^2}{l_{B}}$ from previous studies \cite{levesque1984CrystallizationIncompressibleQuantumfluid,WangLatticeMC}.
By evaluating the guiding center structure factor as well as the Coulomb energy, we conclude that our MC algorithm gives correct result.
\subsection{Finite size scaling and interaction length scaling}
\begin{figure}[h]
    \centering
    \includegraphics[width=0.9\linewidth]{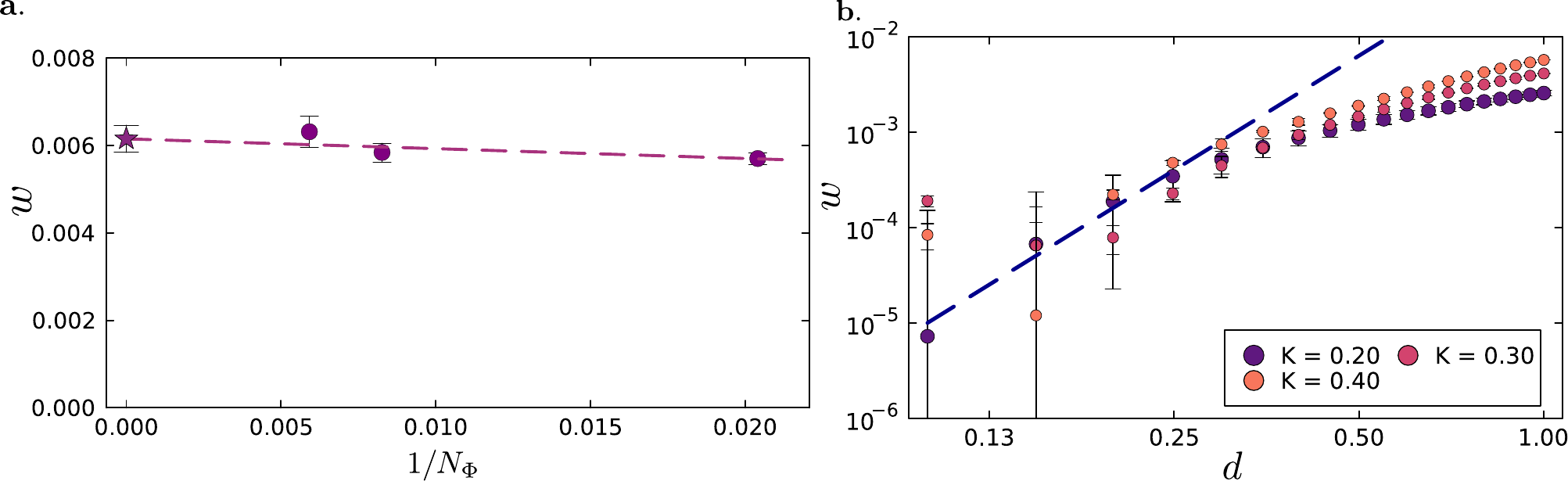}
    \caption{Scaling of the quasihole bandwidth $w$ in first harmonic approximated AC bands with screened Coulomb interaction. \textbf{a}. Finite size scaling for $K=0.4,\ d=1.0$, taking 3 data points $N_e=16,40,56$, correspondingly $N_\Phi=49,121,169$.\textbf{b}. $\log-\log$ plot for $w$ as a function of $d$, the dashed line has a slope of 4, corresponding to the $d^4$ scaling behavior of energy at small $d$.}  
    \label{fig:bandwidthscaling}
\end{figure}
\subsection{Effective magnetic field: check weak field approximation}
As mentioned in the main text Sec.~\ref{Sec:qhberry}, we have been calculating the quasihole quantum geometry in the weak field limit under linearization. 
In practice, the full K\"ahler potential $\R(\bk)$ is not easy to access when the non-uniformity $K$ is not small. This can be understood from the fact that $\Gamma^2=\prod_{l=1}^{N_e}e^{-KQ(\bz_{l})}$ is an $N_e$ body operator which depends on $K$ exponentially. To see this more clearly, we can rewrite $\Gamma$ as 
$\Gamma^2=\prod_{l=1}^{N_e}e^{-KQ(\bz_{l})}=e^{KS}$ 
where $S=\sum_{l=1}^{N_e}Q(\bz_{l})$. The quantity we are evaluating is then $\langle e^{KS}\rangle$. If we assume $S$ can be approximated as a Gaussian with variance $\sigma_{S}^2$. Then the relative variance of the exponential would be given as
\begin{equation}
    \frac{{\rm Var}(\Gamma^2)}{\langle\Gamma^2\rangle^2}=e^{K^2\sigma_{S}^2}-1
\end{equation}
In this simple picture, we see that (i). As long as $K$ is not small enough, there will be exponential amplification of the variance. (ii) $S$ scales with the system size, it will be even more challenging to evaluate the full $\langle\Gamma^2\rangle$ for large system size. Note here we are doing a very naive estimate, numerical methods could be implemented to resolve this issue, but it's beyond the scope of discussion here.

We consider the weak field limit where non-uniformity $K$ is small enough such that we can linearize the problem:
\begin{equation}
    e^{\R(K,\v{\kappa})} \approx 1-K\frac{\int \prod_{i=1}^{N_e} d^2 \v{z}_i  |\psi_\v{\kappa}(\{\bz_i\})|^2\sum_{l=1}^{N_e}  Q(\v{z}_l)}{\int \prod_{i=1}^{N_e} d^2 \v{z}_i |\psi_\v{\kappa}(\{\bz_i\})|^2}
    \label{eq:wkfield}
\end{equation}
And recall that
\begin{equation}
    Q(\bz)=-2 K \sum_{\bb_i}\cos(\bb_{i}\cdot\bz)
\end{equation}
which is just a cosine potential.
Under this approximation, we can first map the K\"ahler potential $\R(K,\v{\kappa})$ defined in the reduced BZ to $\R(K,\v{k})$ defined in the full BZ, which is described only by the first harmonic:
\begin{equation}
    \R(K,\v{k})=Q_{1}\sum_{\ba}e^{i\bk\cdot\ba}
\end{equation}
$\sum_\ba$ sums over the 6 shortest lattice vector in $\Lambda$ and $Q_{1}$ is the corresponding first harmonic amplitude.

We then perform Monte Carlo simulations for a range of system sizes to evaluate Eq.~\ref{eq:wkfield} directly. From the resulting values of $Q_1$
 at different system sizes, we carry out a finite-size scaling analysis to extrapolate to the thermodynamic limit.
\begin{figure}[h]
    \centering
    \includegraphics[width=0.45\linewidth]{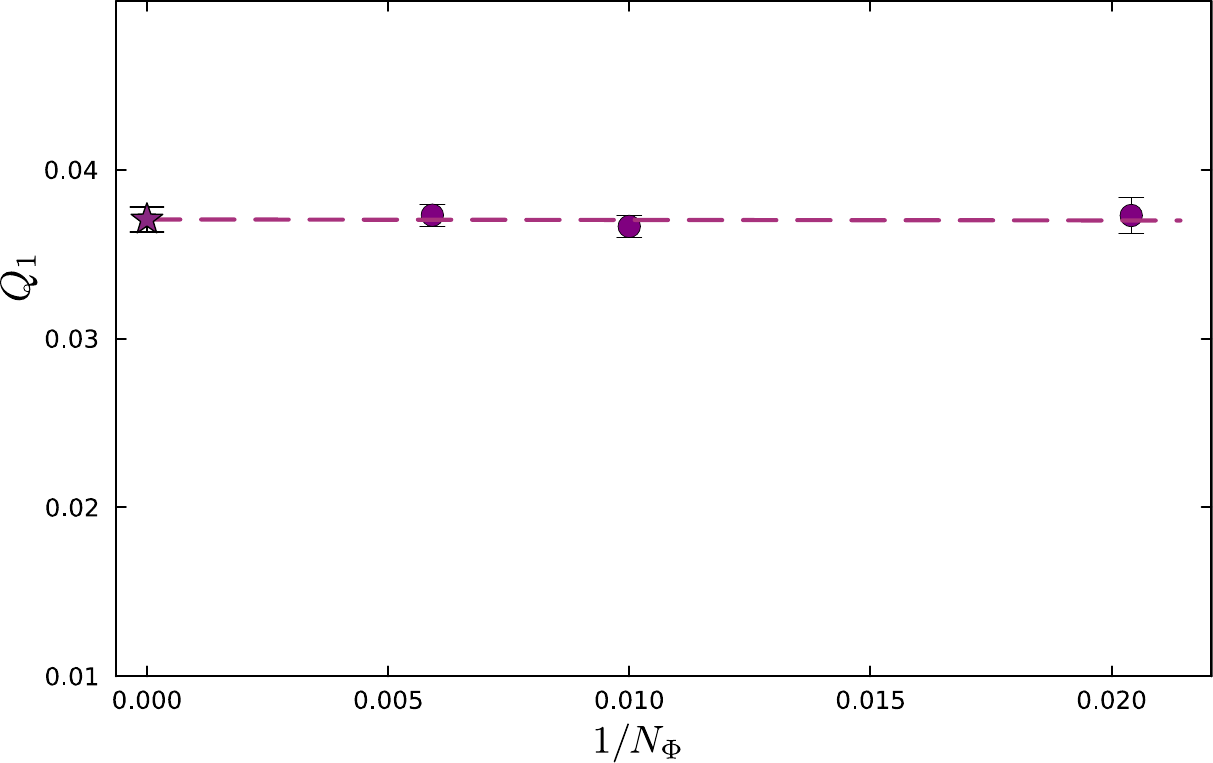}
    \caption{Finite size scaling of the first harmonic $Q_1$, taking 3 data points $N_e=16,40,56$, correspondingly $N_\Phi=49,121,169$}   
    \label{fig:scalingQ1}
\end{figure}

\begin{figure}[h]
    \centering
    \includegraphics[width=0.8\linewidth]{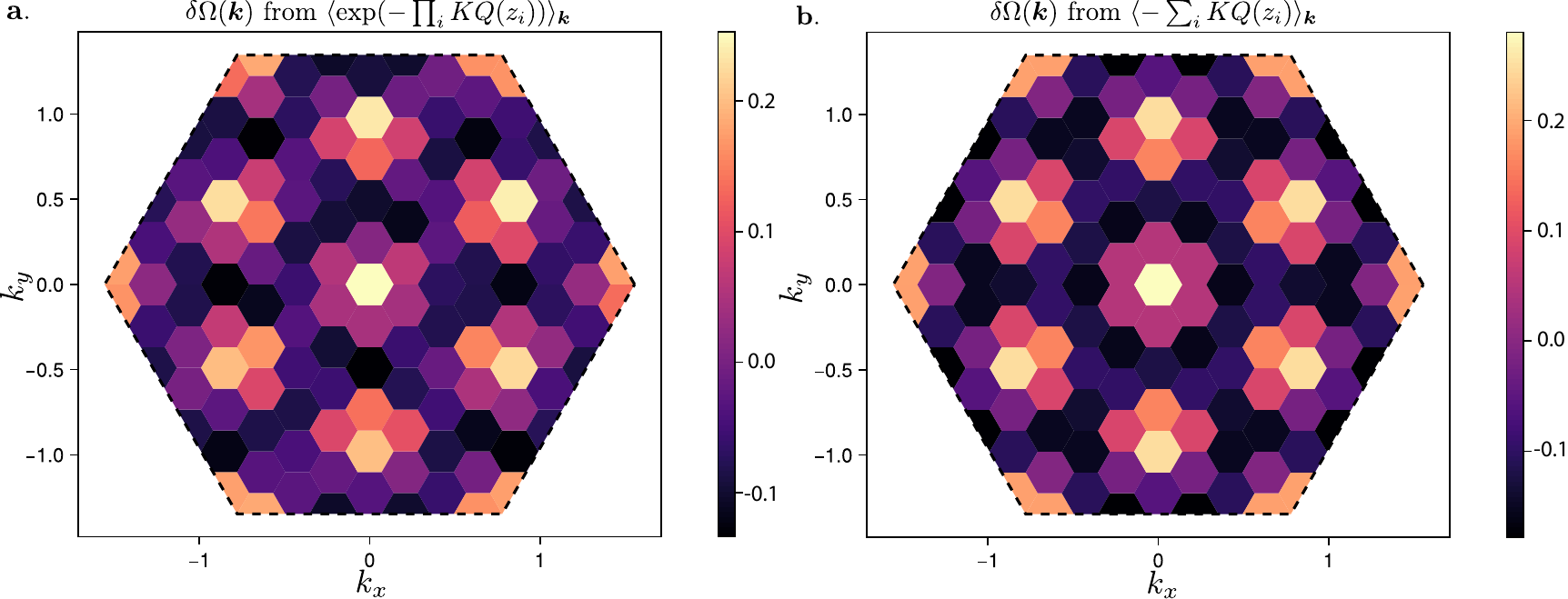}
    \caption{Berry curvature distribution for quasiholes, calculated using $N_e=40,\ K=0.05$.
    \textbf{a-b.} $\delta\Omega(\bk)$ from the full calculation and the weak field approximation respectively.}
    \label{fig:omegakcompare}
\end{figure}

From FIG.~\ref{fig:scalingQ1}, we see that the $Q_1$ dependence on system size is weak which again originates from the fact that quasiholes are local excitations in real space. 
The thermodynamic value of the first harmonic can be extracted as $Q_1= 0.0371 ± 0.0007$ and then used for reconstructing the K\"ahler potential as shown in FIG.~\ref{fig:omegakcompare}.

Now we check the validity of the weak field approximation by directly evaluating the full operator $\Gamma=\prod_{l=1}^{N_e} e^{-K Q(\v{z}_l)}$:
\begin{equation}
    e^{\R^{\rm full}(K,\v{\kappa})} = \frac{\int \prod_{i=1}^{N_e} d^2 \v{z}_i  |\psi_\v{\kappa}(\{\bz_i\})|^2\prod_{l=1}^{N_e} e^{-K Q(\v{z}_l)}}{\int \prod_{i=1}^{N_e} d^2 \v{z}_i |\psi_\v{\kappa}(\{\bz_i\})|^2}
\end{equation}
From the weak field result $\R(k,\bk)$ and the full result $\R^{\rm full}(K,\bk)$, we can evaluate the Berry curvature distribution $\delta\Omega(\bk)$ (uniform part neglected), respectively.

We see from FIG.~\ref{fig:omegakcompare} that at small $K$, there's good agreement between the weak field and the full operator results, thus validating the usage of weak field approximation at small $K$.
\subsection{Real space inversion}
As mentioned in the main text, in order to evaluate the expectation value of an operator in momentum space, we only need the real space information. However, due to the growing Gaussian factor, the numerical noise complicates this procedure. In this section, we discuss some numerical attempts to carry out the real space inversion procedure from Eq.~(\ref{Oxi}).
\begin{equation}
    \O_\v{\xi} = \sum_{\v{\kappa}} |F_{\v{\kappa}}(\v{\xi})|^2 \O_\v{\kappa}
\end{equation}
Their Fourier expansion can be written as:
\begin{equation}
    \O_{\bxi} := \langle \psi_\bxi|\hat \O|\psi_\bxi \rangle = \sum_\ba \O_\ba e^{-i \ba \wedge \bxi}, \qquad \O_{\v{\kappa}} := \langle \psi_{\v{\kappa}}|\hat \O|\psi_{\v{\kappa}} \rangle = \sum_{\v{\alpha}} \O_{\v{\alpha}} e^{-i \v{\alpha} \cdot \v{\kappa}}
\end{equation}

By making use of Eq.~(\ref{Fsquared}), we have the relation between the Fourier modes:
\begin{equation}
    \O_{\v{\alpha}} =  \eta_{\v{\alpha}}\sum_\ba \delta_{\ba, \v{\alpha}/q}  e^{q\frac{\ba^2}{4}} \O_\ba
    \label{OaOkappa}
\end{equation}
So we can start with $\O_\v{\xi}$ to obtain all the Fourier modes $\O_{\ba}$, by sampling different quasihole location $\xi$ in the unit cell. Then we obtain $\O_{\v{\alpha}}$ and can thus reconstruct $\O_{\v{\kappa}}$ .
From Eq.~(\ref{OaOkappa}), we see that when going to momentum space, all the Fourier modes $\O_{\ba}$ from the real space get amplified by the exponential $e^{q\frac{\ba^2}{4}}$, which is one of the origins of the numerical instability.

For simplicity, we work in the weak field limit and consider calculating the quasihole K\"ahler potential $\R(K,\bk)$ for now.
\begin{equation}
    O=K\sum_{l=1}^{N_e}\sum_{i=1}^{3}\cos(\bb_{i}\cdot\br_{l})
\end{equation}
Quasiholes are local excitations at position $\bxi$, we can shift the variable in the integrand and move the $\bxi$ dependence to the cosine potential: 
\begin{equation}
    \langle \psi_\xi|O|\psi_\xi\rangle=K\int d^2\br \rho(\br-\bxi)\sum_{i=1}^{3}\cos(\bb_{i}\cdot\br)=K\int d^2\br \rho(\br)\sum_{i=1}^{3}\cos(\bb_{i}\cdot(\br+\bxi))
\end{equation}
We see that in this limit, quasihole density is the only input we need from MC such that one needs to increase the precision of $\rho(\br)$ as much as possible.
To extract the Fourier modes, we make use of the radial symmetry in the uniform case:
\begin{align}
    \sum_{\ba}O_{\ba}e^{-i\ba\wedge \bxi}&=\int d^{2}\br K\sum_{\pm\bb}e^{i\bb\cdot(\br+\bxi)}\rho(\br)\\
    O_{\ba}=O_{\wedge \bb}&=K\int d^{2}\br e^{i\bb\cdot\br}\rho(\br)=K\rho_{\bb}\\
    \rho_{\bb}&=2\pi\int d r rJ_0(|\bb|r)\rho(r)
\end{align}
where $J_0(r)$ is the zeroth Bessel functions of the first kind.
In the thermodynamic limit, integral $\int dr r J_0(|\bb|r)$ gives a delta function at $\bb=0$, such that the uniform background does not contribute to the Fourier mode, but in finite size system, the integral $\int dr r J_0(|\bb|r)$ is usually not zero. Thus we subtract the uniform background explicitly:
\begin{gather}
    \delta\rho(\br)=\rho(\br)-\bar\rho\\
    \delta\rho_{\bb}=\rho_{\bb}-2\pi\bar\rho \int d r rJ_0(|\bb|r)=2\pi\int d r rJ_0(|\bb|r)\delta\rho(r)\\
    O_{\v{\alpha}}= K\eta_{\v{\alpha}}\sum_\ba \delta_{\ba, \v{\alpha}/q}  e^{q\frac{\ba^2}{4}}\delta_{\ba,\wedge\bb}\delta\rho_{\bb}
\end{gather}
From this procedure, we already see that:
(i). As mentioned before, the Gaussian factor $e^{q\frac{\ba^2}{4}}$ amplifies noise at large momentum (ii). The Bessel function part $rJ_0(r)$ is highly oscillating and approaches $\sqrt{r}$ as $r\to\infty$. This means the integration is very sensitive to the error of the charge density $\rho(r)$ at large $r$.

The charge density of a quasihole can be obtained from MC evaluation and shown below:
\begin{figure}[h]
    \centering
    \includegraphics[width=0.9\linewidth]{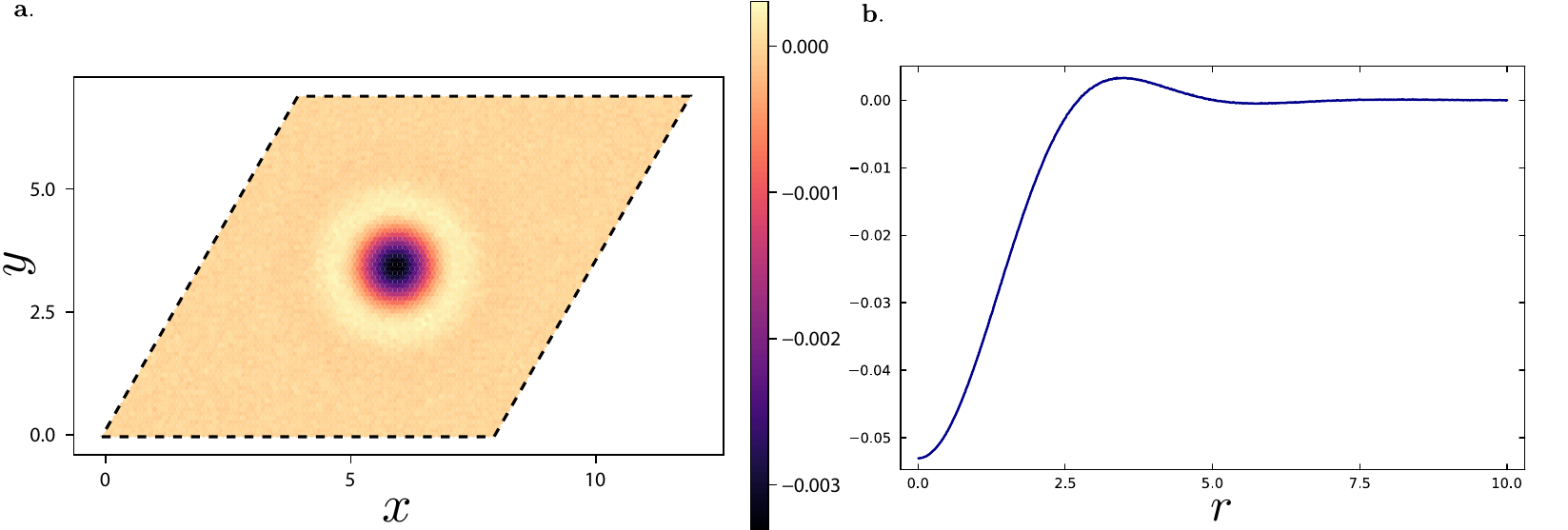}
    \caption{\textbf{a}. Charge density profile of a quasihole $\delta\rho(\br)$ at the center of torus. \textbf{b}. Charge density as a function of radius $r$}   
    \label{fig:rhor}
\end{figure}
We see from FIG.~\ref{fig:rhor} that the charge density $\delta\rho(\br)$ looks smooth and decays to zero within several $\ell_B$, which tells us the size of a quasihole is indeed several $\ell_B$. At large $\br$, the quasihole charge density decays as a Gaussian:
\begin{equation}
    \delta\rho(\br)\propto f(r)e^{-\frac{r^2}{2q}}
\end{equation}
And the integral with Bessel function is very sensitive to the $f(r)$ part.
Thus, in order to estimate the level of noise we have in $\delta\rho(\br)$, we can add the growing Gaussian $e^{\frac{r^2}{2q}}$ to obtain $f(r)$ and determine a cutoff in radius $r_c$:
\begin{figure}[h]
    \centering
    \includegraphics[width=0.9\linewidth]{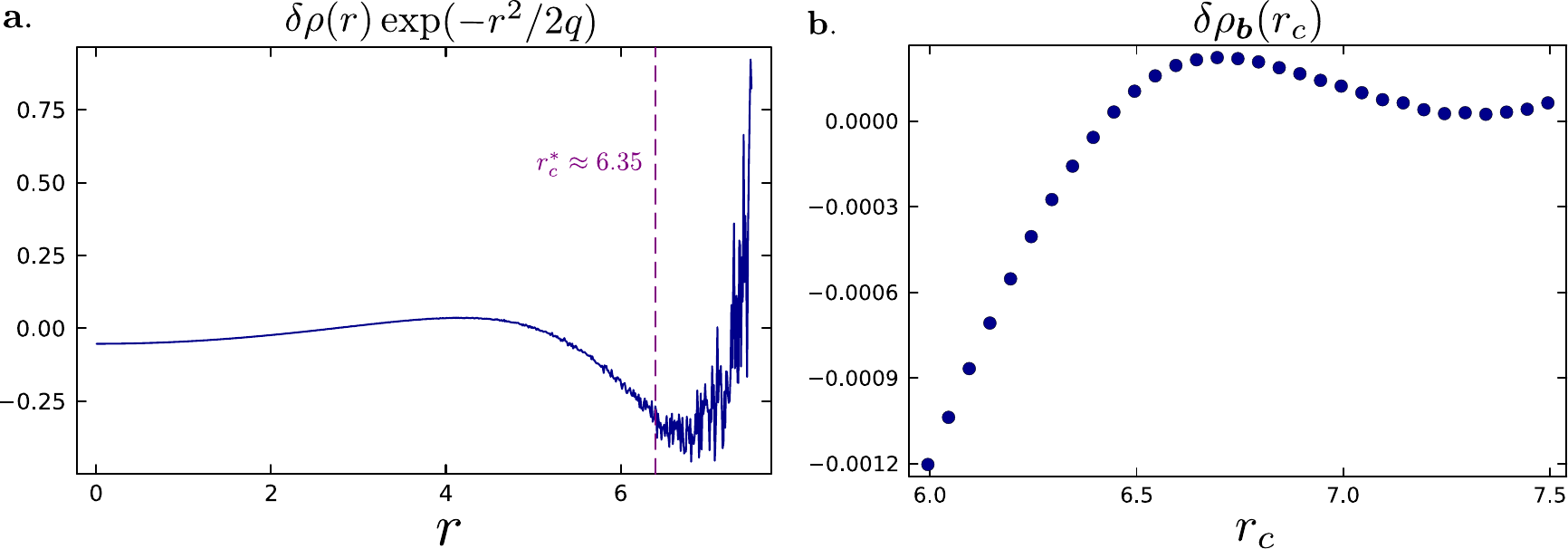}
    \caption{\textbf{a}. Charge density profile of a quasihole by removing the Gaussian decay part: $\delta\rho(\br)e^{\frac{r^2}{2q}}$. Purple vertical line signals roughly where noise starts to dominate. \textbf{b}. The integrated Fourier mode $\delta\rho_{\bb}$ as a function of the cutoff radius $r_c$.}   
    \label{fig:rhor2}
\end{figure}
We see from FIG.~\ref{fig:rhor2} (a). that the $f(r)$ starts smooth at small $r$ and becomes increasingly noisy at large $r$ until one loses the signal entirely. We take the radius $r_c^*\approx6.35$ as a rough estimate for where the noise starts to dominate. The corresponding Fourier mode can be estimated from numerical integration with cutoff radius $r_c^{*}$:
\begin{equation}
    \delta\rho_{\bb}(r_{c}^*)=-1.579\times10^{-4}\Rightarrow O_{\v{\alpha}}^{*}=0.03645
\end{equation}

Compared to the value we extracted from fitting weak field result: $O_{\v{\alpha}}=0.0371 ± 0.0006$, $O_{\v{\alpha}}^{*}$ is very close. However, we can see from FIG.~\ref{fig:rhor2} (b). that because the $rJ_0(|\bb|r)$ is highly oscillating, the integral value changes \emph{a lot} as a function of the cutoff radius $r_c$. So the value extracted $O_{\v{\alpha}}^{*}$ can be quite misleading. 

We see that the real space inversion result has relatively large noise, reasons are the following:
(i). The highly oscillating function $rJ_0(|\bb|r)$ in the integrand makes the numerical integration extremely unstable.
(ii). The Fourier mode in real space $\delta\rho_{\bb}$ is in magnitude small but corresponds to relatively large momentum $\bb=\bb_i$, where $e^{q\bb^2/4}\approx231$ gives a magnification in noise. 

To resolve the numerical integration instability, one can in principle choose an ansatz to expand the charge density profile in a set of basis functions $\rho(\br)=\sum_{i}c_{i}\rho_{i}(\br)$ and perform analytical integral for each term. However, the instability still emerges from the choice of the ansatz. Unless we know exactly the correct basis for quasihole density profile, extracting the Fourier mode remains elusive.

\begin{figure}[h]
    \centering
    \includegraphics[width=0.5\linewidth]{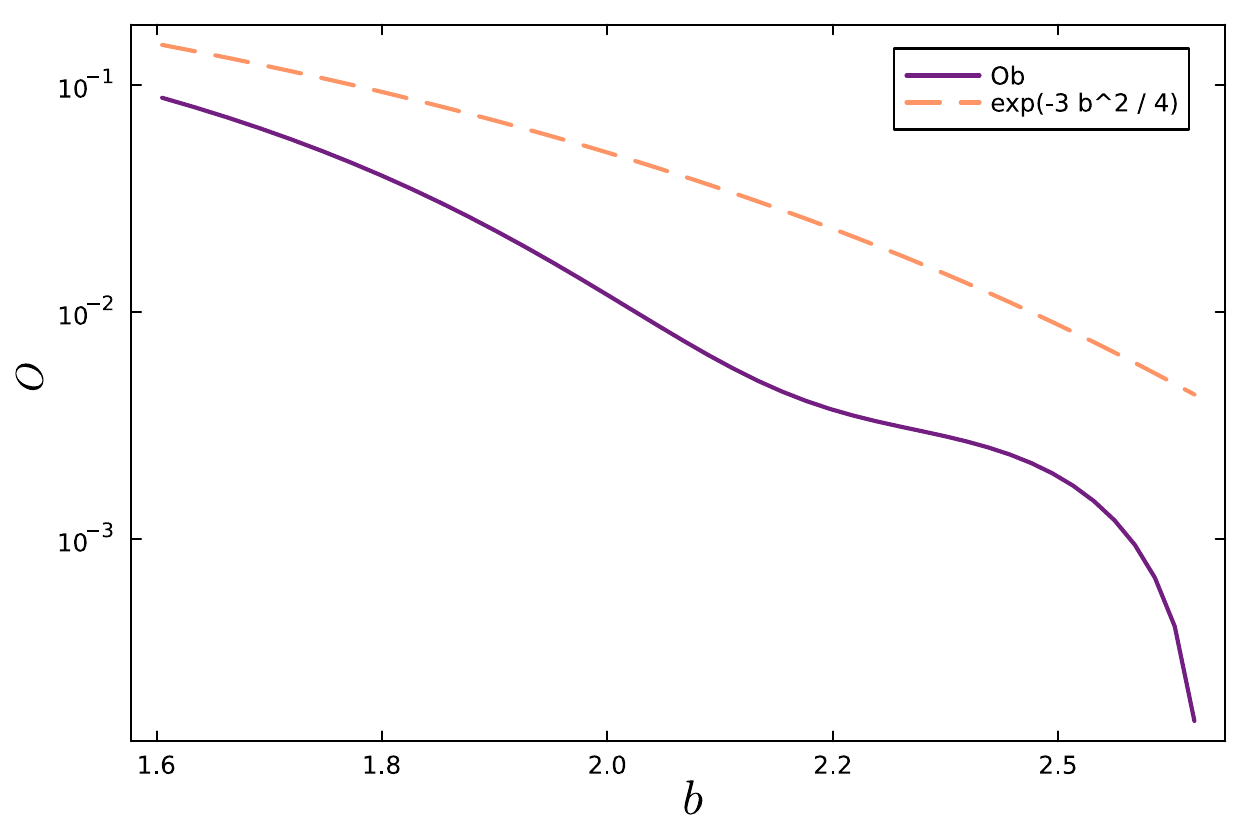}
    \caption{Comparison between the real space mode of charge density $\delta\rho_{\bb}$ and $\exp(-\frac{3}{4}b^2)$} plotted in log-log scale as a function of $\log(b)$.   
    \label{fig:rhorcheck}
\end{figure}
Apart from the difficulty in determining the momentum space results from real space information, we examine the conditions this inversion relation holds: it's clear from Eq.~(\ref{OaOkappa}) that the real space modes $\O_{\bb}$ has to decay faster than $\exp(-qb^2/4)$. While in principle the behavior of the quasihole density can be calculated from correlations of the two dimensional one component plasma, we have not arrived at an exact analytical expression yet. Thus we check this assumption numerically by calculating the real space modes of the density $\delta\rho_{\bb}$ as a function of $\bb$ and look at its decaying behavior.
From FIG.~\ref{fig:rhorcheck}, we see that $\delta\rho_{\bb}$ indeed decays faster than $\exp(-\frac{3}{4}b^2)$ which ensures the validity of the Fourier transform.

From these numerical attempts, we can see that the inversion process is in principle feasible but in practice needs substantially more work for the specific problem of the dispersion. On the other hand, if the momentum appeared in the problem is small instead of reciprocal lattice momentum, then the noise issue should be significantly reduced.

\end{appendix}
\end{document}